\documentclass[aps,twocolumn,showpacs,groupedaddress,nofootinbib]{revtex4-1}
\usepackage{epsfig}
\usepackage{bm}
\usepackage{amsmath,amssymb,times}
\usepackage{url}

\usepackage[colorlinks=true, pdfstartview=FitV, linkcolor=red, citecolor=blue, urlcolor=blue]{hyperref}

\newcommand{\UNIT}[1]{\mbox{$\,{\rm #1}$}}
\newcommand{\MeV}{\UNIT{MeV}}
\newcommand{\GeV}{\UNIT{GeV}}

\newcommand{\fm}{\UNIT{fm}}

\newcommand{\ds}{\displaystyle}
\newcommand{\zerovec}{\vec{0}\,}

\newcommand{\pvec}{\vec{p}\,}

\newcommand{\Gcapvec}{\vec{G}\,}

\newcommand{\Picapvec}{\vec{\Pi}\,}

\newcommand{\sigmavec}{\vec{\sigma}\,}

\newcommand{\tauvec}{\vec{\tau}\,}
\newcommand{\rhovec}{\vec{\rho}\,}
\newcommand{\deltavec}{\vec{\delta}\,}

\newcommand{\be}{\begin{equation}}
\newcommand{\ee}{\end{equation}}
\newcommand{\ba}{\begin{eqnarray}}
\newcommand{\ea}{\end{eqnarray}}

\newcommand{\derivl}{\stackrel{\leftarrow}{\partial}}
\newcommand{\derivr}{\stackrel{\rightarrow}{\partial}}

\newcommand{\fac}{\frac{\kappa}{(2\pi)^{3}}}

\newcommand{\nld}{{\cal D}}

\newcommand{\nldl}{\overleftarrow{{\cal D}}}
\newcommand{\nldr}{\overrightarrow{{\cal D}}}

\newcommand{\calol}{\overleftarrow{\varOmega}}
\newcommand{\calor}{\overrightarrow{\varOmega}}

\newcommand{\pspace}{\int\limits_{|\pvec|\leq p_{F_{i}}}\!\!\!\!\!\! d^{3}p}

\newcommand{\partialr}{\overrightarrow{\partial}}
\newcommand{\partiall}{\overleftarrow{\partial}}

\newcommand{\xil}{\overleftarrow{\xi}}
\newcommand{\xir}{\overrightarrow{\xi}}

\newcommand{\formb}{\frac{1}{1+\sum_{j=1}^{4}\left(\zeta_{j}^{\alpha} \, i\partialr_{\alpha}\right)^{2}}}
\newcommand{\formbnm}{\frac{\Lambda^2}{ \Lambda^2+\vec{p}^{\,2}}}

\newcommand{\Bla}{\Big<}
\newcommand{\Bra}{\Big>}

\begin{document}

\preprint{02-02}

\title{Momentum dependent mean-field dynamics of compressed nuclear matter and neutron stars } 
\author{Theodoros Gaitanos}
\email{theodoros.gaitanos@theo.physik.uni-giessen.de}
\author{Murat M. Kaskulov}
\email{murat.kaskulov@theo.physik.uni-giessen.de}
\affiliation{Institut f\"ur Theoretische Physik, Universit\"at Giessen,
             D-35392 Giessen, Germany}
\date{\today}

\begin{abstract}
Nuclear matter and compact neutron stars are
studied in the framework of the non-linear derivative (NLD) model which accounts for
the momentum dependence of relativistic mean-fields. The
generalized form of the energy-momentum tensor is derived which allows to
consider different forms of the regulator functions in the NLD 
Lagrangian. The thermodynamic consistency of the NLD model is demonstrated for
arbitrary choice of the regulator functions. 
The NLD approach describes the bulk properties of the nuclear matter and 
compares well with microscopic calculations and Dirac phenomenology. 
We further study the high density domain of the 
nuclear equation of state (EoS) relevant for the  matter in $\beta$-equilibrium 
inside neutron stars. It is shown that the low density constraints imposed
on the nuclear EoS and by the momentum dependence of the
Schr\"odinger-equivalent optical potential lead to a maximum mass of the 
neutron stars around $M \simeq 2 M_{\odot}$ which accommodates the  observed 
mass of the J1614-2230 millisecond radio pulsar.
\end{abstract}
\maketitle


\section{\label{sec1}Introduction}

Relativistic hadrodynamics (RHD) of interacting nucleons and mesons provide 
a simple and successful tool for the theoretical description of different 
nuclear systems such as nuclear matter, finite nuclei,  heavy-ion collisions 
and compact neutron stars~\cite{Serot:1984ey}. Starting from the pioneering work of 
Duerr~\cite{Duerr:1956zz}, 
simple RHD Lagrangians have been introduced~\cite{Walecka:1974qa,Serot:1997xg} and since 
then many different extensions of RHD approach, which 
rely on relativistic mean-field (RMF) approximation, have been developed. They describe
the saturation mechanism in nuclear matter and generate a natural 
mechanism for the strong spin-orbit force in nuclei. 
An energy dependence of the Schr\"{o}dinger-equivalent optical
potential~\cite{Cooper:1993nx,Hama:1990vr} is thereby included 
as a consequence of a relativistic description. 
However, when using the standard RHD Lagrangian in RMF
approximation, the nucleon selfenergies become simple functions 
of density only, and do not  depend explicitly on momentum 
of the nucleon. As a consequence a linear energy dependence  of the
Schr\"{o}dinger-equivalent optical potential with a divergent behavior 
at high energies arises~\cite{Weber:1992qc}. This well-known feature contradicts Dirac
phenomenology~\cite{Cooper:1993nx,Hama:1990vr,Typel2002299}. 

To solve this issue one may go beyond the 
mean-field approximation in a quantum field theoretical framework 
by a systematic diagrammatic expansion of nucleon selfenergies. For instance, 
in Dirac-Brueckner-Hartree-Fock
(DBHF)~\cite{Haar:1986ii,Brockmann:1996xy,Muther2000243} calculations 
the nucleon selfenergies indeed depend on both the density and single 
particle momentum. They reproduce the empirical saturation point of 
nuclear matter as well as the energy dependence of the optical potential 
at low energies. However, the DBHF approach has its apparent limitations at 
high energies and densities relevant, for instance, in heavy-ion collisions where its 
application within transport theory turns out to be 
intricate~\cite{Botermans1990115,Buss:2011mx}. Also the thermodynamic
consistency of the DBHF calculations is not obvious~\cite{Hugenholtz:1958}.

As an alternative approach to {\it ab-initio} DBHF calculations for the nuclear 
many-body systems a phenomenological treatment of the problem in the spirit of 
the RMF approximation is still considered as a powerful tool. However, the simple Lagrangian 
of RHD~\cite{Walecka:1974qa,Serot:1997xg} has to be further modified for a 
quantitative description of static nuclear systems such as nuclear matter and/or finite
nuclei. Therefore, it is mandatory to introduce new terms, {\it e.g.}, including 
non-linear self interactions of the scalar~\cite{Boguta:1977xi} and/or 
vector~\cite{Sugahara1994557} meson fields, or to modify existing contributions 
in the Lagrangian, {\it e.g.}, introducing density dependent meson-nucleon 
couplings~\cite{PhysRevLett.68.3408,Fuchs:1995as,Typel:1999yq}. 
The model parameters have to be then fitted to properties 
of nuclear matter and/or atomic nuclei, since, they cannot be derived in a simple manner 
from a microscopic description. 

The momentum dependence of in-medium interactions becomes particularly important in description of 
nuclear collision dynamics such as heavy-ion collisions. Indeed, analyses of proton-nucleus scattering 
data \cite{Cooper:1993nx,Hama:1990vr} show that the proton-nucleus optical potential starts to 
level off already at incident energies of about $300$ MeV. 
Thus, other RMF approaches have been developed by including additional 
non-local contributions, {\it i.e.}, by introducing Fock-terms, on the level of 
the RMF selfenergies leading to a density and momentum dependent
interactions~\cite{Weber:1992qc}. However, such a 
treatment is not covariant and also its numerical 
realization in actual transport calculations is rather difficult~\cite{Weber:1992qc}. 
Another approach has been proposed in~\cite{Zimanyi:1990np} 
and more recently in~\cite{Typel:2002ck,Typel:2005ba} by 
introducing higher order derivative couplings in the Lagrangian of RHD. 
In Ref.~\cite{Zimanyi:1990np} such gradient terms have been studied with the conclusion 
of a softening of the nuclear EoS. In another study of Ref.~\cite{Typel:2005ba} 
both the density dependence of the 
nuclear EoS and the energy dependence of the optical potential have been investigated. While
the modified interactions of meson fields with nucleons explain the 
empirical energy dependence of the optical potential, a stiff EoS at high
densities results from an introduction of an explicit density dependence of the nucleon-meson
couplings with additional parameters. The impact of momentum
dependent RMF models on nuclear matter bulk properties and
particularly on the high density domain of EoS relevant for neutron stars is
presently less understood.

The purpose of the present work is to develop a relativistic and
thermodynamically consistent RMF model, which provides the correct 
momentum dependence of the nucleon selfenergies and agrees well with available
empirical information on nuclear matter ground state,
in a self consistent Lagrangian framework. 
Some steps in this direction have been already done in 
Refs.~\cite{Gaitanos:2011yb,Gaitanos:2011ej,Gaitanos:2009nt} 
where the concept of non-linear derivative meson-nucleon Lagrangian
has been introduced. However, the calculations of 
Refs.~\cite{Gaitanos:2011yb,Gaitanos:2011ej,Gaitanos:2009nt} were based 
on a particular exponential form of the regulators in the RHD Lagrangian and a 
detailed study of nuclear matter ground state properties has not been done. 
In the present work the generalized form of the energy-momentum tensor in the 
NLD model is derived and allows to consider different regulator functions in the 
Lagrangian. The thermodynamic consistency of the NLD model is demonstrated for
arbitrary choice of the regulators. A thorough study of the
properties of nuclear matter around saturation density is further
performed. The model describes 
the bulk properties of the nuclear matter and 
compares well with microscopic calculations and Dirac phenomenology. 
We also investigate the high density region of the 
NLD EoS relevant for the  neutron stars. 
It is found that the low density constraints imposed
on the nuclear matter EoS and by the momentum dependence of the
Schr\"odinger-equivalent optical potential lead to a maximum mass of the 
neutron stars around $M \simeq 2 M_{\odot}$. It is demonstrated 
that the high density pressure-density diagram as extracted from
astrophysical measurements~\cite{Ozel:2010fw,Steiner:2010fz} 
can be well described with nucleonic degrees of freedom only.

\section{\label{sec2}Field theory with higher derivatives}

The non-linear derivative (NLD) model is based on a field-theoretical
formalism which accounts for the higher-order derivative interactions in the RHD Lagrangian.
As a consequence, the conventional RHD mean-field theory based on minimal interaction Lagrangians 
has to be extended to the case of higher-order non-linear derivative functionals. 

For that purpose 
we consider the most general structure of a Lagrangian density $\mathcal{L}$
with higher-order field derivatives, {\it i.e.}
\begin{align}
{\cal L}\left( 
\varphi_{r}(x), \, \partial_{\alpha_{1}}\varphi_{r}(x), 
\, \partial_{\alpha_{1}\alpha_{2}}\varphi_{r}(x),
\cdots\!, 
\partial_{\alpha_{1}\cdots\alpha_{n}}\varphi_{r}(x)
\right)
\,,
\label{EL_0}
\end{align}
where it is supposed that $\mathcal{L}$ has continuous derivatives up to 
order $n$ with respect to all its arguments, that is
\be
\partial_{\alpha_{1}\cdots\alpha_{n}}\varphi_{r}(x) \equiv \frac{\partial}{\partial
x^{\alpha_1}}\cdots \frac{\partial}{\partial
x^{\alpha_n}} \varphi_{r}(x) \equiv
\partial_{\alpha_{1}}\cdots\partial_{\alpha_{n}}\varphi_{r}(x)\,, \nonumber  
\ee
where $\alpha_i$ is a four index and $x$ denotes the coordinates 
in Minkowski space. 
The order $n$ can be a finite number or $n\rightarrow \infty$. 
The subscript $r$ denotes different fields, for instance, 
in the case of the spinor fields one would have $\varphi_{1}=\Psi$ 
and $\varphi_{2}=\overline{\Psi}$. 

The derivation of the generalized Euler-Lagrange equations of motion follows from the 
variation principle for the action $S=\int d^4 x {\cal L}(x)$ with the Lagrangian of 
Eq.~(\ref{EL_0}), where one considers $\varphi_{r}$, 
$\partial_{\alpha_{1}}\varphi_{r}$, 
$\partial_{\alpha_{1}}\partial_{\alpha_{2}}\varphi_{r}$, $\cdots$, 
$\partial_{\alpha_{1}}\!\!\cdots\partial_{\alpha_{n}}\varphi_{r}$ 
as independent generalized coordinates. 
The  Euler-Lagrange equations are obtained from principle of least action
\begin{align}
\delta S=0
\,,
\end{align}
where $\delta S$ is given by
\begin{align}
\delta S = \int d^{4}x \,
\delta{\cal L}\left( 
\varphi_{r}, \, \partial_{\alpha_{1}}\varphi_{r}, 
\, \partial_{\alpha_{1}\alpha_{2}}\varphi_{r}, 
\cdots\!,
\, \partial_{\alpha_{1}\cdots\alpha_{n}}\varphi_{r}
\right)
\label{ELa}
\end{align}
and is obtained by the variation of the generalized coordinates 
\begin{align}
\varphi_{r} & \longrightarrow \varphi_{r} + \delta\,\varphi_{r} ,
\nonumber\\
\partial_{\alpha_{1}}\varphi_{r} & \longrightarrow 
\partial_{\alpha_{1}}\varphi_{r} + \delta\,\partial_{\alpha_{1}}\varphi_{r} ,
\nonumber\\
\partial_{\alpha_{1}\alpha_{2}}\varphi_{r} & \longrightarrow 
\partial_{\alpha_{1}\alpha_{2}}\varphi_{r} + 
\delta\,\partial_{\alpha_{1}\alpha_{2}}\varphi_{r} ,
\nonumber\\
& ,\cdots ,
\nonumber\\
\partial_{\alpha_{1}\cdots\alpha_{n}}\varphi_{r} & \longrightarrow 
\partial_{\alpha_{1}\cdots\alpha_{n}}\varphi_{r} + 
\delta \, \partial_{\alpha_{1}\cdots\alpha_{n}} \varphi_{r} ,
\label{ELc}
\end{align}
with vanishing contributions on the surface of the integration volume as the 
boundary condition. 
The variation of the Lagrangian density  with respect to all degrees of freedom reads
\begin{align}
\delta{\cal L} = & \!\!
\left[
\frac{\partial {\cal L}}{\partial\varphi_{r}}\delta\varphi_{r} 
+ \!
\frac{\partial {\cal L}}{\partial(\partial_{\alpha_{1}}\varphi_{r})}
\partial_{\alpha_{1}}\delta\varphi_{r}
+\!
\frac{\partial {\cal L}}{\partial(\partial_{\alpha_{1}\alpha_{2}}\varphi_{r})}
\partial_{\alpha_{1}\alpha_{2}}\delta\varphi_{r}
\right.
\nonumber\\
&
\left.
+ \cdots
+\frac{\partial {\cal L}}{\partial(\partial_{\alpha_{1}\cdots\alpha_{n}}\varphi_{r})}
\partial_{\alpha_{1}\cdots\alpha_{n}}\delta\varphi_{r}
\right]
\,.
\label{ELd}
\end{align}
As a next step one inserts Eq.~(\ref{ELd}) into Eq.~(\ref{ELa}) and then
performs successively partial integrations, {\it e.g.}, one partial integration for the 
second term in Eq.~(\ref{ELd}), two partial integrations for the third term 
in Eq.~(\ref{ELd}), and $n$ partial integrations for the last term. This 
procedure results in to the following integrand in Eq.~(\ref{ELa})
\begin{align}
\delta{\cal L} = & \!\!
\left[
\frac{\partial {\cal L}}{\partial\varphi_{r}}
- \partial_{\alpha_{1}}
\frac{\partial {\cal L}}{\partial(\partial_{\alpha_{1}}\varphi_{r})}
+\partial_{\alpha_{1}\alpha_{2}}
\frac{\partial {\cal L}}{\partial(\partial_{\alpha_{1}\alpha_{2}}\varphi_{r})}
\right.
\nonumber\\
&
\left.
+ \cdots
+(-)^{n}\partial_{\alpha_{1}\cdots\alpha_{n}}
\frac{\partial {\cal L}}{\partial(\partial_{\alpha_{1}\cdots\alpha_{n}}\varphi_{r})}
\right] \delta\varphi_{r}
\label{ELd2}
\end{align}
up to $4$-divergence terms, which by Gauss law do not contribute to the action 
in Eq.~(\ref{ELa}). Thus, one 
arrives to the following generalized Euler-Lagrange equation 
\begin{align}
\frac{\partial{\cal L}}{\partial\varphi_{r}}
+
\sum_{i=1}^{n} 
(-)^{i}
\partial_{\alpha_{1}\cdots\alpha_{i}}
\frac{\partial{\cal L}}
{\partial(\partial_{\alpha_{1}\cdots\alpha_{i}}\varphi_{r})}
= 0
\;.
\label{Euler0}
\end{align}

The Noether theorem follows from invariance 
principles of the Lagrangian density, Eq.~(\ref{EL_0}), with respect to 
infinitesimal variations of the generalized coordinates 
and their argument $x^{\mu}$ (see for notations Appendix~\ref{app1}). As further shown in 
Appendix~\ref{app2}, the requirement of invariance of the Lagrangian density, 
Eq.~(\ref{EL_0}), with respect to global phase transformations 
\begin{align}
\varphi_{r}(x) \longrightarrow 
\varphi^{\prime}_{r}(x)=e^{-i\epsilon}\varphi_{r}(x)
\label{phaseTrafo}
\end{align}
leads to a continuity equation $\partial_{\mu}J^{\mu}=0$ for a conserved
Noether current $J^{\mu}$. The latter is given by the following expression
\begin{widetext}
\begin{align}
J^{\mu} = -i\left[ 
  {\cal K}^{\mu}_{r}\varphi_{r} 
+ {\cal K}^{\mu\sigma_{1}}_{r}\partial_{\sigma_{1}}\varphi_{r}
+ {\cal K}^{\mu\sigma_{1}\sigma_{2}}_{r}
   \partial_{\sigma_{1}\sigma_{2}}\varphi_{r}
+ \cdots + 
{\cal K}^{\mu\sigma_{1}\cdots\sigma_{n}}_{r}
   \partial_{\sigma_{1}\cdots\sigma_{n}}\varphi_{r}
\right]
\label{current}
\,.
\end{align}
In fact, for $n\to\infty$ the Noether current consists of an infinite sequence of tensors with 
increasing rank order. Furthermore, each of the different tensors 
${\cal K}^{\mu\sigma_{1}\sigma_{2}\cdots}_{r}$ in Eq.~(\ref{current}) 
contains again infinite series terms of higher-order derivatives with respect 
to the Lagrangian density. 
They are given by the following expressions
\begin{align}
{\cal K}^{\mu}_{r} & =  \sum_{i=1}^{n}\;
(-)^{i+1}\; 
\prod_{j=1}^{i-1}\partial_{\alpha_{j}}\;
\frac{\partial {\cal L}}
{\partial (\partial_{\mu\alpha_{j}}\varphi_{r})}\,,
\label{tensors}\\
{\cal K}^{\mu\sigma_{1}}_{r} & =  \sum_{i=1}^{n}\;
(-)^{i+1}\; 
\prod_{j=1}^{i-1}\partial_{\alpha_{j}}\;
\frac{\partial {\cal L}}
{\partial (\partial_{\mu\alpha_{j}\sigma_{1}}\varphi_{r})}\,,
\nonumber\\
{\cal K}^{\mu\sigma_{1}\sigma_{2}}_{r} & = \sum_{i=1}^{n}\;
(-)^{i+1}\; 
\prod_{j=1}^{i-1}\partial_{\alpha_{j}}\;
\frac{\partial {\cal L}}{\partial (\partial_{\mu\alpha_{j}\sigma_{1}\sigma_{2}}\varphi_{r})}\,,
\nonumber\\
\vdots & 
\nonumber\\
{\cal K}^{\mu\sigma_{1}\cdots\sigma_{n}}_{r} & = \sum_{i=1}^{n}\;
(-)^{i+1}\; 
\prod_{j=1}^{i-1}\partial_{\alpha_{j}}\;
\frac{\partial {\cal L}}{\partial (\partial_{\mu\alpha_{j}\sigma_{1}\cdots\sigma_{n}}\varphi_{r})}
\;. \nonumber
\end{align}

The derivation of the energy-momentum tensor proceeds in a similar way, 
see Appendix~\ref{app2}. Now the field arguments  
are transformed, but not the fields them self. In particular, invariance of 
the Lagrangian density~(\ref{EL_0}) with respect to a constant displacement 
$\delta_{\mu}$ of the coordinates $x_{\mu}$
\begin{align}
x_{\mu}\longrightarrow x_{\mu}^{\prime} = x_{\mu} + \delta_{\mu}\,,
\label{phaseTrafo2}
\end{align}
implies a continuity equation $\partial_{\mu}T^{\mu\nu} = 0$  for the 
energy-momentum tensor $T^{\mu\nu}$ which takes the following form 
\begin{align}
T^{\mu\nu} = 
  {\cal K}^{\mu}_{r}\partial^{\nu}\varphi_{r} 
+ {\cal K}^{\mu\sigma_{1}}_{r}\partial_{\sigma_{1}}^{\nu}\varphi_{r}
+ {\cal K}^{\mu\sigma_{1}\sigma_{2}}_{r}
\partial_{\sigma_{1}\sigma_{2}}^{\nu}\varphi_{r}
+ \cdots 
+ {\cal K}^{\mu\sigma_{1}\cdots\sigma_{n}}_{r}
\partial_{\sigma_{1}\cdots\sigma_{n}}^{\nu}\varphi_{r}
- g^{\mu\nu}{\cal L}
\;.
\label{tensor}
\end{align}
The $00$-component of the energy-momentum tensor describes the energy density and 
the spatial diagonal components are related to the pressure
density. These equations form a background for the construction and
application of the NLD formalism presented in the proceeding sections. They will
further provide a thermodynamically consistent framework for the calculation
of the EoS in mean field approximation in terms of energy and pressure densities.

\section{\label{sec3}The non-linear derivative model}

In this section we introduce the non-linear derivative (NLD) model and derive the 
equations of motion for the relevant degrees of freedom. The NLD approach is 
essentially based on the Lagrangian density of 
RHD~\cite{Duerr:1956zz,Walecka:1974qa,Serot:1997xg}, which is given by
\begin{align}
{\cal L} = & \frac{1}{2}
\left[
	\overline{\Psi}\gamma_{\mu} i\partialr^{\mu}\Psi
	- 
	\overline{\Psi} i\partiall^{\mu} \gamma_{\mu} \Psi
\right]
- m\overline{\Psi}\Psi
-\frac{1}{2}m^{2}_{\sigma}\sigma^{2}
+\frac{1}{2}\partial_{\mu}\sigma\partial^{\mu}\sigma
-U(\sigma)
\nonumber\\
+ & \frac{1}{2}m^{2}_{\omega}\omega_{\mu} \omega^{\mu} 
-\frac{1}{4}F_{\mu\nu}F^{\mu\nu}
+\frac{1}{2}m^{2}_{\rho}\rhovec_{\mu}\rhovec^{\mu} 
-\frac{1}{4}\Gcapvec_{\mu\nu}\Gcapvec^{\mu\nu}
-\frac{1}{2}m^{2}_{\delta}\deltavec^{2}
+\frac{1}{2}\partial_{\mu}\deltavec \, \partial^{\mu}\deltavec
+{\cal L}_{int}
\label{NDC-free}
\end{align}
\end{widetext}
where $\Psi=(\Psi_{p},\Psi_{n})^{T}$ denotes the nucleon spinor field 
in the Lagrangian density of a Dirac-type. In a spirit of RHD, the 
interactions between the nucleon fields are described by the exchange of
meson fields. These are the scalar $\sigma$ and vector $\omega^{\mu}$ mesons 
in the isoscalar channel, as well as the scalar $\deltavec$ and vector 
$\rhovec^{\mu}$ mesons in the isovector channel. Their corresponding 
Lagrangian densities are of the Klein-Gordon and Proca types, respectively. 
The term $U(\sigma)=\frac{1}{3}b\sigma^{3}+\frac{1}{4}c\sigma^{4}$ 
contains the usual selfinteractions of the $\sigma$ meson. 
The notations 
for the masses of fields in Eq.~(\ref{NDC-free}) are obvious. The field
strength tensors are defined as 
$F^{\mu\nu}=\partial^{\mu}\omega^{\nu}-\partial^{\nu}\omega^{\mu}$, 
$\Gcapvec^{\mu\nu}=\partial^{\mu}\rhovec^{\nu}-\partial^{\nu}\rhovec^{\mu}$ 
for the isoscalar and isovector fields, respectively. 

In conventional RHD
approaches the interaction Lagrangian $\mathcal{L}_{int}$ is given
by~\cite{Walecka:1974qa,Serot:1997xg}
\begin{align}
{\cal L}_{int} = {\cal L}_{int}^{\sigma}+{\cal L}_{int}^{\omega}
+{\cal L}_{int}^{\rho}+{\cal L}_{int}^{\delta}
\,,
\end{align}
where
\begin{equation}
\label{LintRHD}
{\cal L}_{int}^{\sigma}= 
g_{\sigma}\overline{\Psi}\Psi\sigma \,,
\end{equation}
\begin{equation}
\label{LintRHD2}
{\cal L}_{int}^{\omega} = -g_{\omega}\overline{\Psi}\gamma^{\mu}\Psi\omega_{\mu}\,,
\end{equation}
\begin{equation}
\label{LintRHD3}
{\cal L}_{int}^{\rho} = -g_{\rho}\overline{\Psi}\gamma^{\mu}\tauvec\Psi\rhovec_{\mu}\,,
\end{equation}
\begin{equation}
\label{LintRHD4}
{\cal L}_{int}^{\delta} = g_{\delta}\overline{\Psi}\tauvec\Psi\deltavec \,,
\end{equation}
and ${\cal L}_{int}^{\sigma,\omega,\rho,\delta}$ contains the meson-nucleon 
interactions with coupling strengths $g_{\sigma,\omega,\rho,\delta}$ and 
$\tau$ denotes the isospin Pauli operator.

In the NLD model the momentum dependence of fields is realized by the introduction 
of non-linear derivative operators in the interaction Lagrangian of
conventional RHD. These additional
operators regulate the high momentum components of the RMF fields in the interaction
vertices and can be interpreted as cut-off form factors. This is in spirit of
boson-exchange models
where the phenomenological cut-off is an indispensable part of any
microscopic description of meson-nucleon interaction~\cite{Machleidt:1987hj,Erkelenz:1974uj}.
In the RMF (Hartree) approximation to RHD 
only bare Lorentz structures corresponding to the point-like
meson-nucleon interactions are taken into account and the high momentum 
components of fields are not suppressed due to the
missing nucleon finite size effect. 
The NLD model attempts to account for the suppression
of the high momentum part of the nucleon field in the meson-nucleon interaction on a
field-theoretical level.

The NLD interaction Lagrangians contain the conventional meson-nucleon 
RHD structures, however, they are extended by the inclusion of 
non-linear derivative operators into the meson-nucleon vertices. 
The NLD interaction Lagrangians followed here read
\begin{equation}
{\cal L}_{int}^{\sigma} = \frac{g_{\sigma}}{2}
	\left[
	\overline{\Psi}
	\, \nldl
	\Psi\sigma
	+\sigma\overline{\Psi}
	\, \nldr
	\Psi
	\right]\,,
\end{equation}
\begin{equation}
{\cal L}_{int}^{\omega} = -\frac{g_{\omega}}{2}
	\left[
	\overline{\Psi}
	 \, \nldl
	\gamma^{\mu}\Psi\omega_{\mu}
	+\omega_{\mu}\overline{\Psi}\gamma^{\mu}
	\, \nldr
	\Psi
	\right]\,,
\end{equation}
\begin{equation}
{\cal L}_{int}^{\rho} = 
- \frac{g_{\rho}}{2}
	\left[
	\overline{\Psi}
	 \, \nldl
	\gamma^{\mu}\tauvec\Psi\rhovec_{\mu}
	+\rhovec_{\mu}\overline{\Psi}\tauvec\gamma^{\mu}
	\, \nldr
	\Psi
	\right]\,,
\end{equation}
\begin{equation}
{\cal L}_{int}^{\delta} = 
 \frac{g_{\delta}}{2}
	\left[
	\overline{\Psi}
	\, \nldl \tauvec
	\Psi\deltavec\,
	+\deltavec\overline{\Psi}\tauvec
	\, \nldr
	\Psi
	\right]	\,.
\label{NDCrd}
\end{equation}
As one can see, the only difference with respect to
the conventional RHD interaction Lagrangian is the presence of additional
operators  $\nldr,~\nldl$ which serve to regulate the high momentum
component of the nucleon field. 
The hermiticity of the Lagrangian demands $\nldl=\nldr^{\dagger}$. 
The operator functions (regulators) $\nldr,~\nldl$ are assumed to be generic
functions of partial derivative operator and supposed to 
act on the nucleon spinors $\Psi$ and $\overline{\Psi}$, respectively. 
Furthermore, these regulators are assumed to be smooth functions. 
Therefore, the formal Taylor expansion of the operator functions in terms of partial derivatives
generates an infinite series of higher-order derivative terms
\begin{align}
\nldr := \nld\left( \xir \right) = &
\sum_{j=0}^{n\to\infty}\, 
\frac{\partial^{j}}{\partial\xir^{j}}\nld\vert_{\xir\to 0}
\,\frac{\xir^{j}}{j!} \,, \\
\nldl := \nld \left( \xil \right) = & 
\sum_{j=0}^{n\to\infty}\, \frac{\xil^{j}}{j!}\,
\frac{\partial^{j}}{\partial\xir^{j}}\nld\vert_{\xil\to 0}
\,. \label{ope1}
\end{align}
The expansion coefficients
are given by the partial derivatives of $\nld$ with respect 
to the operator arguments $\xir$ and $\xil$ around the origin. The operators are defined as
$\xir = -\zeta^{\alpha}i\partialr_{\alpha},~
\xil = i\partiall_{\alpha}\zeta^{\alpha}$
where the four vector $\zeta^{\mu}=v^{\mu}/\Lambda$ contains the cut-off $\Lambda$
and $v^{\mu}$ is an auxiliary vector.  The functional form of the regulators is constructed such
that in the limit $\Lambda\to\infty$ the following limit holds $\nldr(\nldl) \to
1$. Therefore, in the limit $\Lambda\to\infty$ the original RHD Lagrangians
are recovered. 

In the most general case the NLD formalism can be extended to the case of
multiple variable regulators. In 
particular, we can assume the non-linear operator to be a multi-variable non-linear 
function of higher-order partial derivatives, which are given by the following
Taylor expansion
\begin{widetext}
\begin{align}
\label{ope0}
\nldr := \nld(\xir_{1},\xir_{2},\xir_{3},\xir_{4}) = & 
\sum_{i_{1}=0}^{n\to\infty}\sum_{i_{2}=0}^{n\to\infty}
\sum_{i_{3}=0}^{n\to\infty}\sum_{i_{4}=0}^{n\to\infty} \, 
\frac{\partial^{i_{1}+i_{2}+i_{3}+i_{4}}}{\partial\xir^{i_{1}}_{1}\partial\xir^{i_{2}}_{2}\partial\xir^{i_{3}}_{3}\partial\xir^{i_{4}}_{4}}
\nld\vert_{\{\xir_{1},\xir_{2},\xir_{3},\xir_{4}\}\to 0}\,
\frac{\xir_{1}^{i_{1}}\xir_{2}^{i_{2}}\xir_{3}^{i_{3}}\xir_{4}^{i_{4}}}{i_{1}!i_{2}!i_{3}!i_{4}!}
\,, \\
\nldl := \nld(\xil_{1},\xil_{2},\xil_{3},\xil_{4}) = & 
\sum_{i_{1}=0}^{n\to\infty}\sum_{i_{2}=0}^{n\to\infty}\sum_{i_{3}=0}^{n\to\infty}
\sum_{i_{4}=0}^{n\to\infty} \, 
\frac{\xil_{1}^{i_{1}}\xil_{2}^{i_{2}}\xil_{3}^{i_{3}}\xil_{4}^{i_{4}}}{i_{1}!i_{2}!i_{3}!i_{4}!}\,
\frac{\partial^{i_{1}+i_{2}+i_{3}+i_{4}}}{\partial\xil^{i_{1}}_{1}\partial\xil^{i_{2}}_{2}\partial\xil^{i_{3}}_{3}\partial\xil^{i_{4}}_{4}}
\nld\vert_{\{\xil_{1},\xil_{2},\xil_{3},\xil_{4}\}\to 0}
\,. \label{ope}
\end{align}
\end{widetext}
Then Eqs.~(\ref{ope0}) and~(\ref{ope})  can be rearranged into the 
terms with increasing order with respect to the partial derivatives, see for details 
Appendix~\ref{app3}. The operators $\xi_{i}$ are defined in a similar way as before
\begin{align}
\xir_{i} = 
-\zeta^{\alpha}_{i}i\partialr_{\alpha}
~,~
\xil_{i} = 
i\partiall_{\alpha}\zeta^{\alpha}_{i}\,,
\label{opee}
\end{align}
with $\zeta^{\mu}_{i}=v^{\mu}_{i}/\Lambda$ ($i=1,2,3,4$) in this case. 
As we will show latter on, this representation allows to generate any desired form of
the regulator function, {\it i.e.}, momentum and/or energy dependent monopole,
dipole {\it etc.}  functions.


The derivation of the equation of motion for the Dirac field 
follows the generalized Euler-Lagrange equations, 
Eq.~(\ref{Euler0}), to the NLD-Lagrangian density using the Taylor form of
the regulators. This obviously  will generate an infinite number of partial
derivative terms in the equations of motions. However,
as shown in detail in Appendix~\ref{app4} these infinite series can be resummed
(up to terms containing the derivatives of the meson fields) to the following 
Dirac equation
\begin{equation}
\left[
	\gamma_{\mu}(i\partial^{\mu}-\Sigma^{\mu}) - 
	(m-\Sigma_{s})
\right]\Psi = 0
\;,
\label{Dirac_nld}
\end{equation}
where the selfenergies $\Sigma^{\mu}$ and $\Sigma_{s}$ are given by 
\begin{eqnarray}
\Sigma^{\mu} & = & g_{\omega}\omega^{\mu}\nldr + 
g_{\rho}\tauvec \cdot \rhovec^{\mu}\nldr+ \cdots~,
\label{Sigmav}\\
\Sigma_{s} & = & g_{\sigma}\sigma\nldr + 
g_{\delta}\tauvec \cdot \deltavec \, \nldr+ \cdots
\;. \label{Sigmas}
\end{eqnarray}
Here both Lorentz-components of the selfenergy, $\Sigma^{\mu}$ and $\Sigma_{s}$, 
show an explicit linear behavior with respect to the meson fields $\sigma$, 
$\omega^{\mu}$, $\vec{\rho\,}^{\mu}$ and $\deltavec$ as in the standard RMF. 
However, they contain an additional dependence on regulator functions. 

The additional terms in Eqs.~(\ref{Sigmav}) and~(\ref{Sigmas})
containing the meson field derivatives are denoted by multiple dots. 
All these contributions can be also resummed.
However, in the mean-field approximation to infinite 
nuclear matter, which will be discussed in the next section, these terms vanish. 
On the other hand, they will be needed in the description of finite systems, such as 
finite nuclei and heavy-ion collisions. 
Therefore, for simplicity we do not consider these terms here, and postpone 
the effect of these terms for future studies.

The derivation of the meson field equations of motion is straightforward, since here one 
has to use the standard Euler-Lagrange equations
\begin{align}
\frac{\partial{\cal L}}{\partial\varphi_{r}}
-
\partial_{\alpha}
\frac{\partial{\cal L}}
{\partial(\partial_{\alpha}\varphi_{r})}
= 0
\;,
\label{EulerMeson}
\end{align}
where now $r=\sigma,\omega,\rho$ and $\delta$.
The following Proca and 
Klein-Gordon equations are obtained

\begin{align}
&
\partial_{\alpha}\partial^{\alpha}\sigma + m_{\sigma}^{2}\sigma 
+ \frac{\partial U}{\partial\sigma} =
\frac{1}{2}g_{\sigma}
\left[
	\overline{\Psi} \, \nldl \Psi + \overline{\Psi}\nldr \Psi
\right] \,,
\label{sigma_meson}\\
&
\partial_{\mu}F^{\mu\nu} + m_{\omega}^{2}\omega^{\nu} =
\frac{1}{2}g_{\omega}
\left[
	\overline{\Psi}\, \nldl \gamma^{\nu}\Psi + \overline{\Psi}\gamma^{\nu}\nldr \Psi
\right] \,,
\label{omega_meson}\\
&
\partial_{\mu}\Gcapvec^{\mu\nu} + m_{\rho}^{2}\rhovec^{\nu} =
\frac{1}{2}g_{\rho}
\left[
\overline{\Psi} \, \nldl \gamma^{\nu}\tauvec \, \Psi + 
\overline{\Psi}\tauvec \, \gamma^{\nu}\nldr \Psi
\right] \,,
\label{rho_meson}\\
&
\partial_{\alpha}\partial^{\alpha}\deltavec + m_{\sigma}^{2}\deltavec =
\frac{1}{2}g_{\delta}
\left[
	\overline{\Psi} \, \nldl \tauvec \Psi + \overline{\Psi}\tauvec\nldr \Psi
\right]
\label{delta_meson}
\,.
\end{align}

Finally, we provide the general expressions for the Noether theorems within the NLD 
formalism. The evaluation of the conserved baryon current results from 
the application of the generalized expression for $J^{\mu}$,
Eq.~(\ref{current}), to the Lagrangian density of the NLD model. 
As shown in detail in Appendix~\ref{app5}, a systematic evaluation 
of the higher-order field derivatives of the NLD Lagrangian and the resummation
procedure result in 
\begin{widetext}
\begin{align}
J^{\mu} = \overline{\Psi}\gamma^{\mu}\Psi 
- & \frac{1}{2}\, g_{\sigma}  \, 
\left[
\overline{\Psi}\, \calol^{\mu} \Psi - 
\overline{\Psi}\, \calor^{\mu}\Psi
\right]\sigma
+ \frac{1}{2}\, g_{\omega}\, 
\left[
\overline{\Psi}\, \calol^{\mu} \gamma^{\alpha}\Psi - 
\overline{\Psi}\gamma^{\alpha}\, \calor^{\mu}\Psi
\right] \omega_{\alpha}
\label{stromNLD} \nonumber\\
+ & \frac{1}{2}\, g_{\rho}\, 
\left[
\overline{\Psi}\, \calol^{\mu} \gamma^{\alpha}\tauvec\Psi  - 
\overline{\Psi}\gamma^{\alpha}\, \calor^{\mu}\tauvec\Psi
\right] \rhovec_{\alpha}
- \frac{1}{2}\, g_{\delta}\, 
\left[
\overline{\Psi}\, \calol^{\mu} \tauvec \Psi - 
\overline{\Psi}\,  \calor^{\mu}\tauvec\Psi
\right] \deltavec + \cdots
\,.
\end{align}
\end{widetext}
The new non-linear derivative operators in Eq.~(\ref{stromNLD}), 
$\calol^{\mu} := \partial\nldl/\partial(i\partiall_{\mu})$ and 
$\calor^{\mu} := \partial\nldr/\partial(i\partialr_{\mu})$, 
denote the derivatives of $\nldl$ and $\nldr$ with respect to their 
operator argument $i\partiall_{\mu}$ and $i\partialr_{\mu}$ 
(see Appendix~\ref{app5}). 
The first term in Eq.~(\ref{stromNLD}) corresponds to the standard 
expression of the RHD models and the additional contributions arise due to 
the additional higher-order field derivatives in the Noether theorem, 
Eq.~(\ref{current}). 

The energy-momentum tensor, $T^{\mu\nu}$, is determined 
according to Eq.~(\ref{tensor}). 
The evaluation procedure, which is similar to that one for the Noether current, 
results in the following NLD expression for $T^{\mu\nu}$ 
\begin{widetext}
\begin{align}
T^{\mu\nu} = &
\frac{1}{2}\, 
\overline{\Psi}\gamma^{\mu} \, i\partialr^{\nu} \Psi - 
\frac{1}{2}\, 
\overline{\Psi} \, i\partiall^{\nu}\gamma^{\mu} \Psi
\label{tensorNLD}\nonumber\\
+ & \frac{1}{2}\, g_{\sigma}\, 
\left[
\overline{\Psi}\, \calor^{\mu}\, i\partialr^{\nu}\, \Psi +
\overline{\Psi}\, i\partiall^{\nu}\calol^{\mu} \, \Psi
\right] \sigma
- \frac{1}{2}\, g_{\omega}\, 
\left[
\overline{\Psi}\gamma^{\alpha}\, \calor^{\mu}\, i\partialr^{\nu}\,\Psi+
\overline{\Psi}\, i\partiall^{\nu}\, \calol^{\mu} \gamma^{\alpha}\Psi 
\right]\omega_{\alpha} \nonumber \\
- & \frac{1}{2}\, g_{\rho}\, 
\left[ 
\overline{\Psi}\tauvec\gamma^{\alpha}\, \calor^{\mu} \,
i\partialr^{\nu}\, \Psi +
\overline{\Psi}\, i\partiall^{\nu}\, \calol^{\mu} \gamma^{\alpha}\tauvec\Psi
\right] \rhovec_{\alpha}
+ \frac{1}{2}\, g_{\delta}\, 
\left[ 
\overline{\Psi}\, \tauvec\calor^{\mu}\, i\partialr^{\nu}\, \Psi +
\overline{\Psi}\, i\partiall^{\nu}\calol^{\mu} \, \tauvec\Psi
\right] \deltavec
\nonumber\\
-& g^{\mu\nu}\, {\cal L} + \cdots
\,.
\end{align}
\end{widetext}
The first line in Eq.~(\ref{tensorNLD}) is just the usual kinetic RHD 
contribution to $T^{\mu\nu}$, while the additional kinetic terms originate 
from the evaluation of the higher-order derivatives in Eq.~(\ref{tensor}). 
These terms will be important for the the thermodynamic consistency of the 
model and the validation of the Hugenholtz-Van Hove 
theorem~\cite{Weisskopf:1957,Hugenholtz:1958}. Again the terms not shown 
in Eqs.~(\ref{stromNLD}) and~(\ref{tensorNLD}) describe the contribution 
of terms containing the derivatives of the meson fields.

\section{\label{sec4}RMF approach to infinite nuclear matter}

In the mean-field approximation the mesons are treated as classical fields. 
Infinite nuclear matter is described by a static homogeneous, isotropic, spin and 
isospin-saturated system of protons and neutrons. In this case, the spatial components 
of the Lorentz-vector meson fields vanish with $\omega^{\mu}\to (\omega^0,~\zerovec)$, 
and in isospin space only the neutral 
component of the isovector fields survive, {\it i.e.}, $\rhovec^{\mu} \to
(\rho^0_3,~\zerovec)$ and $\deltavec \to \delta_3$. For simplicity, we denote
in the following the third
isospin components of the isovector fields as $\rho$ and $\delta$.

The derivation of the RMF equations starts with the usual plane wave \textit{ansatz}
\begin{align}
\Psi_i(s,\pvec) = 
u_i(s,\pvec)e^{-ip^{\mu}x_{\mu}} 
\,,
\label{plane_wave}
\end{align}
where $i$ stands for protons ($i=p$) or neutrons ($i=n$) and $p^{\mu}=(E,\vec{p}\,)$ 
is a single nucleon 4-momentum.
The application of the non-linear derivative operator $\nld$ to the 
plane wave \textit{ansatz} of the spinor fields results in
\begin{equation}
\nld(\xir)\Psi_{i} =  \nld(\xi) u_i(s,\pvec)e^{-ip^{\mu}x_{\mu}} \,,
\label{ope_nm1}
\end{equation}
\begin{equation}
\overline{\Psi}_{i}\nld(\xil) = 
\nld(\xi) \overline{u}_i(s,\pvec)e^{+ip^{\mu}x_{\mu}} \,,
\label{ope_nm}
\end{equation}
where the regulators in the r.h.s. of above equation are now functions of 
the scalar argument $\xi=-\frac{v_{\alpha}p^{\alpha}}{\Lambda}$.

With the help of Eqs.~(\ref{plane_wave}) and~(\ref{ope_nm1}) one gets the Dirac 
equation similar to Eq.~(\ref{Dirac_nld}) with selfenergies given by
\begin{equation}
\Sigma^{\mu}_{vi}  =  g_{\omega}\omega^{\mu}{\cal D}
+g_{\rho} \tau_{i}  \rho^{\mu}{\cal D}~,
\label{Sigmav_nm}
\end{equation}
\begin{equation}
\Sigma_{si}  =  g_{\sigma}\sigma\nld
+g_{\delta} \tau_{i}  \delta\nld~,
\label{Sigmas_nm}
\end{equation}
where now $\tau_{i}=+1$ for protons ($i=p$) and $\tau_{i}=-1$ for neutrons ($i=n$). 
We note again that in the RMF approximation to infinite matter the additional terms including 
the meson field derivatives vanish. This largely simplifies the formalism, 
since these terms which show up in the original Dirac equation, 
see Eq.~(\ref{Dirac_nld}), do not appear any more. 

The solutions of the Dirac equation takes the form
\begin{equation}
u_{i}(s,\pvec) = N_{i}
\left(
\begin{array}{c}
\varphi_{s} \\ \\
\ds \frac{ \sigmavec\cdot\pvec}{E^{*}_{i}+m^{*}_{i}}\varphi_{s}\\
\end{array}
\right)
\; , \label{Spinor}
\end{equation}
with spin eigenfunctions $\varphi_{s}$, the in-medium energy 
\begin{equation}
E^{*}_{i} := E - \Sigma^{0}_{vi}~,
\end{equation}
and the Dirac mass
\begin{equation}
m^{*}_{i} := m - \Sigma_{si}~. 
\end{equation}
For a given momentum the single particle energy $E$ is 
obtained from the in-medium on-shell relation 
\begin{equation}
E^{*2}_{i} - \vec{p\,}^{2} = m^{*2}_{i}~.
\label{onshell}
\end{equation}

The factor $N_{i}$ is determined from the normalization of the 
probability distribution, that is $\int d^{3}x \,J^{0}=1$. 
In the conventional RMF models the baryon density is given by the familiar 
expression $J^{0} = \Psi^{\dagger}\Psi$ and the normalization condition 
$\int d^{3}x \,\Psi^{\dagger}\Psi=1$ would result in  
$N_i=\sqrt{\frac{E^{*}_{i}+m^{*}_{i}}{2E^{*}_{i}}}$. 
In the NLD model one has to use Eq.~(\ref{current}) for the Noether current 
by keeping in mind that the infinite series of meson field derivatives vanish 
in the RMF approach to nuclear matter. In this case, see again
Appendix~\ref{app5} for details, the conserved baryon current $J^{\mu}$ is 
resummed up to infinity and the result reads
\begin{align}
J^{\mu}  = & \!\!
\sum_{i=p,n} \Big[
\Bla
\overline{\Psi}_{i}\gamma^{\mu}\Psi_{i} 
\Bra
\label{current_NLD}\\
& 
\left.
+ g_{\sigma}
\Bla\overline{\Psi}_{i} [\partial^{\mu}_{p}{\cal D}]\Psi_{i}\Bra 
\sigma
- g_{\omega}
\Bla
\overline{\Psi}_{i} [\partial^{\mu}_{p}{\cal D}]
\gamma^{\alpha}\Psi_{i}
\Bra 
\omega_{\alpha}
\right.
\nonumber\\
& 
- g_{\rho}
\tau_{i}
\Bla
\overline{\Psi}_{i} [\partial^{\mu}_{p}{\cal D}]\gamma^{\alpha}\Psi_{i}
\Bra 
\rho_{\alpha}
+ g_{\delta}
\tau_{i}
\Bla
\overline{\Psi}_{i} [\partial^{\mu}_{p}{\cal D}]\Psi_{i}
\Bra 
\delta
\Big]
\nonumber
\; .
\end{align}
The $0$-component of the Noether current describes the conserved nucleon 
density $\rho_{B}=J^{0}$, from which also the relation between the Fermi 
momentum $p_{F}$ and $\rho_{B}$ is uniquely determined.  In particular, 
using the Gordon identity and Eqs.~(\ref{Sigmav_nm}) and~(\ref{Sigmas_nm}) 
for the RMF selfenergies, one obtains
\begin{align}
J^{\mu} & = 
\fac \; \sum_{i=p,n} \; \pspace \, N_{i}^{2}
\nonumber\\
& \times \left[ 
\frac{p^{*\mu}_{i}}{E^{*}_{i}} 
+  
\Big( \partial_{p}^{\mu}\Sigma_{si} \Big) \frac{m^{*}_{i}}{E^{*}_{i}}
-  
\left( \partial_{p}^{\mu}\Sigma^{\beta}_{vi} \right) 
\frac{p^{*}_{i\beta}}{E^{*}_{i}}
\right]
\nonumber
\,,
\end{align}
where $\kappa=2$ is a spin degeneracy factor, $p_{F_{i}}$ stands for the 
proton or neutron Fermi-momentum and the effective momentum is given by
\begin{equation}
p^{*\mu}_{i}=p^{\mu}-\Sigma^{\mu}_{vi}. 
\end{equation}
One defines now a new in-medium 
$4$-momentum $\Pi^{\mu}_i$ as
\begin{align}
\Pi^{\mu}_{i} = p^{*\mu}_{i}+ m^{*}_{i}\Big(\partial_{p}^{\mu}\Sigma_{si} \Big)
- \Big(\partial_{p}^{\mu}\Sigma^{\beta}_{vi} \Big) p^{*}_{i\beta}
\label{bigPi}
\,,
\end{align}
and arrives to the following expression
\begin{align}
J^{\mu} = 
\fac \, \sum_{i=p,n} \, \pspace \, N_{i}^{2} \, 
\frac{\Pi^{\mu}_{i}}{E^{*}_{i}} 
\label{nldeq7}
\,.
\end{align}
On the other hand, the general definition of the baryon current results from the 
covariant superposition of all the occupied in-medium on-shell nucleon 
positive energy states up to the proton or neutron Fermi momentum~\cite{Weber:1992qc} 
\begin{align}
J^{\mu} = &
\fac  \, \sum_{i=p,n} \, \int\limits_{|\pvec|\leq p_{F_{i}}}\!\!\!\!\!\!  d^4p
\label{nldeq8}\\
\times & \Pi^{\mu}_{i} \,
\delta\left( p^{*\mu}_{i}p_{i\mu}^{*}-m^{*2}_{i}\right) \, 2\Theta(p^{0})
\nonumber
\,.
\end{align}
In the NLD approach the mean-field selfenergies depend explicitly on the 
single-particle momentum $p^{\mu}$. Therefore, using the properties of the 
$\delta$-function the time-like $dp^{0}$  component can be integrated out
explicitly. The result reads
\begin{align}
J^{\mu} = \fac  \, \sum_{i=p,n} \, \pspace \, \frac{\Pi^{\mu}_{i}}{\Pi^{0}_{i}}
\label{nldeq12}
\,.
\end{align}
Comparing Eq.~(\ref{nldeq12}) with the equation for the NLD current, 
Eq.~(\ref{nldeq7}), one gets the following result for the normalization 
\begin{align}
N_{i}=\sqrt{\frac{E^{*}_{i}+m^{*}_{i}}{2E^{*}_{i}}}
\sqrt{\frac{E^{*}_{i}}{\Pi^{0}_{i}}}
\label{SpinorNorm}
\,,
\end{align}
and the bilinear products between the in-medium spinors of protons and 
neutrons are given by
\begin{align}
\overline{u}_i(p)u_i(p) = & \frac{m^{*}_i}{\Pi^{0}_i},
\label{norms}\\
\overline{u}_i(p)\gamma^{0}u_i(p) = & 1
\label{normv}
\,.
\end{align}
Eq.~(\ref{normv}) ensures also the proper normalization of the 
probability distribution, {\it i.e.}, $\int \Psi^{\dagger}\Psi=1$. 

In our first work~\cite{Gaitanos:2009nt}, where the non-linear derivative model 
has been proposed, 
the correction terms proportional to the partial derivatives of the 
selfenergies with respect to the single-particle momentum $p^{\mu}$ in 
Eq.~(\ref{bigPi}) were not taken into account. 
Even if 
their contributions are small at low densities, these 
terms will be included in the present calculations which attempt to consider
also the high density domain of the EoS in neutron stars. On the other hand, 
the inclusion of these terms is crucial for a fully thermodynamically consistent 
formalism and is independent of the particular form of the
cut-off functions.  Note that, the additional cut-off dependent terms in the baryon 
and energy densities
of Ref.~\cite{Gaitanos:2009nt} are now canceled by the proper normalization
conditions.

The energy-momentum tensor in NLD is obtained by applying the Noether theorem 
for translational invariance. In nuclear matter the
resummation procedure results in the following expression 
\begin{widetext}
\begin{align}
T^{\mu\nu} = &
\sum_{i=p,n}\bigg[
\Bla \overline{\Psi}_{i} \gamma^{\mu} p^{\nu} \Psi_{i} \Bra
\bigg.
\label{tensornm} \nonumber \\
&
\bigg. 
+ g_{\sigma}
\Bla 
\overline{\Psi}_{i} [\partial^{\mu}_{p}{\cal D}] p^{\nu} \Psi_{i}
\Bra 
\sigma
- g_{\omega}
\Bla  
\overline{\Psi}_{i} [\partial^{\mu}_{p}{\cal D}]\gamma^{\alpha}p^{\nu}\Psi_{i}
\Bra  \omega_{\alpha}
- g_{\rho}
\tau_{i}
\Bla 
\overline{\Psi}_{i}[\partial^{\mu}_{p}{\cal D}]\gamma^{\alpha}p^{\nu}\Psi_{i}
\Bra 
\rho_{\alpha}
+ g_{\delta}
\tau_{i}
\Bla 
\overline{\Psi}_{i}[\partial^{\mu}_{p}{\cal D}]p^{\nu}\Psi_{i}
\Bra 
\delta_{\alpha}
\bigg]
\nonumber\\
& 
- g^{\mu\nu}\langle{\cal L}\rangle
\,.
\end{align}
The evaluation of the expectation values in Eq.~(\ref{tensornm}) can be done
in a similar way as for the current with the result
\begin{align}
T^{\mu\nu} = 
\sum_{i=p,n} \fac \pspace \, \frac{p^{\nu}}{\Pi^{0}_{i}} \left[
p^{*\mu}_{i} + m^{*}_{i} \Big( \partial_{p}^{\mu}\Sigma_{si} \Big) 
- p^{*\alpha}_{i} \Big( \partial_{p}^{\mu}\Sigma_{\alpha i} \Big) 
\, \right]
- g^{\mu\nu}\langle{\cal L}\rangle
\label{nldeq14}
\,.
\end{align}
\end{widetext}
Using Eq.~(\ref{bigPi}) one arrives to the final expression for the energy-momentum 
tensor in the NLD formalism, which can be written in the following form
\begin{align}
T^{\mu\nu} = 
\sum_{i=p,n} \fac \pspace \,
\frac{\Pi^{\mu}_{i} p^{\nu}}{\Pi^{0}_{i}} - g^{\mu\nu}\langle{\cal L}\rangle
\label{nldeq17}
\,,
\end{align}
from which the energy density $\varepsilon\equiv T^{00}$ and the pressure $P$ can
be calculated, {\it i.e.},
\begin{align}
\varepsilon = &
\sum_{i=p,n} \fac \pspace \, E(\pvec) - \langle{\cal L}\rangle ~,
\label{nldeq18a}\\
P = & \frac{1}{3}\sum_{i=p,n} \fac \pspace \,
\frac{\Picapvec_i \cdot \pvec}{\Pi^{0}_{i}} + \langle{\cal L}\rangle
\label{nldeq18b}
\,.
\end{align}
Eqs.~(\ref{nldeq18a}) and~(\ref{nldeq18b})  look similar as the familiar expressions of 
the usual RMF models. However, the non-linear effects induced by the
regulators show up through the generalized 
momentum $\Pi^{\mu}_i$ and through the dispersion relation for the single-particle 
energy $E(\pvec)$. Note the different form of the generalized momentum, $\Pi^{\mu}_i$, 
when one chooses energy or momentum dependent cut-off functions. Indeed, in the latter 
case the spatial derivatives in $\vec{\Pi}_i$ contribute in the pressure, while for 
energy-dependent cut-off functions they vanish and $\Picapvec_i=\vec{p}$ 
holds. In any case, the expressions for the energy-density and pressure within the 
conventional RMF models are recovered by simple replacement 
$\Pi^{\mu}_i\rightarrow p^{*\mu}_i$, which is just equivalent to taking the limiting case  
$\Lambda\rightarrow\infty$ in the NLD expressions.

Finally, the NLD meson-field equations in the RMF approach to nuclear matter 
read
\begin{align}
m_{\sigma}^{2}\sigma + \frac{\partial U}{\partial\sigma} = & g_{\sigma}
\sum_{i=p,n}\,\Bla \overline{\Psi}_{i}{\cal D}\Psi_{i}\Bra
= g_{\sigma}\rho_{s} ~, \\
m_{\omega}^{2}\omega = & 
g_{\omega}
\sum_{i=p,n}\,\Bla \overline{\Psi}_{i} \gamma^{0}{\cal D}\Psi_{i}\Bra
= g_{\omega}\rho_{0} ~,\\
m_{\rho}^{2}\rho = & 
g_{\rho}
\sum_{i=p,n}\,\tau_{i}\Bla \overline{\Psi}_{i} \gamma^{0} {\cal D}\Psi_{i}\Bra
= g_{\rho}\rho_{I} ~,\\
m_{\delta}^{2}\delta = & 
g_{\delta}
\sum_{i=p,n}\,\tau_{i}\Bla \overline{\Psi}_{i} {\cal D}\Psi_{i}\Bra
= g_{\delta}\rho_{IS} ~.
\label{mesonsNM}
\end{align}
Using Eqs.~(\ref{norms}) and (\ref{normv}), 
the evaluation of the source terms of the 
meson-field equations is straightforward. In particular, the scalar-isoscalar $\rho_{s}$, 
vector-isoscalar $\rho_{0}$, vector-isovector $\rho_{I}$ and scalar-isovector 
$\rho_{IS}$ are given
\begin{equation}
\rho_{s} =  
\fac\sum_{i=p,n} \; \pspace \,
\frac{m^{*}_{i}}{\Pi^{0}_{i}} \, {\cal D}(p)
~,
\label{dens_s}
\end{equation}
\begin{equation}
\rho_{0} = 
\fac\sum_{i=p,n} \; \pspace \, \frac{E^{*}_{i}}{\Pi^{0}_{i}} \, {\cal D}(p)
\,,
\label{dens_0}
\end{equation}
\begin{equation}
\rho_{I} =  \rho_{0p} - \rho_{0n} \,,
\label{dens_i}\
\end{equation}
\begin{equation}
\rho_{IS} = \rho_{sp} - \rho_{sn} \,.
\end{equation}
The meson-field equations of motion  show a similar structure 
as those of the standard RMF approximation. For example, the scalar-isoscalar 
density $\rho_{s}$ is suppressed with respect to the vector density $\rho_{0}$ 
by the factor $m^{*}_i/\Pi^{0}_{i}$, in a similar way as in the conventional Walecka 
models~\cite{Walecka:1974qa}. However, the substantial difference between NLD and 
conventional RMF appears in the source terms which now contain in addition the 
momentum-dependent regulator $\nld$.

\section{\label{sec5}Results}

\subsection{\label{sec5a}Model parameters}

The non-linear derivative operators, $\nld$, in the NLD Lagrangian are not 
constrained from first principles and allows us to consider different functional 
forms of $\nld$. In nuclear matter, these regulators can be chosen as functions of 
the single-particle energy or momentum, depending on the choice of the auxiliary 
multi-dimensional parameters $\zeta^{\mu}_{i}$. The available constraints are the 
bulk properties of nuclear matter and the empirically known energy dependence of the 
in-medium optical potential. It is also well 
known~\cite{Haar:1986ii,TerHaar:1987ce,Plohl:2005fn} that the selfenergies
should decrease or saturate as function of baryon density and single-particle
4-momentum. In the NLD model all these features of the relativistic
mean-fields can be realized using energy or momentum dependent form
factors which regulate the high energy (momentum) behavior of the nucleon
4-momentum. 

\begin{table*}
\begin{tabular}{c|cccccccc}
\hline\hline  \\
Model & $\rho_{sat}$ & $E_{b}$ & $K$ & $a_{sym}$ & $L$ & $K_{sym}$ & $K_{asy}$ &
Ref. \\
&&&&&&&& \\
 & $[fm^{-3}]$ & $[\MeV/A]$ & $[\MeV]$ & $[\MeV]$ & $[\MeV]$ & $[\MeV]$ & $[\MeV]$ & \\
&&&&&&&& \\
\hline\hline \\
NLD & $0.156$ & $-15.30$ & $251$ & $30$ & $81$ & $-28$ & $-514$ & this work
\\ &&&&&&&& \\
\hline
\\
NL3* & $0.150$ & $-16.31$ & $258$ & $38.68$ & $125.7$ & $104.08$ & $-650.12$ & \cite{Lalazissis:2009zz}
\\ &&&&&&&& \\
DD & $0.149$ & $-16.02$ & $240$ & $31.60$ & $56$ & $-95.30$ & $-431.30$ & \cite{Typel:1999yq}
\\ &&&&&&&& \\
D$^3$C & $0.151$ & $-15.98$ & $232.5$ & $31.90$ & $59.30$ & $-74.7$ & $-430.50$ & \cite{Typel:2005ba}
\\ &&&&&&&& \\
\hline \\
DBHF & $0.185$ & $-15.60$ & $290$ & $33.35$ & $71.10$ & $-27.1$ & $-453.70$ & \cite{Li:1992zza,Brockmann:1990cn}
\\ &&&&&&&& \\
     & $0.181$ & $-16.15$ & $230$ & $34.20$ & $71$ & $87.36$ & $-340$ & \cite{GrossBoelting:1998jg}
\\ &&&&&&&& \\ 
\hline
\\ 
empirical & $~~0.167\pm 0.019$ & $-16\pm 1$ & $230\pm~10$ & 
$31.1\pm~1.9$ & $88\pm~25$ & -- & $-550\pm~100$ & \\
 & $\mbox{\small{\cite{Myers:1969zz,Myers:1984zz,Blaizot:1980tw}}}$ & $\mbox{\cite{Myers:1969zz,Myers:1984zz,Blaizot:1980tw}}$ & $\mbox{\cite{Blaizot:1980tw,PhysRevLett.82.691,PhysRevC.56.2518}}$ & 
$\mbox{\cite{Li:2008gp}}$ & $\mbox{\cite{Li:2008gp}}$ & -- & 
$\mbox{\cite{Li:2007bp}}$ &

\\ &&&&&&&& \\ 
\hline\hline
\end{tabular} 
\caption{Bulk saturation properties of nuclear matter, i.e.,  saturation density 
$\rho_{sat}$, binding energy per nucleon $E_{b}$, compression modulus $K$, asymmetry 
parameter $a_{sym}$, slope and curvature parameters $L$ and $K_{sym}$, respectively, 
and the observable $K_{asy}$ in the NLD model. Our results are compared with the 
non-linear Walecka parametrization NL$3^{*}$, the density dependent DD and the 
derivative coupling D$^{3}$C models as well as with two versions of the 
microscopic DBHF approach. The empirical values are shown too.}
\label{tab2}
\end{table*}

\begin{table*}
\begin{tabular}{c|c|c|cc|ccccc|ccc}
\hline\hline &&&&&&&&&&&& \\
 & $\nldr$ & cut-off & $~~\Lambda_s$ & $\Lambda_v~~$ & $~~g_{\sigma}$   & $g_{\omega}$ & $g_{\rho}$ & $b$          & $c$      & $~~m_{\sigma}$ & $m_{\omega}$ & $m_{\rho}$ \\
 &         &         & $[\GeV]$      & $[\GeV]$    &                &              &            & $[\fm^{-1}]$ &          & $[\GeV]$       & $[\GeV]$     & $[\GeV]$ 
\\&&&&&&&&&&&&\\ \hline\hline &&&&&&&&&&&&\\
NLD & $\displaystyle~~\formb~~$ & $\displaystyle ~~\formbnm~~$ & $0.95$ & $1.125~~$ & $~~10.08$ & $10.13$ & $3.50$ & $15.341$ & $-14.735~~$ & $~~0.592$ & $0.782$ & $0.763$ 
\\&&&&&&&&&&&& \\
\hline\hline
\end{tabular} 
\caption{The parameters of the NLD model. First and second columns show the form of the 
non-linear operator and the regulator in nuclear matter, respectively. The other columns 
show the values of the parameters, as resulted from the fit to nuclear matter bulk properties.}
\label{tab1}
\end{table*}

Various choices of the regulator functions are possible. We have done
calculations for different forms of $\nld$ and found that the simplest
momentum dependent monopole form factor provides the best description of the
low and high density nuclear matter properties and agrees very well with the 
empirical momentum dependence of the in-medium Schr\"odinger-equivalent 
optical potential. A momentum-dependent cut-off of a monopole form 
can be obtained by the following choice 
\begin{align}
\nldr = 
\left[
\frac{1}{1+\sum_{j=1}^{4}\left(\zeta_{j}^{\alpha} \, i\partialr_{\alpha}\right)^{2}}
\right] \,,
\end{align}
with  $v_{1}^{\alpha}=(0,0,0,0)$, $v_{2}^{\alpha}=(0,1,0,0)$, $v_{3}^{\alpha}=(0,0,1,0)$ and 
$v_{4}^{\alpha}=(0,0,0,1)$. In nuclear matter this results in
\begin{equation}
\nld = \frac{\Lambda^2}{\Lambda^2 + \vec{p}^{\,2}}\,.
\end{equation}
Furthermore, in our fit to bulk properties of nuclear matter we use different cut-off 
parameters $\Lambda_{s}\equiv\Lambda_{\sigma}$ and 
$\Lambda_{v}\equiv\Lambda_{\omega}=\Lambda_{\rho}$ for the scalar and vector 
meson-nucleon vertices, respectively, and we neglect in the following the contribution 
of the $\delta$-meson. 

The NLD model contains in total eight parameters. These are the meson-nucleon couplings 
$g_{\sigma}$, $g_{\omega}$ and $g_{\rho}$, the parameters $b$ and $c$ of the 
selfinteractions of the $\sigma$-meson, the mass $m_\sigma$ of the $\sigma$-meson, 
and the cut-offs $\Lambda_{s}$ and $\Lambda_{v}$. 
In principle, the masses of the $\omega$- and $\rho$-mesons should be also included in 
the fit. In all the calculations concerning the fit the results for these 
two masses turned out to be always around their free values. Therefore, 
we keep for $m_\omega$ and $m_\rho$ the bare masses. 
Seven parameters, {\it i.e.}, $g_{\sigma}$, $g_{\omega}$, $b$ and $c$, 
$\Lambda_{s}$, $\Lambda_{v}$ and $m_\sigma$ are adjusted to the bulk properties of 
symmetric nuclear matter and to the empirical energy dependence of the in-medium optical 
potential. The remaining parameter $g_{\rho}$ is determined by the experimentally known symmetry 
energy value at saturation density.
The constraints in symmetric nuclear matter  are the binding energy per nucleon 
$E_{b}\equiv\varepsilon/\rho_{B}-m$ and the compression modulus 
$K=9\rho_{sat}^{2}\frac{\partial^{2}E_{b}}{\partial\rho_{B}^{2}}\vert_{\rho_{B}=\rho_{sat}}$
at the ground-state or saturation density $\rho_{sat}$, and the saturation density 
$\rho_{sat}$ itself. 
Furthermore, the momentum dependence is fixed by 
the optical potential $U_{opt}$ at ground-state density and at two kinetic energies. 
First, at $E_{kin}=205$ $\MeV$ (where $U_{opt}=0$ $\MeV$) 
and at $E_{kin}=1000$ $\MeV$. 

The numerical calculations for the fit procedure have been performed using the 
Nelder-Mead minimization 
algorithm NELMIN~\cite{Nelder:1965zz}. Furthermore, an adaptive non-linear least-square 
package NL2SOL~\cite{Nl2sol:ref1,Nl2sol:ref2} has been supplemented in order to test robustness 
of the fit procedure. 
The experimental data used in the fit are supplemented by the corresponding errors
as provided by the empirical information in Table~\ref{tab2}.
At each iteration step of the minimization routines, the coupled set 
of the NLD equations has been solved with a Powell-hybrid method as provided by the 
HYBRD routine~\cite{HYBRD:ref1}. The integrals which appear in the source terms of the meson-field 
equations have been treated numerically using an adaptive Gauss algorithm. 
Note that, in the NLD model with the momentum-dependent regulator functions the solution 
of the dispersion relation for the single particle energy $E$, Eq.~(\ref{onshell}), 
does not involve additional complex root-finding algorithms, because the RMF 
selfenergies depend on the nucleon momentum $p$ only, and not on the particle energy 
as in our previous works~\cite{Gaitanos:2011yb,Gaitanos:2011ej,Gaitanos:2009nt}.

The functional form of the non-linear operator, its regulator (cut-off) in nuclear matter 
and the fit parameters of the NLD model are shown in Table~\ref{tab1}. Table~\ref{tab2} 
shows the extracted bulk properties of nuclear matter, 
{\it i.e.}, the binding energy, the compression modulus, 
the asymmetry parameter $a_{sym}=E_{sym}(\rho_{sat})$ ($E_{sym}$ is the symmetry 
energy), the slope and curvature parameters, 
$L=3\rho_{sat}\frac{\partial E_{sym}}{\partial\rho_{B}}\vert_{\rho_{B}=\rho_{sat}}$ and 
$K_{asy}=K_{sym}-6L$ (with 
$K_{sym}=9\rho_{sat}^{2}\frac{\partial^{2}
  E_{sym}}{\partial\rho_{B}^{2}}\vert_{\rho_{B}=\rho_{sat}}$ 
being the compressibility of the symmetry energy), respectively, in the 
NLD model, and in comparison with other RMF models widely used in the literature. These are 
the NL$3^{*}$ parametrization~\cite{Lalazissis:2009zz}, the density dependent (DD) and derivative 
coupling (D$^{3}$C) models~\cite{Typel:1999yq,Typel:2005ba}. The bulk nuclear matter properties 
in the NLD model are comparable with those results in the NL$3^{*}$, DD and 
D$^{3}$C parameterizations, while for the saturation density and the slope and 
curvature parameters of the symmetry energy the NLD calculation is closer to the empirical 
data. It is interesting that the NLD results are also comparable and in some cases even 
better than the DBHF calculations. 

\begin{figure}[t]
\begin{center}
\includegraphics[clip=true,width=1\columnwidth,angle=0.]
{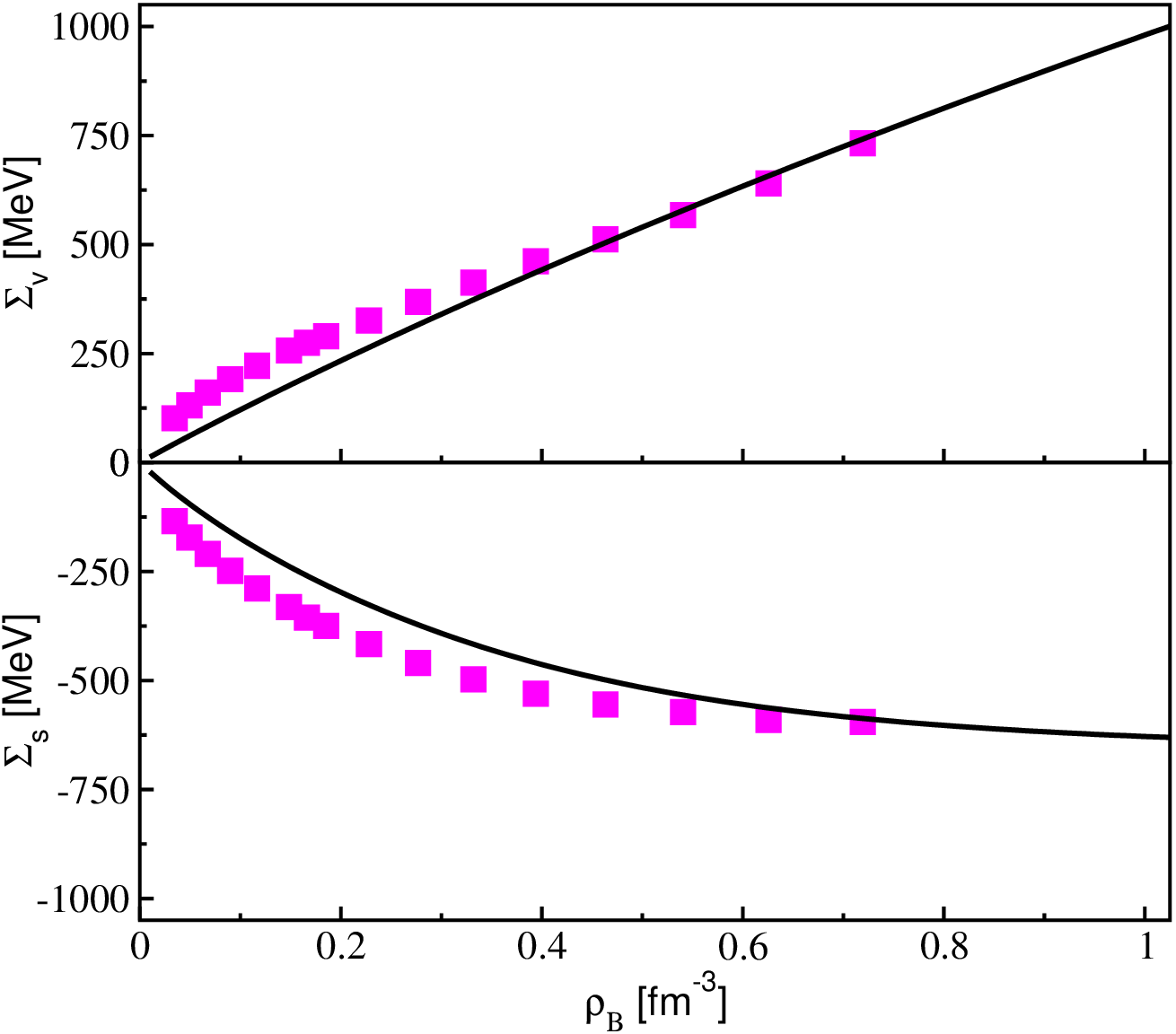}
\caption{\label{fig1} 
Vector (upper panel) and scalar (lower panel) selfenergies 
as function of the baryon density $\rho_{B}$ in the NLD model (solid curves) 
and in the DBHF approach (filled squares)~\cite{Brockmann:1990cn}. 
The calculations refer to isospin-symmetric ($\alpha=0$) nuclear matter.
\vspace{-0.3cm}
}
\end{center}
\end{figure}
As we will discuss later on in more details, the NLD model with the same parameters 
of Table~\ref{tab1} describes remarkably well  the empirical energy dependence of the 
Schr\"odinger-equivalent optical potential. This is not the case in standard RMF approaches, 
such as the NL$3^{*}$ and DD models, except if one uses supplementary derivative interactions 
(D$^{3}$C), however, with the cost of many additional parameters~\cite{Typel:2005ba}.

\subsection{\label{sec5b}The NLD equation of state}

We start the discussion of the NLD calculations with the density dependence of the 
relativistic mean-fields. Fig.~\ref{fig1} shows the density dependence of the 
nucleon selfenergies for isospin-symmetric nuclear matter in the NLD model. 
Our results are compared also with DBHF calculations, widely used in the 
literature~\cite{Brockmann:1990cn}. It is remarkable that a simple folding 
of the meson-nucleon couplings $g_i$ ($i=\sigma,\omega,\rho$) with the cut-off 
function $\nld$ results in a very similar density dependence of the of NLD 
selfenergies as compared to the DBHF selfenergies, 
in particular, for densities above $\rho_{B}>0.3-0.4~fm^{-3}$. 
Note again that for momentum dependent cut-off functions the standard normalization 
of the spinors, {\it i.e.}, $N_i=\sqrt{\frac{E^{*}_{i}+m^{*}_{i}}{2E^{*}_{i}}}$, 
is not affected in this case, 
since $\Pi^{0}_{i}=E^{*}_{i}$ holds, and according to Eq.~(\ref{bigPi}) only the pressure 
is modified, as shown in Eq.~(\ref{nldeq18b}). 

\begin{figure}[t]
\begin{center}
\includegraphics[clip=true,width=1\columnwidth,angle=0.]
{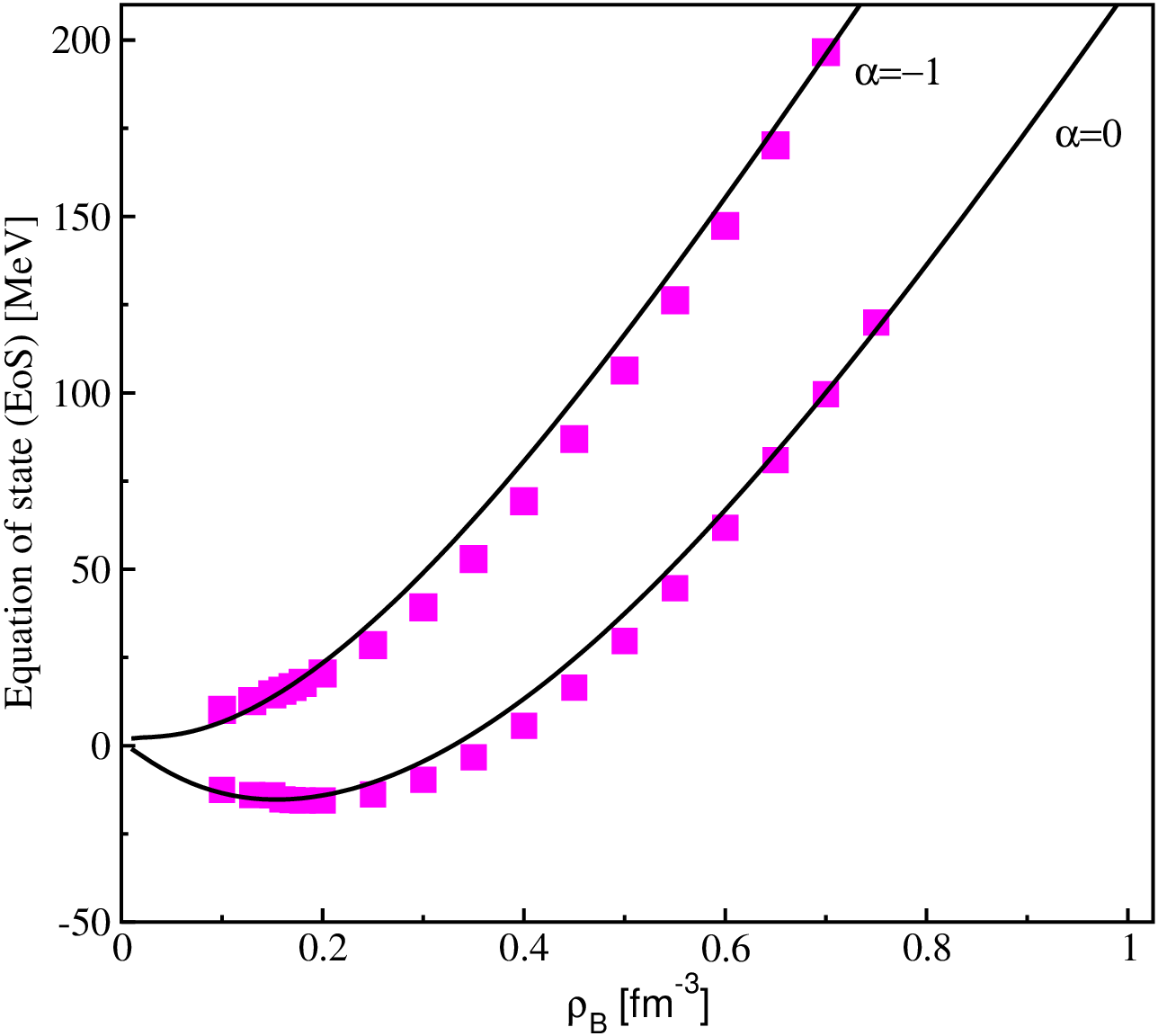}
\caption{\label{fig2}
Equation of state (EoS) in terms of the binding energy per nucleon as function of the 
baryon density $\rho_{B}$ for isospin-symmetric ($\alpha=0$, lower curve and symbols) and 
pure neutron matter ($\alpha=-1$, upper curve and upper symbols). The NLD results (solid curves) 
are compared with DBHF calculations (filled squares)~\cite{Li:1992zza}.
\vspace{-0.3cm}
}
\end{center}
\end{figure}

Fig.~\ref{fig2} shows the equation of state in terms of the binding energy per nucleon 
as function of baryon density for isospin-symmetric ($\alpha=0$) and pure neutron 
matter ($\alpha=-1$). The isospin-asymmetry parameter $\alpha$ is defined in the 
usual way as $\alpha = \frac{J^{0}_{p}-J^{0}_{n}}{J^{0}_{p}+J^{0}_{n}}$, where 
$J^{0}_{p,n}$ denote the proton and neutron densities, respectively. 
The momentum-dependent monopole form-factor of the NLD model regulates the 
high-density dependence of the fields such that the NLD EoS fits very well the 
DBHF calculations, for both, symmetric nuclear and pure neutron matters. Note that 
the NLD parameters are not fitted to the calculations of DBHF models, but to the 
empirical information at ground state density only. 

\begin{figure}[t]
\begin{center}
\includegraphics[clip=true,width=1\columnwidth,angle=0.]
{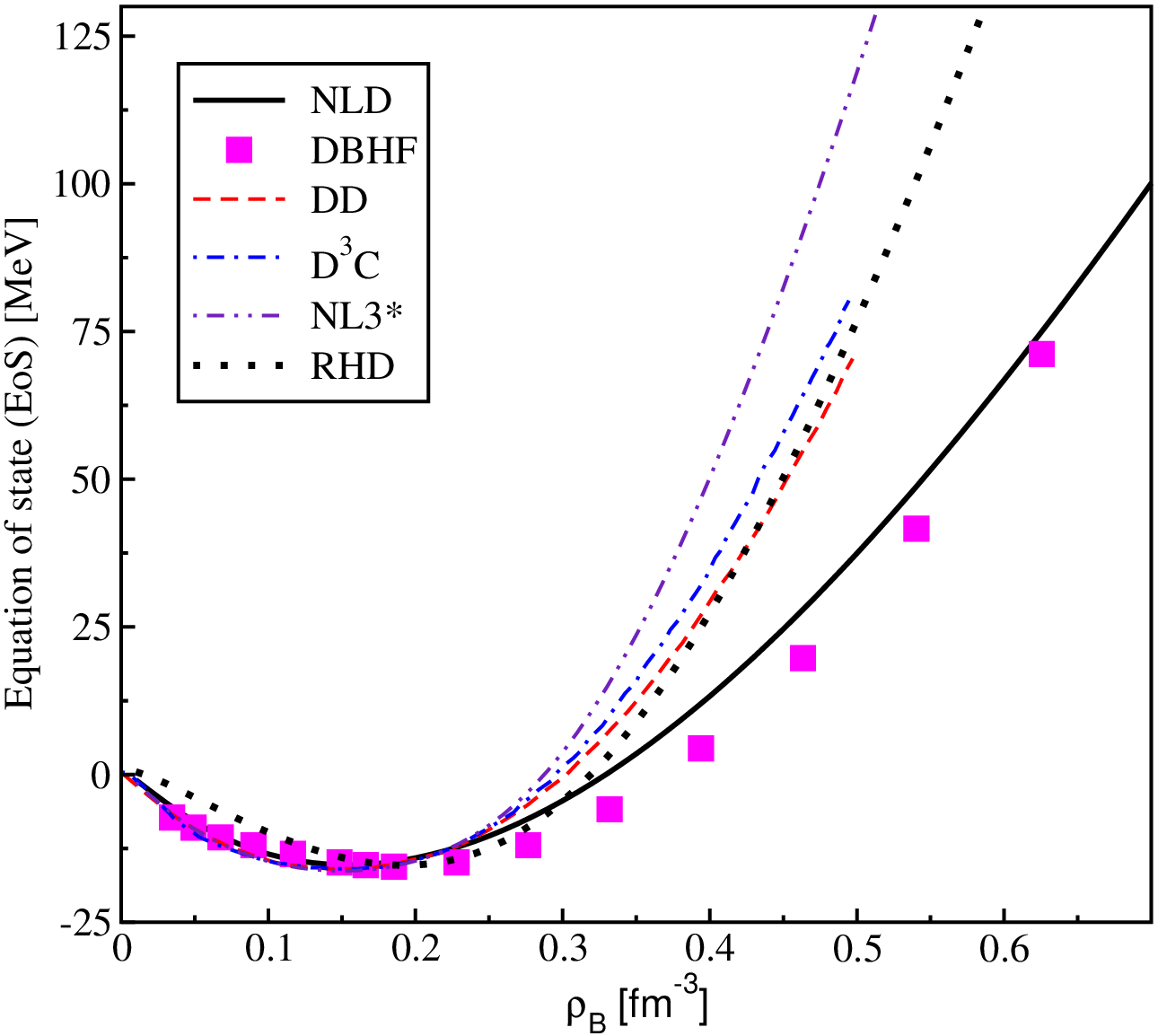}
\caption{\label{fig3} 
Same as in Fig.~\ref{fig2}, but now the NLD model (solid curve) 
is compared with various RMF calculations. These are the NL3$^{*}$ 
parametrization (dashed-dotted-dotted curve)~\cite{Lalazissis:2009zz}, 
the density-dependent (DD, dashed curve) and derivative-coupling 
(D$^{3}$C, dashed-dotted curve) models~\cite{Typel:1999yq,Typel:2005ba} 
and the linear Walecka model (RHD, dotted curve)~\cite{Walecka:1974qa}. 
The DBHF calculations (filled squares) are shown too. All the 
calculations refer to isospin-symmetric ($\alpha=0$) nuclear matter.
\vspace{-0.3cm}
}
\end{center}
\end{figure}

In Fig.~\ref{fig3} the density dependence of the EoS is displayed again, but now 
we compare the NLD calculation with other RMF models, which have been widely used 
in studies of finite nuclei and of nuclear matter. Even if all models give similar results 
for the saturation point of nuclear matter, the differences between the NLD model and 
the other RMF approaches at densities beyond the 
ground state are large. In particular, the NL3$^{*}$ parametrization of 
Lalazissis {\it et. al.}~\cite{Lalazissis:2009zz}, which is based on the Walecka model 
with non-linear 
selfinteractions predicts the stiffest EoS, while the RMF approaches DD and D$^{3}$C 
with density-dependent meson-nucleon couplings give a softer density behavior at high 
densities. However, none of these RMF approaches can reproduce the microscopic 
calculations of the DBHF model at such high densities. In general, the non-linear 
density dependence of the NLD model, induced by the cut-off form factor, result in a
much softer EoS at high densities, and it is in agreement with the DBHF 
calculations. Note that heavy-ion studies at energies just below the kaon production 
threshold ($E_{beam}\leq 1$ $GeV/A$) on collective nucleon 
flows~\cite{Danielewicz:2002pu,Sahu:1998vz,Gaitanos:2001hv} 
and on produced mesons (positive charged 
kaons)~\cite{Fuchs:2005zg,Fuchs:2000kp,Hartnack:2005tr,Hartnack:2011cn} 
support a rather soft EoS at densities around $\rho_{B}\simeq (2-3)\rho_{sat}$.

Furthermore, an interesting issue concerns the inclusion of the non-linear 
selfinteractions of the $\sigma$-field in the RMF descriptions. It is well known, 
that such terms may cause divergences at very high densities, where one intends 
to study compact neutron stars. Indeed, many RMF approaches such as the 
NL3$^{*}$~\cite{Lalazissis:2009zz} or the NL$\rho$ and 
NL$\rho\delta$~\cite{Gaitanos:2003zg} 
parameterizations, lead to a non-physical behavior of the $\sigma$-field at 
supra-normal densities. This is not the case in the NLD model, where the non-linear 
selfinteraction terms of the $\sigma$-meson does not cause any divergences of the 
$\sigma$-field even at supra-normal densities. This novel NLD feature arises from 
the suppression of the scalar sources by the cut-off function $\nld$, which in the 
limiting case of large densities $\rho_{B}$ tend to vanishes. 

\begin{figure}[t]
\begin{center}
\includegraphics[clip=true,width=1\columnwidth,angle=0.]
{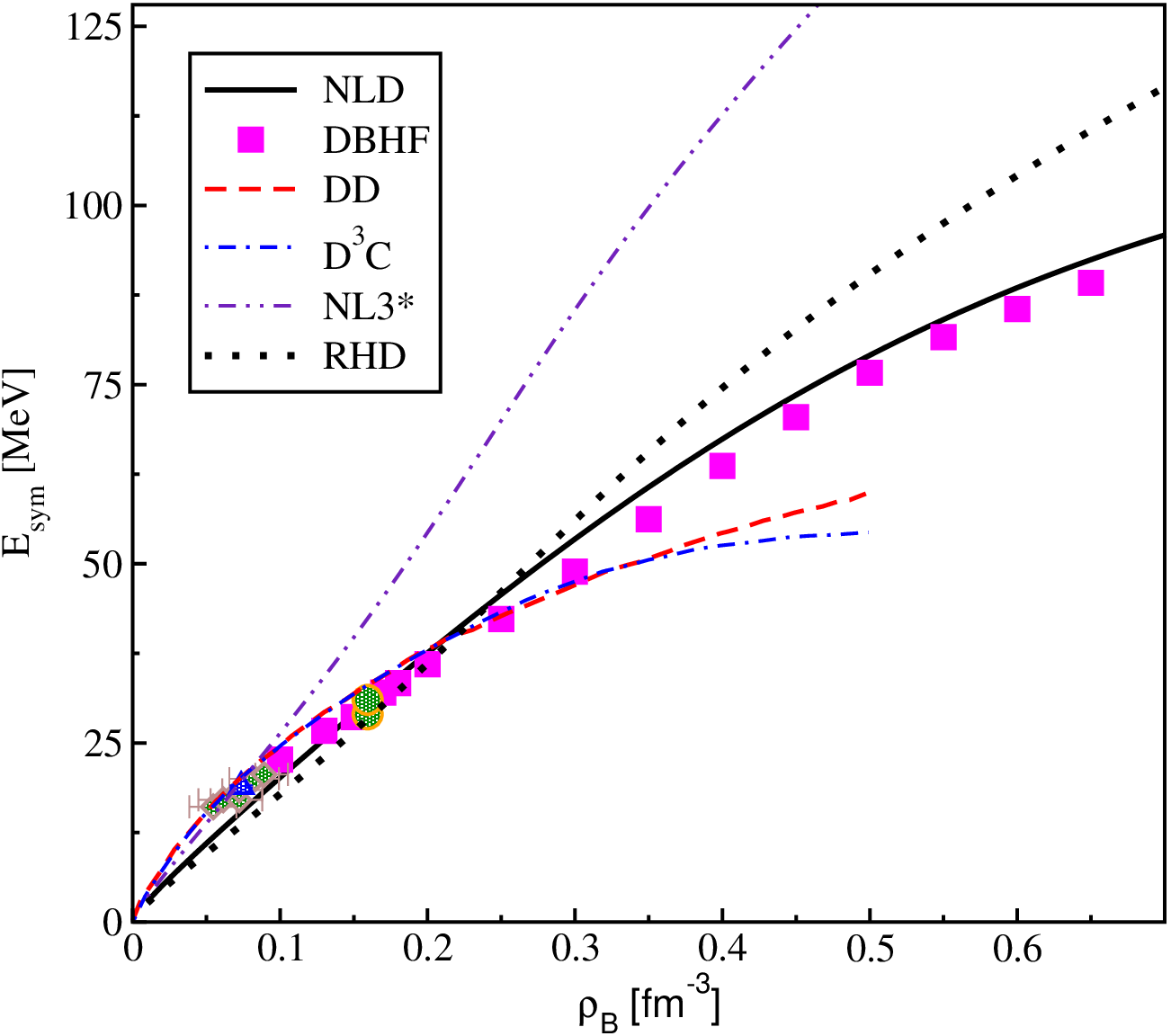}
\caption{\label{fig4}
Same as in Fig.~\ref{fig3}, but now for the 
symmetry energy $E_{sym}$ as function of baryon density $\rho_B$. 
The additional symbols (diamonds, triangle and circles) around and below 
the saturation density refer to 
empirical data~\cite{Shetty:2010ib,Shetty:2007zg}.
\vspace{-0.3cm}
}
\end{center}
\end{figure}

Another quantity of interest is the symmetry energy. This quantity is important 
for astrophysical applications, since it directly influences the density dependence 
of the proton-fraction in the neutron star interior. It is 
extracted as the difference between the EoS's from pure neutron and symmetric 
matters. Fig.~\ref{fig4} shows the symmetry energy as function of the 
baryon density in the NLD model, in other RMF models and in the DBHF approach. 
Note that the NLD results differ from the previous 
works~\cite{Gaitanos:2011yb}. This is because now the NLD model consistently includes the 
renormalization of the spinors. This is essential for the thermodynamic consistency of the 
formalism, which was not taken into account in Ref.~\cite{Gaitanos:2011yb}. Furthermore, 
a different form of the regulator is used here and the parameters 
of the NLD model are fitted using low-density observables at saturation. 
At first, all models describes fairly well the empirically known region around the 
saturation density, as it was already shown in Table~\ref{tab2}. However, the 
differences between the models become more pronounced with increasing density. The 
most stiffer symmetry energy is obtained again in the NL3$^{*}$ and the RHD calculations, 
because in these cases $E_{sym}$ is just proportional to the $\rho$-meson, which 
linearly increases with density. Furthermore, the symmetry energy is considerably 
softened in the DD and D$^{3}$C approaches, due to the exponentially decreasing 
density dependence of the $\rho$-nucleon vertices. 
On the other hand, the NLD model leads to a softer density behavior of the
symmetry energy relative to the standard NL and RHD parametrization but to a 
stiffer dependence than the density-dependent approaches. It is again
an interesting feature that the NLD calculations predict almost perfectly the calculations 
of the microscopic DBHF calculations~\cite{Li:1992zza}.

\subsection{\label{sec5c}In-medium nucleon optical potentials}

Apart the density dependence, which arises from both, the source terms of the 
meson-field equations and the cut-off functions, the NLD selfenergies 
depend explicitly on momentum. 
The momentum (or energy) behavior of the RMF fields is described by the 
in-medium Schr\"odinger-equivalent optical potential. This quantity results from the 
reduction of the Dirac equation and reads~\cite{Jaminon:1989wj}
\begin{align}
U_{opt} = -S + \frac{E}{m}V + 
\frac{1}{2m}\left( S^{2}-V^{2}\right)
\label{Uopt}
\,,
\end{align}
where the selfenergies $S=\Sigma_{si}(\rho_{B},p)$,~$V=\Sigma_{vi}^{0}(\rho_{B},p)$ 
refer to a proton ($i=p$) or a neutron ($i=n$) with a particular momentum $|\vec{p}\,|=p$ 
relative to nuclear matter at rest at a given density $\rho_{B}$ and isospin-asymmetry 
$\alpha$. 
\begin{figure}[t]
\begin{center}
\includegraphics[clip=true,width=1\columnwidth,angle=0.]
{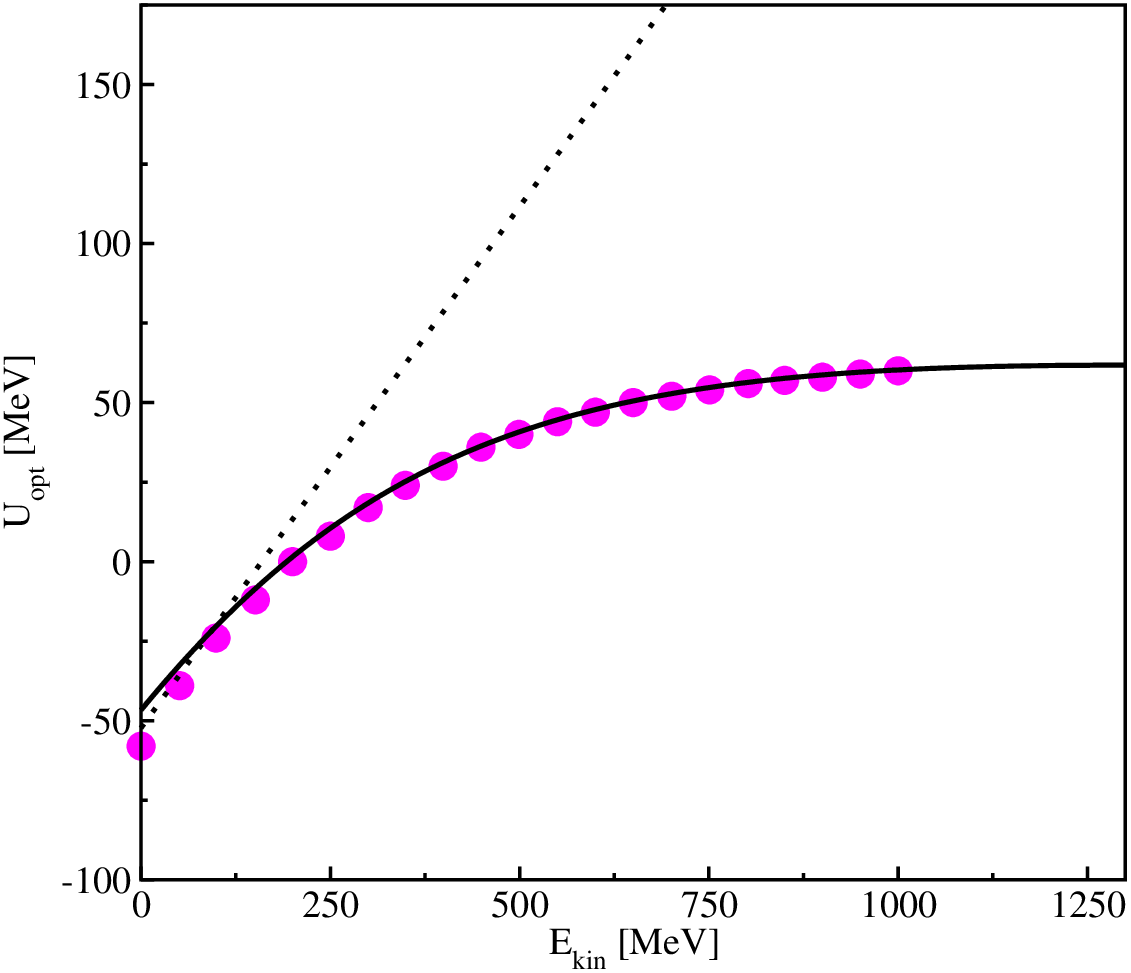}
\caption{\label{fig5} 
(Color online) 
The in-medium proton optical potential $U_{opt}$ according Eq.~(\ref{Uopt}) 
as function of the in-medium single-particle kinetic energy, Eq.~(\ref{Ekin}), 
within the NLD model (solid curve) and the the RHD~\cite{Walecka:1974qa} 
approach at saturation density. The symbols (filled circles) refers to 
the results of the Dirac phenomenology~\cite{Cooper:1993nx,Hama:1990vr}. 
\vspace{-0.3cm}
}
\end{center}
\end{figure}
The single-particle 
energy $E$ is calculated from the in-medium dispersion relation
\begin{align}
E = \sqrt{m^{*2}+p^{2}} + V
\label{Ekin}
\end{align}
and the kinetic energy reads $E_{kin}=E-m$~\cite{Typel:2005ba,Haar:1986ii,TerHaar:1987ce}. 
Here we consider the real part of $U_{opt}$ only, which enters the momentum dependent 
mean-field dynamics of a nucleon in nuclear matter. The imaginary part, which is beyond 
the scope of the present work, can be further accounted for in the collision integral 
within the relativistic transport equation, see for instance Ref.~\cite{Botermans1990115}, 
when applying the NLD approach to the description of  heavy-ion collisions.  
Alternatively, one can use a dispersion theoretical framework to calculate the 
imaginary part of the optical potential using as an input the real part of the 
optical potential   in the dispersion integral, see Ref.~\cite{Gaitanos:2011ej}  
as an example of such approach to the imaginary part of $U_{opt}$. 
In this context we would like to note, that the imaginary contributions of the
relativistic mean-fields do not influence essentially the real part of the 
optical potential, see for example Refs.~\cite{Weber:1992qc,Typel:2005ba}. 
The reason is, that the 
imaginary contributions of the selfenergies, as obtained from empirical 
studies, are of similar magnitude. They enter only through the terms quadratic 
in the selfenergies in the expression for the real part of the 
Schr\"odinger-equivalent optical potential, Eq.~(\ref{Uopt}), 
and therefore they almost cancel each other.

Fig.~\ref{fig5} shows the results of our calculations in comparison with the empirical 
data extracted from Dirac phenomenology~\cite{Cooper:1993nx,Hama:1990vr} for the in-medium 
proton optical potential in symmetric ($\alpha=0$) nuclear matter at saturation density 
$\rho_{B}=\rho_{sat}$. The optical 
potential saturates with increasing single-particle energy in the NLD model and it 
is in agreement with the experimental data. The saturation mechanism is 
induced by the explicit momentum dependence of the cut-off functions, which drop 
with increasing momentum. 

\begin{figure}[t]
\begin{center}
\includegraphics[clip=true,width=1\columnwidth,angle=0.]
{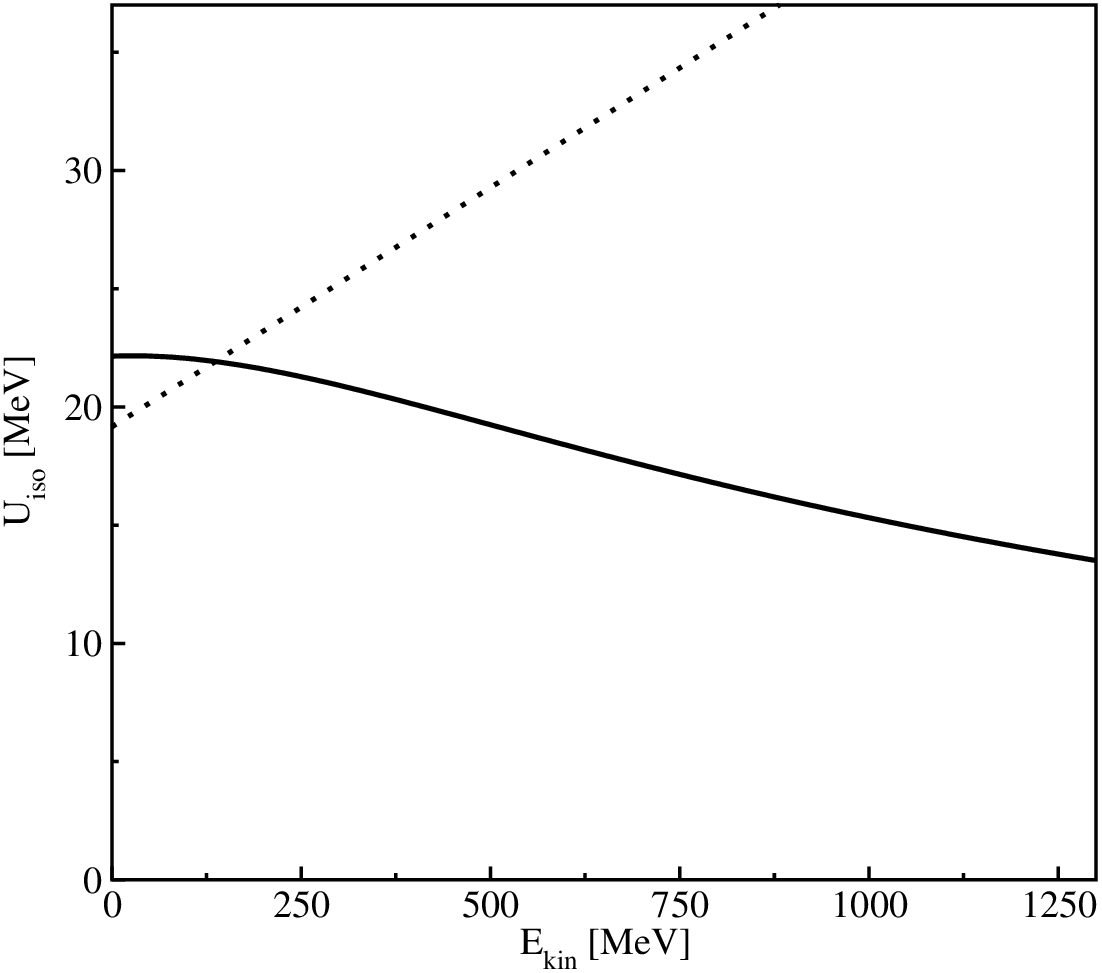}
\caption{\label{fig6} 
Energy dependence of the Lane-type optical potential $U_{iso}$, Eq.~(\ref{Uiso}), 
for asymmetric ($\alpha=0.4$) nuclear matter at saturation density. 
Calculation in the RHD~\cite{Walecka:1974qa} (dotted curve) and NLD (solid curve) models 
are shown.
\vspace{-0.3cm}
}
\end{center}
\end{figure}
It is a novel feature of the NLD model, that the regulators with a simple 
momentum-dependent monopole form are sufficient to describe accurately the bulk 
properties of nuclear matter and at the same time the empirical energy dependence of the 
optical potential. In fact, this issue has been a long-standing problem in nuclear 
matter studies, when one attempted to describe heavy-ion reactions within RMF 
models~\cite{Blaettel:1993uz}. As also shown in Fig.~\ref{fig5}, in standard RMF models, 
such as the widely used NL-parameterizations~\cite{Lalazissis:2009zz,Lalazissis:1996rd}, 
which describe excellent the 
saturation properties and also the properties of finite nuclei, cannot reproduce the 
correct energy dependence of the optical potential. Moreover, they strongly diverge with increasing
energy. Similar conclusions are obtained in the RMF approaches with density-dependent 
vertex functions~\cite{Typel:1999yq}, except if one includes additional
energy-dependent terms~\cite{Typel:2005ba}. 
Note that, the microscopic DBHF models~\cite{Haar:1986ii,TerHaar:1987ce,Plohl:2005fn} 
describe satisfactory the empirical data at low energies bellow the pion production threshold.

For isospin-asymmetric nuclear matter the key quantity is the so-called Lane 
potential~\cite{Li:2008gp,Baran:2004ih}, which is defined as the difference between 
the neutron and proton optical potentials
\begin{align}
U_{iso} = \frac{U_{n}-U_{p}}{2|\alpha|}
\label{Uiso}
\,.
\end{align}
Fig.~\ref{fig6} displays the energy dependence of the Lane potential. In contrast 
to the case of isospin-symmetric matter, empirical information is here less known. 
The studies from Ref.~\cite{Li:2008gp} predict a decrease of the Lane potential with 
increasing momentum, but other analysis~\cite{Baran:2004ih} conclude the opposite 
trend with the result of an increasing Lane potential. The microscopic DBHF calculations
predict also  a decreasing  optical potential, which is in agreement with results of 
other BHF calculations~\cite{Zuo:2005hw,vanDalen:2005sk}. 
Furthermore, the standard RHD model leads to an 
almost quadratic (or linear in energy) dependence in momentum, just because of the missing 
momentum dependence in the RHD selfenergies (similar results are obtained in the 
NL${}^{*}$ and DD parameterizations). The NLD calculations predict a weakly 
decreasing Lane potential, which is in qualitative agreement with the (D)BHF results.

\begin{figure}[t]
\begin{center}
\includegraphics[clip=true,width=1\columnwidth,angle=0.]
{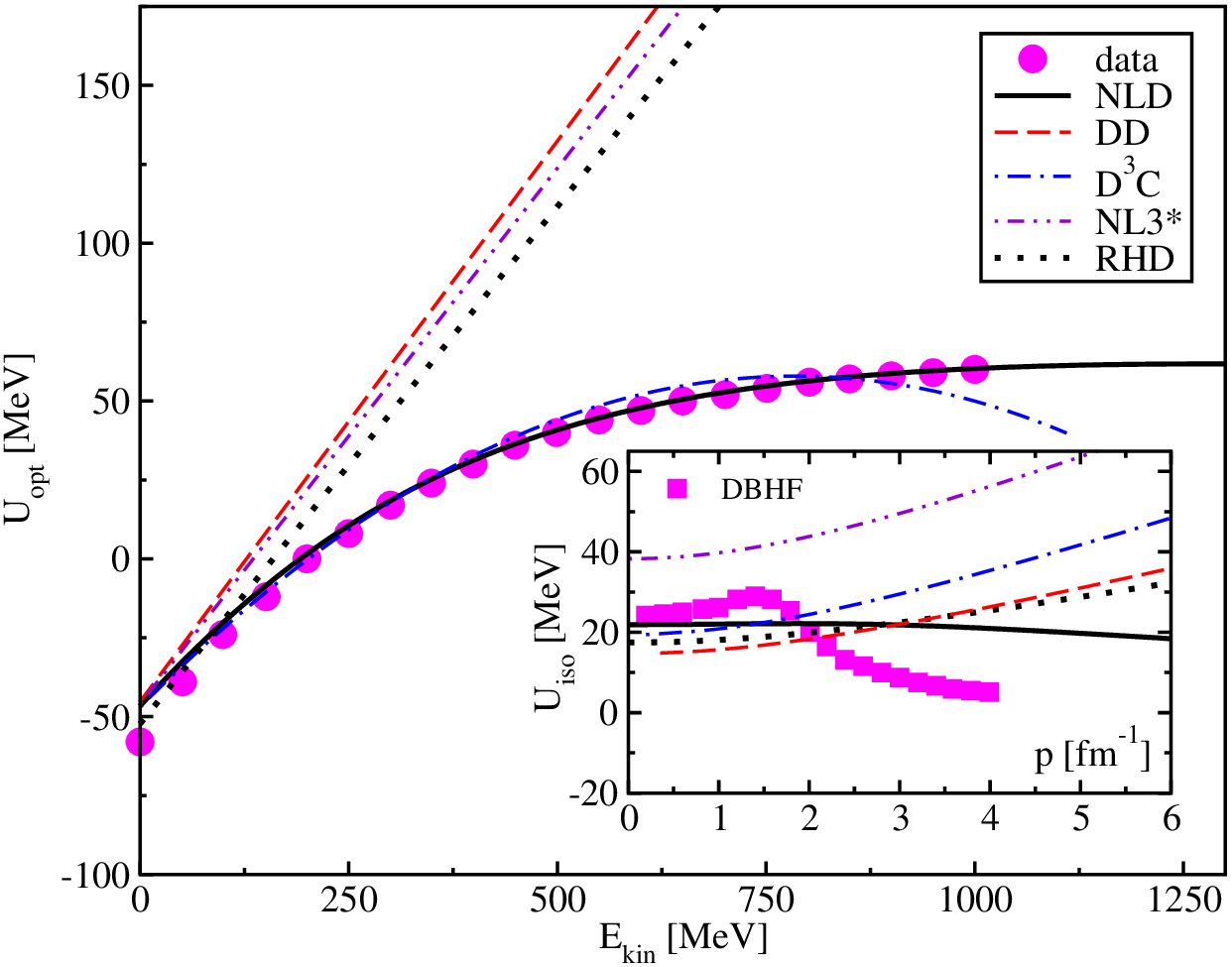}
\caption{\label{fig6a} 
Energy (momentum) dependence of the optical $U_{opt}$ (Lane) potentials 
in the main (inserted) panel. The curves have the same meaning as in Figs.~\ref{fig3} 
and~\ref{fig4}. 
Here the D$^{3}$C results are taken from~\cite{typelpc}.
\vspace{-0.3cm}
}
\end{center}
\end{figure}
We compare now the NLD results separately in Fig.~\ref{fig6a} 
for both, the in-medium proton optical potential for symmetric nuclear matter and the 
Lane potential, with the same RMF approaches as in Figs.~\ref{fig3} and~\ref{fig4}. Not 
only the original linear Walecka model (RHD), but also more modern RMF approaches, such as 
the non-linear NL3 model in its updated version (NL3$^{*}$) and the density-dependent RMF 
model (DD) predict an energy dependence, which is not consistent with phenomenology 
(filled circles). This circumstance was improved in a modified DD model (D$^{3}$C) by the 
inclusion of terms linearly proportional to the single-particle energy. The NLD model 
provides here a smoother energy dependence. 

The situation for the Lane potential is shown in the insert of Fig.~\ref{fig6a}. The interesting 
issue here is, that not only the standard RMF models (RHD, NL3$^{*}$ and DD), but also the 
energy-dependent RMF approach (D$^{3}$C), show a common behavior in momentum. This is due 
to the fact that the isovector channel in the D$^{3}$C RMF model does not include any 
explicit momentum dependence~\cite{typelpc}. On the other hand, in the NLD model 
also the isovector mean-field depends on momentum and predicts a decrease of the Lane potential 
with increasing momentum. This NLD trend seems to be supported by microscopic DBHF (filled 
symbols in the insert of Fig.~\ref{fig6a}), at least on a qualitative level.

\begin{figure}[t]
\begin{center}
\includegraphics[clip=true,width=1\columnwidth,angle=0.]
{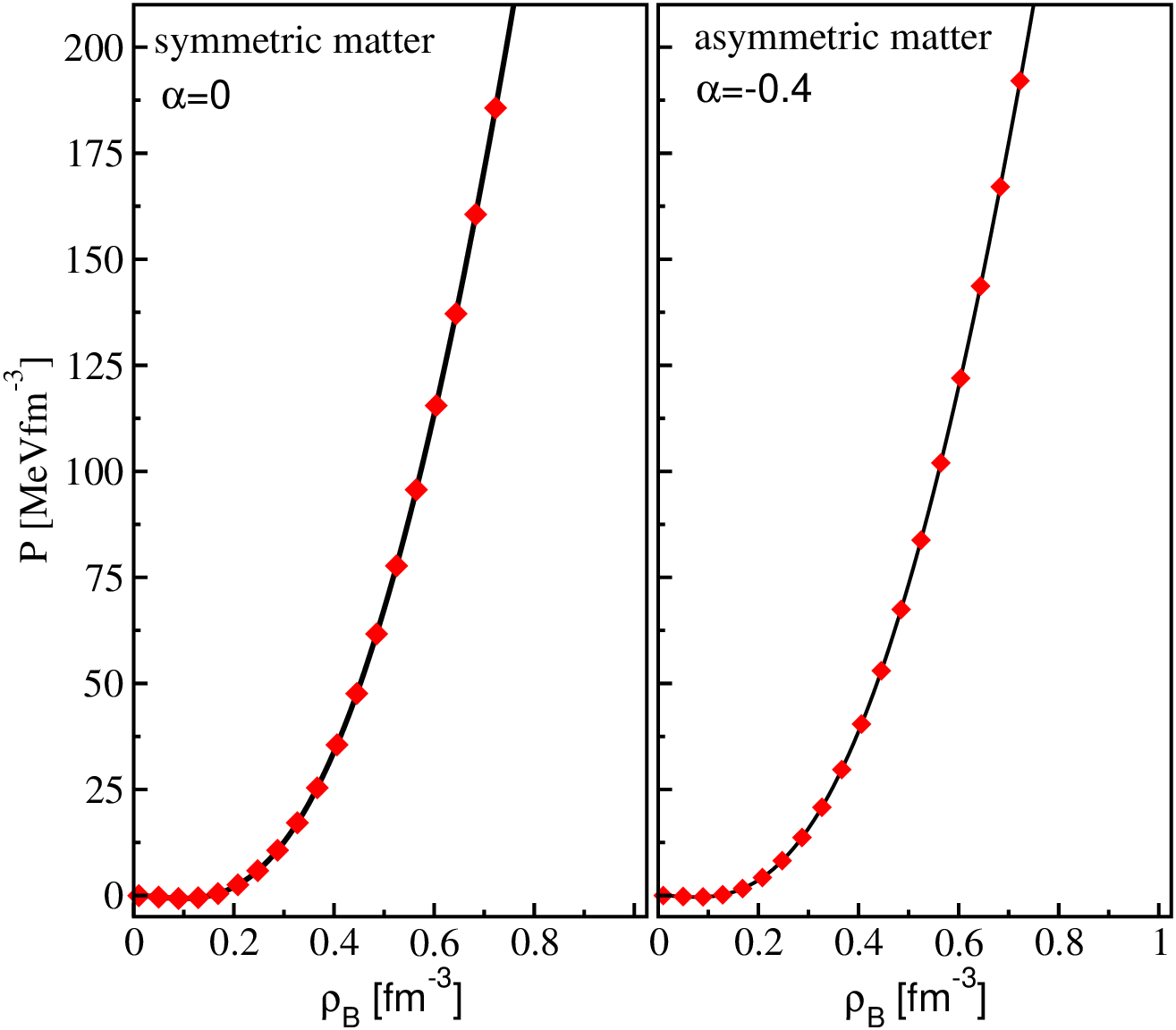}
\caption{\label{fig7}
Thermodynamic consistency of the NLD model 
for pure symmetric nuclear ($\alpha=0$, left panel) and asymmetric 
($\alpha=-0.4$, right panel) matters. 
The pressure densities $P$ as function of the baryon density $\rho_{B}$ are shown, as 
calculated within the perfect fluid formula (solid curves) using the l.h.s. 
of Eq.~(\ref{press}) and from the thermodynamic definition (filled diamonds) using 
the r.h.s. of Eq.~(\ref{press}).
\vspace{-0.3cm}
}
\end{center}
\end{figure}

\subsection{\label{sec5d}Thermodynamic consistency of NLD model}

As showed by Weisskopf~\cite{Weisskopf:1957} in an independent-particle model, the single 
particle energy at the Fermi surface must be equal the average energy per particle 
at saturation density. Hugenholtz and Van~Hove~\cite{Hugenholtz:1958} proved this also 
for an interacting Fermi gas at zero temperature.

For the thermodynamic consistency of the RMF model it is sufficient to prove Euler's theorem
\begin{align}
\varepsilon = -P + \sum_{i=p,n}\mu_{i}\rho_{B_{i}}
\label{euler}
\,,
\end{align}
with the chemical potentials $\mu_{i}=\frac{\partial\varepsilon}{\partial\rho_{B_{i}}}$
and the thermodynamic definition of the pressure
\begin{align}
P=\rho_{B}^{2}\frac{\partial(\varepsilon/\rho_{B})}{\partial\rho_{B}}\,.
\end{align}

\begin{figure}[t]
\begin{center}
\includegraphics[clip=true,width=1\columnwidth,angle=0.]
{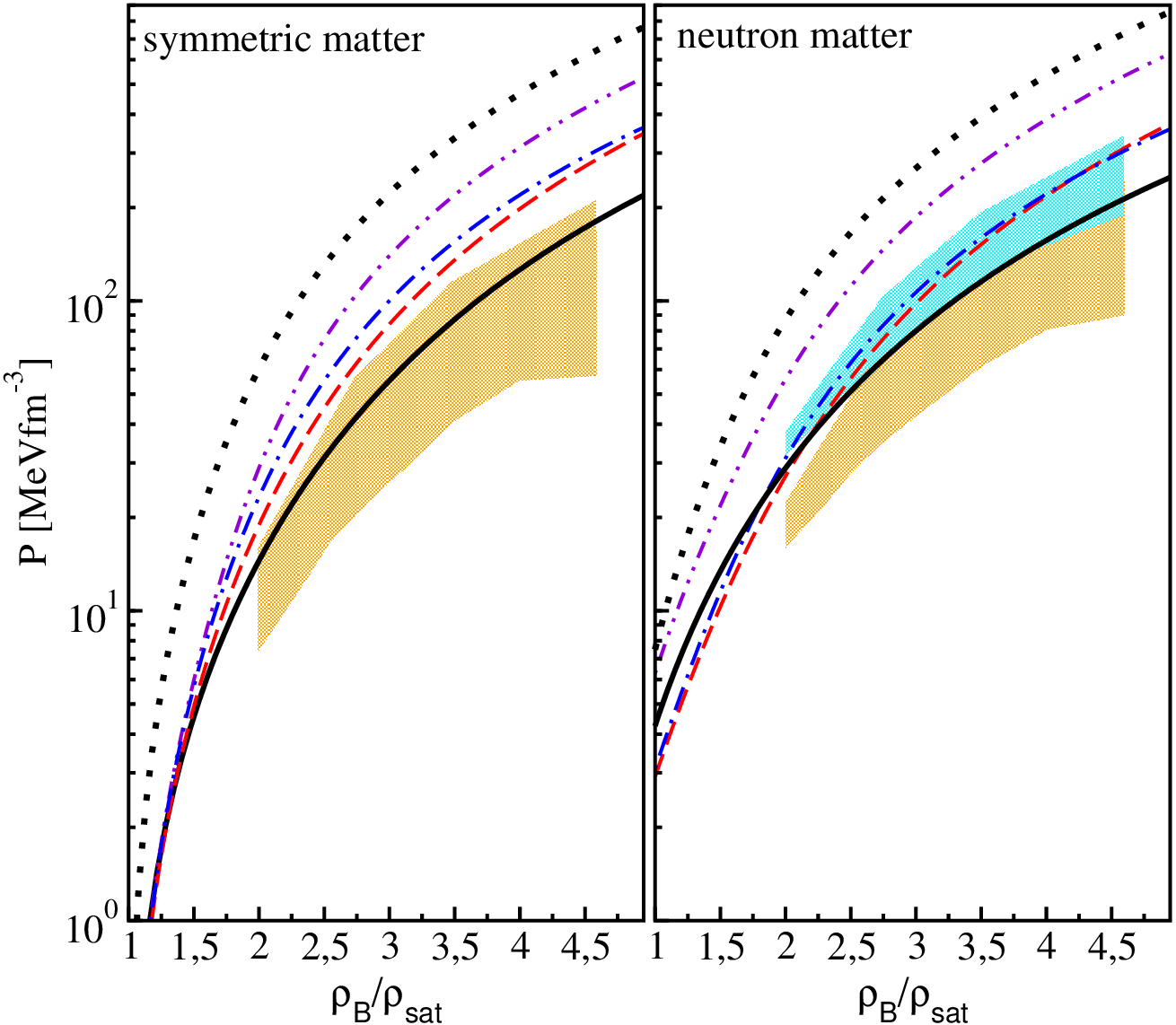}
\caption{\label{fig8} 
Equation of state in terms of the pressure densities as function of the baryon density 
(normalized to the corresponding saturation values $\rho_{sat}$). 
The shaded areas denote possible experimental regions, as extracted from studies on 
heavy-ion collisions~\cite{Danielewicz:2002pu}. The different curves have the same meaning 
as in Figs.~\ref{fig3} and~\ref{fig4}.
\vspace{-0.3cm}
}
\end{center}
\end{figure}

The expression in Eq.~(\ref{euler}) allow to examine 
the internal consistency of the model. For this purpose it is sufficient 
to check the equality between the pressure obtained from the assumption of nuclear 
matter as a prefect-fluid system and the pressure obtained from the thermodynamic 
definition. That is 
\begin{align}
P=\frac{1}{3}\; (T^{xx}+T^{yy}+T^{zz}) = \rho_{B}^{2}
\frac{\partial(\varepsilon/\rho_{B})}{\partial\rho_{B}}
\label{press}
\,,
\end{align}
where $\varepsilon$ and $T^{ii}$ ($i=x,y,z$) denote the energy density and 
the spatial diagonal components of the energy-momentum tensor 
tensor $T^{\mu\nu}$, respectively. 

Conventional RMF approaches with bare meson-nucleon couplings or density-dependent 
meson-nucleon vertices are thermodynamically 
consistent~\cite{Lalazissis:1996rd,Serot:1997xg,Typel:2002ck,Boguta:1981yn}. 
However, in the 
case of explicit energy or momentum-dependent mean-fields the situation is more 
complex. In fact, as we have examined here, one has to take care of the proper 
renormalization of the Dirac fields.  This issue was not taken into 
account in previous studies~\cite{Gaitanos:2009nt} resulting in a deviation between the l.h.s. 
and r.h.s. of Eq.~(\ref{press}) by $\simeq~5-10\%$ at high baryon densities. 

We have checked that 
the NLD model presented here satisfies thermodynamic consistency exactly for 
any kind of energy- or momentum-dependent cut-off form factors. This is demonstrated in 
Fig.~\ref{fig7} for the monopole form factor, where the pressure as function of density 
for symmetric nuclear matter (left panel) and pure neutron matter (right panel) are shown. 
Indeed, the pressures obtained from both definitions of Eq.~(\ref{press}) agree exactly. 
Furthermore, the chemical potentials at zero temperature are equal to the corresponding 
Fermi energies of protons and neutrons (not shown here) and therefore Euler's theorem is 
evidently fulfilled.

We can now consider the high-density behavior of the pressure by 
comparing our results with available empirical informations. Fig.~\ref{fig8} shows 
the density dependence of the pressure densities in symmetric matter and pure neutron 
matter. The conventional RHD approach leads again to a stiff behavior for all the densities. 
Similar conclusions are drawn for the NL3$^*$, DD and D${}^{3}$C approaches 
for symmetric matter, where for pure neutron matter the density-dependent 
models come closer to the empirical HIC data. A more detailed discussion on these 
approaches can be found in Ref.~\cite{Klahn:2006ir}. 
The pressure in the NLD model show generally a softer density dependence and agree better with 
the estimated experimental regions, in particular, at densities up to 
$\rho_{B}\simeq 4\rho_{sat}$ for the symmetric case and for all densities for pure neutron 
matter. Note that for larger densities conclusions on the nuclear 
matter EoS from heavy-ion studies are more ambiguous, because at such high densities (or 
corresponding beam energies larger than 4 $\GeV$ per nucleon) a large fraction of the initial 
energy is converted into new degrees of freedom~\cite{Danielewicz:2002pu,Klahn:2006ir}.

\subsection{\label{sec5e}High density observables and Neutron Stars}

\begin{figure}[t]
\begin{center}
\includegraphics[clip=true,width=1\columnwidth,angle=0.]
{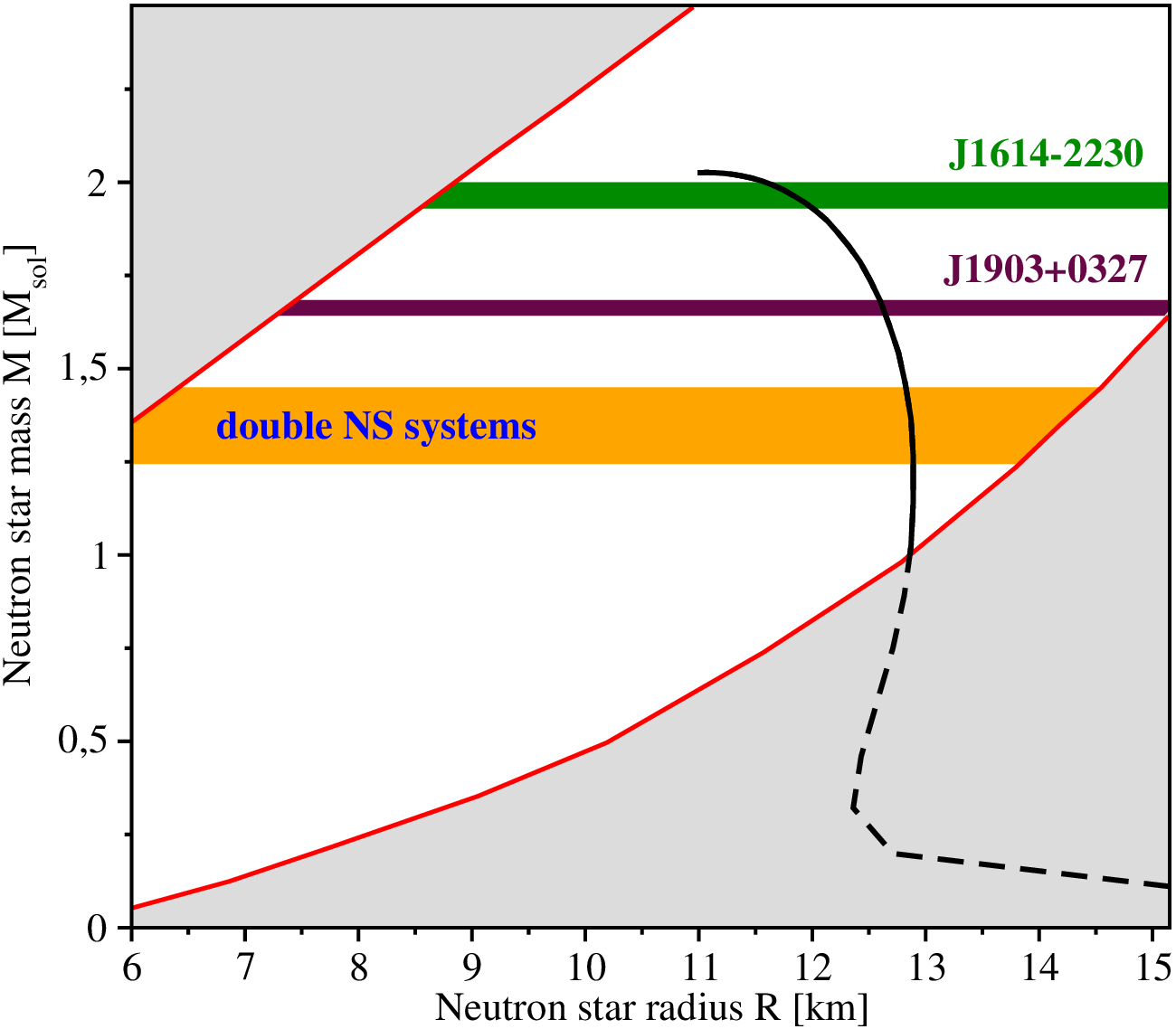}
\caption{\label{fig9} 
Relation between Neutron star mass $M$ (in units of the solar mass $M_{\odot}$) 
versus radius $R$ in the NLD model (dashed and solid curves). 
The three horizontal shaded bands refer to astrophysical measurements 
from double neutron star (NS) systems~\cite{Lattimer:2004pg,Lattimer:2006xb} 
and from the pulsars PSR J1903+0327~\cite{Freire:2010tf} and 
PSR J1614-2230~\cite{Demorest:2010bx}, as indicated. The other shaded areas 
bordered by thick curves indicate parameter space excluded by general 
relativity, causality (shaded area on the top-left) and rotational 
constraints (shaded area on the bottom-right)~\cite{Lattimer:2004pg,Lattimer:2006xb}. 
\vspace{-0.3cm}
}
\end{center}
\end{figure}

We test now the high density domain of the NLD equation of state. Compact neutron 
stars offer such an opportunity to gain deeper insight into compressed baryonic 
matter. Of particular interest are recent measurements on the binary millisecond 
pulsar J1614-2230 with a mass of $(1.97\pm 0.04)~M_{\odot}$~\cite{Demorest:2010bx}. 
The latter is much heavier than the average mass 
 of the binary radio pulsars $M=1.35\pm 0.04~M_{\odot}$~\cite{Thorsett:1998uc}
 and provides a strong constraint on high density EoS.
Therefore, we apply the NLD model to spherical, non-rotating stars in $\beta$-equilibrium 
between protons, neutrons and electrons including crustal effects on the 
EoS~\cite{Baym:1971pw}. The star structure is calculated by solving the 
Tolman-Oppenheimer-Volkov (TOV) equation~\cite{Glendenning:2005ix,Sagert:2005fw,Weber:2004kj}.

The results for neutron stars are shown in Fig.~\ref{fig9} in terms 
of the mass-radius relation. The various 
astrophysical measurements of NS masses~\cite{Lattimer:2004pg,Lattimer:2006xb} 
can be arranged in the three 
horizontal shaded areas as displayed in Fig.~\ref{fig9}. The lowest band 
around an average value of $1.44~M_{\odot}$ refer to the well established 
measurements on double neutron star systems and the middle one around 
$1.67~M_{\odot}$ on the extracted mass of the pulsar PSR J1903+0327. The band 
on the top represents by far the highest precisely observed neutron star mass 
$1.97\pm 0.04~M_{\odot}$
of the pulsar PSR J1614-2230~\cite{Demorest:2010bx}. There are two regions in
Fig.~\ref{fig9} excluded~\cite{Lattimer:2004pg,Lattimer:2006xb} by 
general relativity, causality (shaded area on the top-left) and rotational 
constraints (shaded area on the bottom-right). 

\begin{figure}[t]
\begin{center}
\includegraphics[clip=true,width=1\columnwidth,angle=0.]
{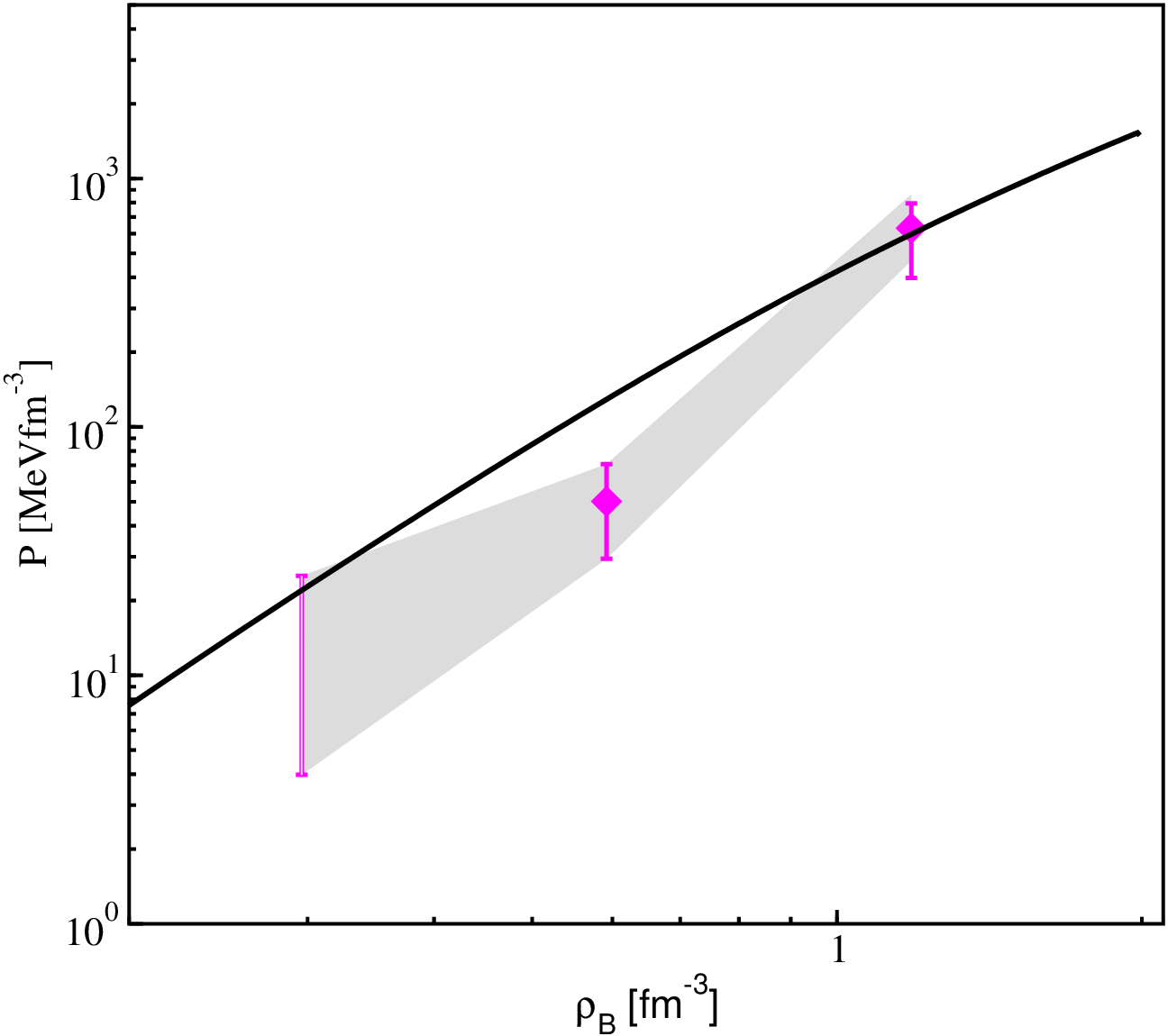}
\caption{\label{fig10} Pressure densities as function 
of the baryon density for nuclear matter in $\beta$-equilibrium within 
the NLD (solid curve) model. The shaded areas together with the three 
error bars indicate the pressure-density region extracted directly from 
neutron-star measurements~\cite{Ozel:2010fw}.
\vspace{-0.3cm}
}
\end{center}
\end{figure}

The neutron stars mass-radius relation in the NLD model is shown by the solid/dotted curve
in~Fig.~\ref{fig9}. The dotted part of the NLD curve is 
excluded by rotational constraints. The solid curve crosses the low-mass 
constraints, and arrives to a maximum neutron star mass of $M=2.03~M_{\odot}$ at 
a radius of $R=11.07$ km and a corresponding central density of 
$\rho_{c}\simeq 7~\rho_{sat}$. The NLD prediction for the maximum value 
of neutron star masses crosses also the constraint provided  by the pulsar 
PSR J1614-2230, and therefore this recent mass measurement is accommodated by the 
NLD model. 

Other possible constraints on the high-density EoS are obtained by statistical 
Bayesian analyses, which rely on neutron star 
measurements~\cite{Ozel:2010fw,Steiner:2010fz}. 
They provide the most probable distribution of the equation of state, 
as shown in Fig.~\ref{fig10} (shaded area)
for highly compressed matter in $\beta$-equilibrium. 
The RHD model (not shown in this figure) leads to a too stiff density
dependence and it overestimates this empirical region particularly at high densities. 
The NLD calculations, where only nucleonic degrees 
of freedom are accounted for, describe fairly well the most probable region 
of the pressure at high densities. 
Note again that the parameters of the NLD model have been adjusted just to the saturation 
properties of nuclear matter and to the energy dependence of the in-medium proton optical 
potential at saturation, without the consideration of any other high density 
observables in the fit procedure. We conclude here that the NLD describes well the available 
low- and also high-density constraints on EoS of nuclear matter and 
neutron stars.

\section{\label{sec6}Summary}

In summary, in the present work 
the generalized form of the energy-momentum tensor in the NLD model was
derived which allowed us to consider different forms of the regulator
functions in the NLD Lagrangian. The thermodynamic consistency of the NLD
model was further demonstrated for
arbitrary choice of the regulator functions. A thorough study of the
properties of nuclear matter around saturation density has been
performed. We have shown that the NLD approach describes well 
the saturation properties of the nuclear matter and 
compares remarkably well with microscopic calculations and Dirac phenomenology. 
We have investigated the high density part of the 
NLD EoS. This is relevant for the  neutron stars in $\beta$-equilibrium. 
We found that the low density constraints imposed
on the nuclear matter EoS and by the momentum dependence of the
Schr\"odinger-equivalent optical potential lead to a maximum mass of the 
neutron stars around $M \simeq 2 M_{\odot}$. The latter mass accommodates 
the  observed mass in the J1614-2230 millisecond radio pulsar system.
We further studied the EoS of matter in $\beta$-equilibrium and
find that the high density pressure-density diagram as extracted from
astrophysical measurements can be well described in the NLD model
which rely on nucleonic degrees of freedom only.

The EoS proposed here can be used in transport theoretical studies of  nuclear
collisions, since, it describes very well both, the low energy (density) and
the high energy (density) regions of the nuclear phase diagram. The model 
predicts saturation of the optical potential at high energies and results 
in saturations of the symmetry energy. 
An interesting finding is that the momentum dependent 
interaction  make the EoS softer at low densities, however, it is still stiff
enough at supra-normal densities to account for the recent measurements of
the neutron star masses. Furthermore, the model can be applied to the transport
description of the anti-nucleon optical potential as well as to the study of
dynamics of compressed matter in reactions induced by heavy-ions and 
anti-proton beams at the future FAIR facility.

\begin{acknowledgments}
This work was supported by DFG and by DFG through TR16. We are greatful to
Dr. Thomas von Chossy for discussions concerning the numerical implementation of
the minimization routines. We also acknowledge the correspondence with 
Prof. Dr. Lie-Wen Chen and Bao-Jun Cai on the thermodynamic consistency of the 
NLD model.
\end{acknowledgments}

\begin{appendix}

\section{\label{app1}Notations and infinetisemal variations}

We use following abbreviations for higher-order 
partial derivatives
\begin{align}
\partial_{\alpha_{1}\cdots\alpha_{n}} := & 
\partial_{\alpha_{1}}\partial_{\alpha_{2}}\cdots\partial_{\alpha_{n}} \,,
\nonumber\\
\partial^{\alpha_{1}\cdots\alpha_{n}} := & 
\partial^{\alpha_{1}}\partial^{\alpha_{2}}\cdots\partial^{\alpha_{n}} \,,
\nonumber\\
\partial_{\alpha_{1}\cdots\alpha_{n}}^{\beta_{1}\cdots\beta_{m}} := & 
\partial_{\alpha_{1}}\cdots\partial_{\alpha_{n}}
\partial^{\beta_{1}}\cdots\partial^{\beta_{m}}
\nonumber
\,.
\end{align}

In the following appendices we will need various definitions of infinitesimal 
variations, which are specified here. The total variation of a field 
$\varphi_{r}(x)$ with $x^{\mu} = (x^{0},\vec{x}\,)$ is defined as
\begin{align}
\delta_{T}\varphi_{r}(x) :=  \varphi^{\prime}_{r}(x^{\prime})-\varphi_{r}(x) 
\, ,
\end{align}
with the variation with respect to the $4$-coordinate given by
\begin{align}
x^{\prime\,\mu} := x^{\mu} + \delta x^{\mu} = 
x^{\mu} + \Delta\vartheta^{\mu\nu} x_{\nu} + \epsilon^{\mu}
\,,
\end{align}
where $\delta x^{\mu}$ denotes an infinitesimal transformation, 
\textit{e.g.}, a constant 
translation, $\epsilon^{\mu}$, and/or a rotation, 
$\Delta\vartheta^{\mu\nu}x_{\mu}$, $\vartheta^{\mu\nu}$ 
is an infinitesimal antisymmetric tensor and 
$\Delta\vartheta^{\mu\nu}=-\Delta\vartheta^{\nu\mu}$. 
We define the infinitesimal transformation of the field at a fixed 
argument as
\begin{align}
\delta\varphi_{r}=\varphi^{\prime}_{r}-\varphi_{r}
\,.
\end{align}
For the derivation of the Noether theorem we will need not only the infinitesimal 
variation of a field at fixed argument only, but also of its higher-order 
derivatives, {\it i.e.}, 
$\delta(\partial_{\alpha_{1}}\varphi_{r})$, 
$\delta(\partial_{\alpha_{1}\alpha_{2}}\varphi_{r})$, $\cdots$, 
$\delta(\partial_{\alpha_{1}\cdots\alpha_{n}}\varphi_{r})$. For such variations 
the commutation between symbols $\delta$ and $\partial$ is holds, that is 
\begin{align}
\delta\left( \partial_{\alpha_{1}}\varphi_{r} \right)  = &
\partial_{\alpha_{1}}\varphi^{\prime\,}_{r}(x) - \partial_{\alpha_{1}}\varphi_{r}(x) 
\nonumber\\
= & 
\partial_{\alpha_{1}}\left( \varphi^{\prime\,}_{r}(x) - \varphi_{r}(x) \right) = 
\partial_{\alpha_{1}}\delta\varphi_{r}(x) \,,
\nonumber\\
\delta\left( \partial_{\alpha_{1}\alpha_{2}}\varphi_{r} \right)  = &
\partial_{\alpha_{1}\alpha_{2}}\varphi^{\prime\,}_{r}(x) - 
\partial_{\alpha_{1}\alpha_{2}}\varphi_{r}(x) 
\nonumber\\
= & \partial_{\alpha_{1}\alpha_{2}}\left( \varphi^{\prime\,}_{r}(x) - 
\varphi(x) \right) = 
\partial_{\alpha_{1}\alpha_{2}}\left(\delta\varphi_{r}(x)\right)
\,, \nonumber
\end{align}
and obviously for higher-order fields. 
The total variation of a field $\varphi_{r}(x)$ can be written 
in a more handleable way as
\begin{align}
\delta_{T}\varphi_{r}(x) := &  \varphi^{\prime}_{r}(x^{\prime})-\varphi_{r}(x) 
\nonumber\\
= & \left[ 
	\varphi^{\prime}_{r}(x^{\prime}) - \varphi_{r}(x^{\prime})
\right] 
+
\left[ 
	\varphi_{r}(x^{\prime}) - \varphi_{r}(x)
\right] 
\,,
\label{var1}
\end{align}
where the first term is just the variation $\delta\varphi_{r}$ at fixed argument. 
The second term in Eq.~(\ref{var1}) is the variation with respect to the argument. 
Eq.~(\ref{var1}) reduces in first order to
\begin{equation}
\delta_{T}\varphi_{r}(x) = 
\delta\varphi_{r}(x) + \partial_{\alpha}\varphi_{r}(x)\delta x^{\alpha}
\quad .
\label{var2}
\end{equation}

\begin{widetext}

\section{\label{app2}Derivation of the Noether theorem for higher-order Lagrangians}

For the derivation of the Noether theorem we start with the following 
Lagrangian density 
\begin{align}
{\cal L} = 
{\cal L}\left[ \varphi_{r}(x), \partial_{\alpha_{1}}\varphi_{r}(x), \cdots,
\partial_{\alpha_{1}\cdots\alpha_{n}}\varphi_{r}(x)
\right]
\label{La}
\,.
\end{align}
Invariance of the Lagrangian density, (\ref{La}), with respect to an infinitesimal 
transformation of all the fields and their coordinates implies
\begin{align}
{\cal L}\left[ \varphi^{\prime}_{r}(x^{\prime\,}), 
\partial_{\alpha_{1}}^{\prime}\varphi^{\prime}_{r}(x^{\prime\,}), \cdots,
\partial_{\alpha_{1}\cdots\alpha_{n}}^{\prime}\varphi^{\prime}_{r}(x^{\prime\,})
\right] = 
{\cal L}\left[ \varphi_{r}(x), \partial_{\alpha_{1}}\varphi_{r}(x), \cdots,
\partial_{\alpha_{1}\cdots\alpha_{n}}\varphi_{r}(x)
\right]
\label{Lvar0}
\,.
\end{align}
In terms of the total variation Eq.~(\ref{Lvar0}) results in 
\begin{align}
\delta_{T}{\cal L} = 
{\cal L}\left[ \varphi^{\prime}_{r}(x^{\prime\,}), 
\partial_{\alpha_{1}}^{\prime}\varphi^{\prime}_{r}(x^{\prime\,}), \cdots,
\partial_{\alpha_{1}\cdots\alpha_{n}}^{\prime}\varphi^{\prime}_{r}(x^{\prime\,})
\right] - 
{\cal L}\left[ \varphi_{r}(x), \partial_{\alpha_{1}}\varphi_{r}(x), \cdots,
\partial_{\alpha_{1}\cdots\alpha_{n}}\varphi_{r}(x)
\right] = 0
\label{Lvar}
\,,
\end{align}
where $\ds \partial_{\mu}\equiv \frac{\partial}{\partial x^{\mu}}$
and $\ds \partial_{\mu}^{\prime}\equiv \frac{\partial}{\partial x^{'\mu}}$. 
Our goal is to arrive from Eq.~(\ref{Lvar}) to a continuity equation of the form 
\begin{equation}
\partial_{\mu}f^{\mu} = 0
\,,
\label{cont}
\end{equation}
with $f^{\mu}$ being a conserved current to be determined in the following. 
We will work out here all the analytical evaluations up to third order in the 
partial derivatives of the Lagrangian density, and we will give all the terms 
up to infinity in the final equations. 

At first, the total variation of the Lagrangian density, Eq.~(\ref{Lvar}), 
can be written as
\begin{align}
\delta_{T}{\cal L} = & 
{\cal L}[\varphi^{\prime}_{r}(x^{\prime}),
\partial_{\alpha}\varphi^{\prime}_{r}(x^{\prime}), 
\partial_{\alpha\beta}\varphi^{\prime}_{r}(x^{\prime}),\ldots ]
 -
{\cal L}[\varphi_{r}(x^{\prime}),
\partial_{\alpha}\varphi_{r}(x^{\prime}), 
\partial_{\alpha\beta}\varphi_{r}(x^{\prime}),\ldots ]
\label{var3}\\
+ & 
{\cal L}[\varphi_{r}(x^{\prime}),
\partial_{\alpha}\varphi_{r}(x^{\prime}), 
\partial_{\alpha\beta}\varphi_{r}(x^{\prime}),\ldots ]
 -
{\cal L}[\varphi_{r}(x),
\partial_{\alpha}\varphi_{r}(x), 
\partial_{\alpha\beta}\varphi_{r}(x),\ldots ] = 0
\,, \nonumber
\end{align}
where we used 
$\partial_{\mu}^{\prime}\varphi^{\prime}(x^{\prime\,}) = 
\partial_{\nu}\varphi^{\prime}(x^{\prime\,})\partial_{\mu}^{\prime}x^{\nu} = 
\partial_{\nu}\varphi^{\prime}(x^{\prime\,})\, \delta_{\mu}^{\nu} = 
\partial_{\mu}\varphi^{\prime}(x^{\prime\,})$ where  $\delta_{\mu}^{\nu}$ is a
Kronecker symbol.
The first line in Eq. (\ref{var3}) is just the variation of ${\cal L}$ 
with respect to the fields 
$\varphi_{r}(r^{\prime}),\partial_{\alpha}\varphi_{r}(r^{\prime}),\cdots$ 
at fixed argument, whereas the terms in the second line 
give the variation of the Lagrangian with respect to the argument $x^{\mu}$. 
Since we consider infinitesimal transformations only, it is sufficient to 
evaluate latter quantity up to first order with respect to the argument. 
Therefore, Eq.~(\ref{var3}) can be written as
\begin{align}
\delta_{T}{\cal L}  = 
\delta{\cal L} + \partial_{\alpha}{\cal L}\delta x^{\alpha} = 0
\,.
\label{var4} 
\end{align}
The variation at fixed argument, $\delta{\cal L}$, can be evaluated 
in the usual way according
\begin{align}
\delta{\cal L} = 
\sum_{r} \left[
 \frac{\partial{\cal L}}{\partial\varphi_{r}} \delta\varphi_{r}
+ 
 \frac{\partial{\cal L}}{\partial(\partial_{\alpha_{1}}\varphi_{r})} 
 \delta(\partial_{\alpha_{1}}\varphi_{r})
+
 \frac{\partial{\cal L}}{\partial(\partial_{\alpha_{1}\alpha_{2}}\varphi_{r})} 
 \delta (\partial_{\alpha_{1}\alpha_{2}}\varphi_{r})
+ 
\frac{\partial{\cal L}}{\partial(\partial_{\alpha_{1}\alpha_{2}\alpha_{3}}\varphi_{r})} 
 \delta (\partial_{\alpha_{1}\alpha_{2}\alpha_{3}}\varphi_{r})
+ \cdots
\right]
\,.
\label{var5} 
\end{align}
Replacing the first term in Eq. (\ref{var5}) with the help of the Euler-Lagrange 
equations of motion, Eq.~(\ref{Euler0}), results in
\begin{align}
\delta{\cal L} = 
\sum_{r} & \left[
 \partial_{\alpha_{1}}\frac{\partial{\cal L}}{\partial(\partial_{\alpha_{1}}\varphi_{r})}\delta\varphi_{r}
-
 \partial_{\alpha_{1}\alpha_{2}}
\frac{\partial{\cal L}}{\partial(\partial_{\alpha_{1}\alpha_{2}}\varphi_{r})}\delta\varphi_{r}
+
 \partial_{\alpha_{1}\alpha_{2}\alpha_{3}}
\frac{\partial{\cal L}}{\partial(\partial_{\alpha_{1}\alpha_{2}\alpha_{3}}\varphi_{r})}\delta\varphi_{r}
-  \cdots
\right.
\label{deltaL_a}\\
& 
\left.
+ \frac{\partial{\cal L}}{\partial(\partial_{\alpha_{1}}\varphi_{r})} 
 \delta(\partial_{\alpha_{1}}\varphi_{r})
+
 \frac{\partial{\cal L}}{\partial(\partial_{\alpha_{1}\alpha_{2}}\varphi_{r})} 
 \delta (\partial_{\alpha_{1}\alpha_{2}}\varphi_{r})
+
 \frac{\partial{\cal L}}{\partial(\partial_{\alpha_{1}\alpha_{2}\alpha_{3}}\varphi_{r})} 
 \delta (\partial_{\alpha_{1}\alpha_{2}\alpha_{3}}\varphi_{r})
+
\cdots
\right]
\,. \nonumber
\end{align}

We use the commutation between the variation at fixed argument and 
the partial derivative (see appendix~\ref{app1}) and apply once the product rule 
as follows for the term proportional to $\partial_{\alpha_{1}\alpha_{2}}\delta\varphi$
\begin{align}
\frac{\partial{\cal L}}{\partial(\partial_{\alpha_{1}\alpha_{2}}\varphi_{r})}
\partial_{\alpha_{1}\alpha_{2}}\delta\varphi_{r} =  
- \partial_{\alpha_{1}}
  \frac{\partial{\cal L}}{\partial(\partial_{\alpha_{1}\alpha_{2}}\varphi_{r})}
  \partial_{\alpha_{2}}\delta\varphi_{r}
+ \partial_{\alpha_{1}}
  \left[
  \frac{\partial{\cal L}}{\partial(\partial_{\alpha_{1}\alpha_{2}}\varphi_{r})}
  \partial_{\alpha_{2}}\delta\varphi_{r}
  \right]
\,.
\label{product1}
\end{align}
For the term proportional to $\partial_{\alpha_{1}\alpha_{2}\alpha_{3}}\delta\varphi$ 
we apply the product rule twice
\begin{align}
\frac{\partial{\cal L}}{\partial ( \partial_{\alpha_{1}\alpha_{2}\alpha_{3}}\varphi_{r} ) } 
\partial_{\alpha_{1}\alpha_{2}\alpha_{3}}\delta\varphi_{r} = & 
\partial_{\alpha_{1}\alpha_{2}}
\frac{\partial{\cal L}}{\partial(\partial_{\alpha_{1}\alpha_{2}\alpha_{3}}\varphi_{r})}
\partial_{\alpha_{3}}\delta\varphi_{r}
\nonumber\\
& - 
\partial_{\alpha_{2}}
  \left[
  \partial_{\alpha_{1}}
  \frac{\partial{\cal L}}{\partial(\partial_{\alpha_{1}\alpha_{2}\alpha_{3}}\varphi_{r})}
  \partial_{\alpha_{3}}\delta\varphi_{r}
  \right]
+
\partial_{\alpha_{1}}
  \left[
  \frac{\partial{\cal L}}{\partial(\partial_{\alpha_{1}\alpha_{2}\alpha_{3}}\varphi_{r})}
  \partial_{\alpha_{2}\alpha_{3}}\delta\varphi_{r}
  \right]
\,,
\label{product2}
\end{align}
and so forth for the terms proportional to higher-order derivatives. 
In total, this procedure leads to a series of terms 
proportional to a $4$-divergences only. 
It is more convenient to arrange these terms such to obtain several 
infinite series for each derivative field 
$\delta\varphi$, $\partial_{\alpha_{1}}\delta\varphi$, 
$\partial_{\alpha_{1}\alpha_{2}}\delta\varphi$, 
$\partial_{\alpha_{1}\cdots\alpha_{n}}\delta\varphi$. After 
insertion of Eqs.~(\ref{product1}) and~(\ref{product2}) into the 
Eq.~(\ref{deltaL_a}) and after their rearrangement we obtain as an 
intermediate result
\begin{align}
\delta{\cal L} = \!
\partial_{\mu} \!
\Big \{ & \!
\left[
\frac{\partial{\cal L}}{\partial(\partial_{\mu}\varphi_{r})}
- \partial_{\alpha_{1}}
\frac{\partial{\cal L}}{\partial(\partial_{\mu\,\alpha_{1}}\varphi_{r})}
+ \partial_{\alpha_{1}\alpha_{2}}
\frac{\partial{\cal L}}{\partial(\partial_{\mu\,\alpha_{1}\alpha_{2}}\varphi_{r})}
-
\cdots
+ (-)^{n}
\partial_{\alpha_{1}\cdots\alpha_{n}}
\frac{\partial{\cal L}}{\partial(\partial_{\mu\,\alpha_{1}\cdots\alpha_{n}}\varphi_{r})}
\right]\delta\varphi_{r}
\Big .
\label{delta_L}\\
\Big . 
+ & \!
\left[
\phantom{
\frac{\partial{\cal L}}{\partial(\partial_{\mu}\varphi_{r})}
- \partial_{\alpha_{1}}
}
\frac{\partial{\cal L}}{\partial(\partial_{\mu\,\sigma_{1}}\varphi_{r})}
- 
\phantom{{}_{\alpha_{2}}}
\partial_{\alpha_{1}}
\frac{\partial{\cal L}}{\partial(\partial_{\mu\,\sigma_{1}\alpha_{1}}\varphi_{r})}
+\cdots
+ (-)^{n}
\partial_{\alpha_{1}\cdots\alpha_{n}}
\frac{\partial{\cal L}}{\partial(\partial_{\mu\,\sigma_{1}\,\alpha_{1}\cdots\alpha_{n}}\varphi_{r})}
\right]\partial_{\sigma_{1}}\delta\varphi_{r}
\Big .
\nonumber\\
 \Big . 
+ & \!
\left[
\phantom{
\frac{\partial{\cal L}}{\partial(\partial_{\mu}\varphi_{r})}
- \partial_{\alpha_{1}}
\frac{\partial{\cal L}}{\partial(\partial_{\mu\,\alpha_{1}}\varphi_{r})}
+ \partial_{\alpha_{1}\alpha_{2}}
}
\frac{\partial{\cal L}}{\partial(\partial_{\mu\,\sigma_{1}\sigma_{2}}\varphi_{r})}
+\cdots
+ (-)^{n}
\partial_{\alpha_{1}\cdots\alpha_{n}}
\frac{\partial{\cal L}}{\partial(\partial_{\mu\,\sigma_{1}\sigma_{2}\,\alpha_{1}\cdots\alpha_{n}}\varphi_{r})}
\right]\partial_{\sigma_{1}\sigma_{2}}\delta\varphi_{r}
\Big .
\nonumber\\
\Big .
& \vdots
\Big .
\nonumber\\
\Big . \!
+ & 
\left[ 
\phantom{~~~~~~~~~}
\frac{\partial{\cal L}}{\partial(\partial_{\mu\,\sigma_{1}\cdots\sigma_{n}}\varphi_{r})}
- \partial_{\alpha_{1}}
\frac{\partial{\cal L}}{\partial(\partial_{\mu\,\sigma_{1}\cdots\sigma_{n}\,\alpha_{1}}\varphi_{r})}
+ 
\cdots
+ (-)^{n}
\partial_{\alpha_{1}\cdots\alpha_{n}}
\frac{\partial{\cal L}}{\partial(\partial_{\mu\,\sigma_{1}\cdots\sigma_{n}\,\alpha_{1}\cdots\alpha_{n}}\varphi_{r})}
\right]\partial_{\sigma_{1}\cdots\sigma_{n}}\delta\varphi_{r}
\Big \}
\nonumber
\,.
\end{align}
As next step we replace $\delta\varphi_{r}$ by the total variation, 
$\delta_{T}\varphi$, and insert the resulting equation into the total 
variation for the Lagrangian in Eq.~(\ref{var4}). Furthermore, we use 
\begin{align}
\partial_{\alpha}{\cal L}\delta x^{\alpha} 
= \partial_{\mu}\left( g^{\mu}_{\alpha}{\cal L}\delta x^{\alpha}\right)
\label{trick}
\,,
\end{align}
where $ g^{\mu}_{\alpha}$ is a metric tensor. Eq.~(\ref{trick}) obviously
holds, when the infinitesimal transformation for 
the coordinates, $\delta x^{\mu}$, concerns a constant displacement of the 
$4$-vector $x^{\mu}$. In case of rotations, where $\delta x^{\mu}$ depends 
on the coordinate $x^{\mu}$ its self, Eq.~(\ref{trick}) still applies due to 
the antisymmetry of the tensor $\vartheta^{\mu\nu}$. 
These steps give us the final and general expression for the Noether theorem 
with respect to variations of the different fields and their coordinates
\begin{align}
\delta{\cal L} = \!
\partial_{\mu} \!
\Bigg \{ & \!
\left[
\frac{\partial{\cal L}}{\partial(\partial_{\mu}\varphi_{r})}
- \partial_{\alpha_{1}}
\frac{\partial{\cal L}}{\partial(\partial_{\mu\,\alpha_{1}}\varphi_{r})}
+ \partial_{\alpha_{1}\alpha_{2}}
\frac{\partial{\cal L}}{\partial(\partial_{\mu\,\alpha_{1}\alpha_{2}}\varphi_{r})}
-
\cdots
+ (-)^{n}
\partial_{\alpha_{1}\cdots\alpha_{n}}
\frac{\partial{\cal L}}{\partial(\partial_{\mu\,\alpha_{1}\cdots\alpha_{n}}\varphi_{r})}
\right]
\Bigg .
\nonumber\\
\Bigg.
& 
\times 
\left( \delta_{T}\varphi_{r} - \partial_{\alpha}\varphi_{r}\delta x^{\alpha} \right)
\Bigg .
\nonumber\\
\Bigg . 
+ & \!
\left[
\phantom{
\frac{\partial{\cal L}}{\partial(\partial_{\mu}\varphi_{r})}
- \partial_{\alpha_{1}}
}
\frac{\partial{\cal L}}{\partial(\partial_{\mu\,\sigma_{1}}\varphi_{r})}
- 
\phantom{{}_{\alpha_{2}}}
\partial_{\alpha_{1}}
\frac{\partial{\cal L}}{\partial(\partial_{\mu\,\sigma_{1}\alpha_{1}}\varphi_{r})}
+\cdots
+ (-)^{n}
\partial_{\alpha_{1}\cdots\alpha_{n}}
\frac{\partial{\cal L}}{\partial(\partial_{\mu\,\sigma_{1}\,\alpha_{1}\cdots\alpha_{n}}\varphi_{r})}
\right]
\Bigg .
\nonumber\\
\Bigg.
& 
\times 
\partial_{\sigma_{1}}
\left( \delta_{T}\varphi_{r} - \partial_{\alpha}\varphi_{r}\delta x^{\alpha} \right)
\Bigg .
\nonumber\\
 \Bigg . 
+ & \!
\left[
\phantom{
\frac{\partial{\cal L}}{\partial(\partial_{\mu}\varphi_{r})}
- \partial_{\alpha_{1}}
\frac{\partial{\cal L}}{\partial(\partial_{\mu\,\alpha_{1}}\varphi_{r})}
+ \partial_{\alpha_{1}\alpha_{2}}
}
\frac{\partial{\cal L}}{\partial(\partial_{\mu\,\sigma_{1}\sigma_{2}}\varphi_{r})}
+\cdots
+ (-)^{n}
\partial_{\alpha_{1}\cdots\alpha_{n}}
\frac{\partial{\cal L}}{\partial(\partial_{\mu\,\sigma_{1}\sigma_{2}\,\alpha_{1}\cdots\alpha_{n}}\varphi_{r})}
\right]
\Bigg .
\nonumber\\
\Bigg.
& 
\times 
\partial_{\sigma_{1}\sigma_{2}}
\left( \delta_{T}\varphi_{r} - \partial_{\alpha}\varphi_{r}\delta x^{\alpha} \right)
\Bigg .
\nonumber\\
\Bigg .
& \vdots
\Bigg .
\nonumber\\
\Bigg . \!
+ & 
\left[ 
\phantom{~~~~~~~~~}
\frac{\partial{\cal L}}{\partial(\partial_{\mu\,\sigma_{1}\cdots\sigma_{n}}\varphi_{r})}
- \partial_{\alpha_{1}}
\frac{\partial{\cal L}}{\partial(\partial_{\mu\,\sigma_{1}\cdots\sigma_{n}\,\alpha_{1}}\varphi_{r})}
+ 
\cdots
+ (-)^{n}
\partial_{\alpha_{1}\cdots\alpha_{n}}
\frac{\partial{\cal L}}{\partial(\partial_{\mu\,\sigma_{1}\cdots\sigma_{n}\,\alpha_{1}\cdots\alpha_{n}}\varphi_{r})}
\right]
\Bigg .
\nonumber\\
\Bigg.
& 
\times 
\partial_{\sigma_{1}\cdots\sigma_{n}}
\left( \delta_{T}\varphi_{r} - \partial_{\alpha}\varphi_{r}\delta x^{\alpha} \right)
\Bigg .
\nonumber\\
 \Bigg . 
& - g^{\mu\alpha}{\cal L}\delta x_{\alpha}
\Bigg \}
\label{delta22_L}
\,.
\end{align}

We consider now global phase transformations ($\epsilon \ll 1$)
\begin{align}
\delta x^{\mu}=0~~,~~\varphi^{\prime}_{r}(x^{\prime})=e^{-i\epsilon}\varphi_{r}(x)
\Rightarrow \delta\varphi_{rT} = \delta\varphi_{r} = -i\epsilon \varphi_{r}
\label{disymm}
\end{align}
and obtain the following relations for global phase transformations
\begin{align}
\delta\varphi_{r} = & -i\epsilon \varphi_{r}\,,
\nonumber\\
\partial_{\alpha_{1}}\delta\varphi_{r} = & -i\epsilon \partial_{\alpha_{1}}\varphi_{r}\,,
\nonumber\\
\partial_{\alpha_{1}\alpha_{2}}\delta\varphi_{r} 
= & 
-i\epsilon \partial_{\alpha_{1}\alpha_{2}}\varphi_{r}\,,
\nonumber\\
 \cdots & \,,
\nonumber\\
\partial_{\alpha_{1}\alpha_{2}\cdots\alpha_{n}}\delta\varphi_{r} 
= & 
-i\epsilon \partial_{\alpha_{1}\alpha_{2}\cdots\alpha_{n}}\varphi_{r}
\,.
\end{align}
Therefore, 
the invariance of the Lagrangian under global phase transformations results 
to the continuity equation $\partial_{\mu}J^{\mu} = 0$ with the current given by
\begin{align}
J^{\mu} = \! -i
\Bigg \{ & \!
\left[
\frac{\partial{\cal L}}{\partial(\partial_{\mu}\varphi_{r})}
- \partial_{\alpha_{1}}
\frac{\partial{\cal L}}{\partial(\partial_{\mu\,\alpha_{1}}\varphi_{r})}
+ \partial_{\alpha_{1}\alpha_{2}}
\frac{\partial{\cal L}}{\partial(\partial_{\mu\,\alpha_{1}\alpha_{2}}\varphi_{r})}
-
\cdots
+ (-)^{n}
\partial_{\alpha_{1}\cdots\alpha_{n}}
\frac{\partial{\cal L}}{\partial(\partial_{\mu\,\alpha_{1}\cdots\alpha_{n}}\varphi_{r})}
\right]\varphi_{r}
\Bigg .
\label{Noether-Current}\\
\Bigg . 
+ & \!
\left[
\phantom{
\frac{\partial{\cal L}}{\partial(\partial_{\mu}\varphi_{r})}
- \partial_{\alpha_{1}}
}
\frac{\partial{\cal L}}{\partial(\partial_{\mu\,\sigma_{1}}\varphi_{r})}
- 
\phantom{{}_{\alpha_{2}}}
\partial_{\alpha_{1}}
\frac{\partial{\cal L}}{\partial(\partial_{\mu\,\sigma_{1}\alpha_{1}}\varphi_{r})}
+\cdots
+ (-)^{n}
\partial_{\alpha_{1}\cdots\alpha_{n}}
\frac{\partial{\cal L}}{\partial(\partial_{\mu\,\sigma_{1}\,\alpha_{1}\cdots\alpha_{n}}\varphi_{r})}
\right]\partial_{\sigma_{1}}\varphi_{r}
\Bigg .
\nonumber\\
 \Bigg . 
+ & \!
\left[
\phantom{
\frac{\partial{\cal L}}{\partial(\partial_{\mu}\varphi_{r})}
- \partial_{\alpha_{1}}
\frac{\partial{\cal L}}{\partial(\partial_{\mu\,\alpha_{1}}\varphi_{r})}
+ \partial_{\alpha_{1}\alpha_{2}}
}
\frac{\partial{\cal L}}{\partial(\partial_{\mu\,\sigma_{1}\sigma_{2}}\varphi_{r})}
+\cdots
+ (-)^{n}
\partial_{\alpha_{1}\cdots\alpha_{n}}
\frac{\partial{\cal L}}{\partial(\partial_{\mu\,\sigma_{1}\sigma_{2}\,\alpha_{1}\cdots\alpha_{n}}\varphi_{r})}
\right]\partial_{\sigma_{1}\sigma_{2}}\varphi_{r}
\Bigg .
\nonumber\\
\Bigg .
& \vdots
\Bigg .
\nonumber\\
\Bigg . \!
+ & 
\left[ 
\phantom{~~~~~~~~~}
\frac{\partial{\cal L}}{\partial(\partial_{\mu\,\sigma_{1}\cdots\sigma_{n}}\varphi_{r})}
- \partial_{\alpha_{1}}
\frac{\partial{\cal L}}{\partial(\partial_{\mu\,\sigma_{1}\cdots\sigma_{n}\,\alpha_{1}}\varphi_{r})}
+ 
\cdots
+ (-)^{n}
\partial_{\alpha_{1}\cdots\alpha_{n}}
\frac{\partial{\cal L}}{\partial(\partial_{\mu\,\sigma_{1}\cdots\sigma_{n}\,\alpha_{1}\cdots\alpha_{n}}\varphi_{r})}
\right]\partial_{\sigma_{1}\cdots\sigma_{n}}\varphi_{r}
\Bigg \}
\nonumber
\,.
\end{align}

Using Eqs.~(\ref{tensors}) for the tensors 
${\cal K}^{\mu\sigma_{1}\sigma_{2}\cdots}_{r}$ one arrives to the Expression in 
Eq.~(\ref{current}).

The energy-momentum tensor is derived again with the help of Eq.~(\ref{delta22_L}) 
for the case of constant $4$-translations
$x^{\prime\,\mu} := x^{\mu} + \delta x^{\mu}$. This means 
$\delta_{T}\varphi_{r}(x)=0$ and following expression is obtained 
\begin{align}
\partial_{\mu} 
\Bigg \{ &
  \left[
\frac{\partial{\cal L}}{\partial(\partial_{\mu}\varphi_{r})}
- \partial_{\alpha_{1}}
\frac{\partial{\cal L}}{\partial(\partial_{\mu\,\alpha_{1}}\varphi_{r})}
+ \partial_{\alpha_{1}\alpha_{2}}
\frac{\partial{\cal L}}{\partial(\partial_{\mu\,\alpha_{1}\alpha_{2}}\varphi_{r})}
-
\cdots
+ (-)^{n}
\partial_{\alpha_{1}\cdots\alpha_{n}}
\frac{\partial{\cal L}}{\partial(\partial_{\mu\,\alpha_{1}\cdots\alpha_{n}}\varphi_{r})}
  \right]\partial^{\nu}\varphi_{r}
\Bigg .
\nonumber\\
\Bigg .. 
+ &
\left[
\phantom{
\frac{\partial{\cal L}}{\partial(\partial_{\mu}\varphi_{r})}
- \partial_{\alpha_{1}}
}
\frac{\partial{\cal L}}{\partial(\partial_{\mu\,\sigma_{1}}\varphi_{r})}
- 
\phantom{{}_{\alpha_{2}}}
\partial_{\alpha_{1}}
\frac{\partial{\cal L}}{\partial(\partial_{\mu\,\sigma_{1}\alpha_{1}}\varphi_{r})}
+\cdots
+ (-)^{n}
\partial_{\alpha_{1}\cdots\alpha_{n}}
\frac{\partial{\cal L}}{\partial(\partial_{\mu\,\sigma_{1}\,\alpha_{1}\cdots\alpha_{n}}\varphi_{r})}
  \right]
\partial_{\sigma_{1}}^{\nu}\varphi_{r}
\Bigg ..
\nonumber\\
 \Bigg . 
+ &
\left[
\phantom{
\frac{\partial{\cal L}}{\partial(\partial_{\mu}\varphi_{r})}
- \partial_{\alpha_{1}}
\frac{\partial{\cal L}}{\partial(\partial_{\mu\,\alpha_{1}}\varphi_{r})}
+ \partial_{\alpha_{1}\alpha_{2}}
}
\frac{\partial{\cal L}}{\partial(\partial_{\mu\,\sigma_{1}\sigma_{2}}\varphi_{r})}
+\cdots
+ (-)^{n}
\partial_{\alpha_{1}\cdots\alpha_{n}}
\frac{\partial{\cal L}}{\partial(\partial_{\mu\,\sigma_{1}\sigma_{2}\,\alpha_{1}\cdots\alpha_{n}}\varphi_{r})}
  \right]\partial_{\sigma_{1}\sigma_{2}}^{\nu}\varphi_{r}
\Bigg .
\nonumber\\
 \Bigg . 
& 
\vdots
\Bigg.
\nonumber\\
\Bigg . \!
+ & 
\left[ 
\phantom{~~~~~~~~~}
\frac{\partial{\cal L}}{\partial(\partial_{\mu\,\sigma_{1}\cdots\sigma_{n}}\varphi_{r})}
- \partial_{\alpha_{1}}
\frac{\partial{\cal L}}{\partial(\partial_{\mu\,\sigma_{1}\cdots\sigma_{n}\,\alpha_{1}}\varphi_{r})}
+ 
\cdots
+ (-)^{n}
\partial_{\alpha_{1}\cdots\alpha_{n}}
\frac{\partial{\cal L}}{\partial(\partial_{\mu\,\sigma_{1}\cdots\sigma_{n}\,\alpha_{1}\cdots\alpha_{n}}\varphi_{r})}
\right]\partial_{\sigma_{1}\cdots\sigma_{n}}^{\nu}\varphi_{r}
\Bigg .
\nonumber\\
\Bigg .
 - & g^{\mu\nu}{\cal L}
\Bigg \}\delta x_{\nu} = 0
\,. \label{currapp}
\end{align}
This leads to the continuity equation 
\begin{align}
\partial_{\mu}\, T^{\mu\nu} = 0
\end{align}
with the energy-momentum tensor $T^{\mu\nu}$ given by
\begin{align}
T^{\mu\nu} = \! -i
\Bigg \{ & \!
\left[
\frac{\partial{\cal L}}{\partial(\partial_{\mu}\varphi_{r})}
- \partial_{\alpha_{1}}
\frac{\partial{\cal L}}{\partial(\partial_{\mu\,\alpha_{1}}\varphi_{r})}
+ \partial_{\alpha_{1}\alpha_{2}}
\frac{\partial{\cal L}}{\partial(\partial_{\mu\,\alpha_{1}\alpha_{2}}\varphi_{r})}
-
\cdots
+ (-)^{n}
\partial_{\alpha_{1}\cdots\alpha_{n}}
\frac{\partial{\cal L}}{\partial(\partial_{\mu\,\alpha_{1}\cdots\alpha_{n}}\varphi_{r})}
\right]\partial^{\nu}\varphi_{r}
\Bigg .
\label{tensorapp}\\
\Bigg . 
+ & \!
\left[
\phantom{
\frac{\partial{\cal L}}{\partial(\partial_{\mu}\varphi_{r})}
- \partial_{\alpha_{1}}
}
\frac{\partial{\cal L}}{\partial(\partial_{\mu\,\sigma_{1}}\varphi_{r})}
- 
\phantom{{}_{\alpha_{2}}}
\partial_{\alpha_{1}}
\frac{\partial{\cal L}}{\partial(\partial_{\mu\,\sigma_{1}\alpha_{1}}\varphi_{r})}
+\cdots
+ (-)^{n}
\partial_{\alpha_{1}\cdots\alpha_{n}}
\frac{\partial{\cal L}}{\partial(\partial_{\mu\,\sigma_{1}\,\alpha_{1}\cdots\alpha_{n}}\varphi_{r})}
\right]\partial_{\sigma_{1}}^{\nu}\varphi_{r}
\Bigg .
\nonumber\\
 \Bigg . 
+ & \!
\left[
\phantom{
\frac{\partial{\cal L}}{\partial(\partial_{\mu}\varphi_{r})}
- \partial_{\alpha_{1}}
\frac{\partial{\cal L}}{\partial(\partial_{\mu\,\alpha_{1}}\varphi_{r})}
+ \partial_{\alpha_{1}\alpha_{2}}
}
\frac{\partial{\cal L}}{\partial(\partial_{\mu\,\sigma_{1}\sigma_{2}}\varphi_{r})}
+\cdots
+ (-)^{n}
\partial_{\alpha_{1}\cdots\alpha_{n}}
\frac{\partial{\cal L}}{\partial(\partial_{\mu\,\sigma_{1}\sigma_{2}\,\alpha_{1}\cdots\alpha_{n}}\varphi_{r})}
\right]\partial_{\sigma_{1}\sigma_{2}}^{\nu}\varphi_{r}
\Bigg .
\nonumber\\
\Bigg .
& \vdots
\Bigg .
\nonumber\\
\Bigg .. \!
+ & 
\left[ 
\phantom{~~~~~~~~~}
\frac{\partial{\cal L}}{\partial(\partial_{\mu\,\sigma_{1}\cdots\sigma_{n}}\varphi_{r})}
- \partial_{\alpha_{1}}
\frac{\partial{\cal L}}{\partial(\partial_{\mu\,\sigma_{1}\cdots\sigma_{n}\,\alpha_{1}}\varphi_{r})}
+ 
\cdots
+ (-)^{n}
\partial_{\alpha_{1}\cdots\alpha_{n}}
\frac{\partial{\cal L}}{\partial(\partial_{\mu\,\sigma_{1}\cdots\sigma_{n}\,\alpha_{1}\cdots\alpha_{n}}\varphi_{r})}
\right]\partial_{\sigma_{1}\cdots\sigma_{n}}^{\nu}\varphi_{r}
\Bigg \}
\nonumber\\ 
- & g^{\mu\nu}{\cal L} 
\,.
\nonumber
\end{align}
Above expression for the energy-momentum tensor can be written also in a more 
compact form resulting to Eq.~(\ref{tensor}).

\section{\label{app3}Preliminaries for the NLD formalism}
For the derivation of the Dirac equation, the current and the 
energy-momentum tensor in the NLD model we need the 
derivatives of the NLD Lagrangian with respect to the spinor fields 
and their higher-order derivatives. We evaluate them here up to second order 
and provide all terms of infinite series in the final expressions. 
Since we are interesting on the derivatives with respect to the spinor 
fields only, we start with the NLD Lagrangian without 
the meson-field contributions. That is 
\begin{align}
{\cal L} = & \frac{1}{2}
\left[
	\overline{\Psi}i\gamma_{\mu}\partial^{\mu}\Psi
	- 
	(i\partial^{\mu}\overline{\Psi}) \gamma_{\mu}\Psi
\right]
- \overline{\Psi}\Psi m
+ 
\frac{g_{\sigma}}{2}
	\left[
	\overline{\Psi}
	\, \nldl
	\Psi\sigma
	+\sigma\overline{\Psi}
	\, \nldr
	\Psi
	\right]
-  \frac{g_{\omega}}{2}
	\left[
	\overline{\Psi}
	 \, \nldl
	\gamma^{\mu}\Psi\omega_{\mu}
	+\omega_{\mu}\overline{\Psi}\gamma^{\mu}
	\, \nldr
	\Psi
	\right]
\nonumber\\	
 - & \frac{g_{\rho}}{2}
	\left[
	\overline{\Psi}
	 \, \nldl
	\gamma^{\mu}\vec{\tau}\Psi\vec{\rho}_{\mu}
	+\vec{\rho}_{\mu}\overline{\Psi}\vec{\tau}\gamma^{\mu}
	\, \nldr
	\Psi
	\right]
 +  \frac{g_{\delta}}{2}
	\left[
	\overline{\Psi}
	 \, \nldl
	\vec{\tau}\Psi\vec{\delta}
	+\vec{\delta}\,\overline{\Psi}\vec{\tau}
	\, \nldr
	\Psi
	\right]
\label{app4-1}
\,.
\end{align}
The application of the various higher-order partial derivatives with respect to 
the spinor fields $\overline{\Psi}$ and $\Psi$ to the Lagrangian density 
in Eq.~(\ref{app4-1}) proceeds with the help of the multiple Taylor expansions, see 
Eqs.~(\ref{ope}). It is convenient to rearrange these series in 
ascending order with respect to the partial derivatives. 
With $\xir_{j} = -\zeta^{\alpha}_{j}\, i\partialr_{\alpha}$ and 
$\xil_{j} = i\partiall_{\alpha}\,\zeta^{\alpha}_{j}~(j=1,2,3,4)$ 
where $\zeta^{\mu}_{j}=v^{\mu}_{j}/\Lambda$ one obtains for the expansion up to order $n$
\begin{align}
\nldr = & 
d^{(0)} - 
\frac{1}{1!}\, d^{(1)}_{i_{1}} \, \zeta^{\alpha_{1}}_{i_{1}}\, i\partialr_{\alpha_{1}} + 
\frac{1}{2!}\, d^{(2)}_{i_{1}i_{2}} \, 
\zeta^{\alpha_{1}}_{i_{1}}i\partialr_{\alpha_{1}} \, 
\zeta^{\alpha_{2}}_{i_{2}}i\partialr_{\alpha_{2}} - \cdots +
(-)^{n}\frac{1}{n!}\, d^{(n)}_{i_{1}\cdots i_{4}} \, 
\left( \zeta^{\alpha_{1}}_{i_{1}}i\partialr_{\alpha_{1}} \right)^{n_{1}} \, 
\cdots 
\left( \zeta^{\alpha_{4}}_{i_{4}}i\partialr_{\alpha_{4}} \right)^{n_{4}} \,,
\label{expr}\\
\nldl = & 
d^{(0)} + 
\frac{1}{1!}\, i\partiall_{\alpha_{1}}\zeta^{\alpha_{1}}_{i_{1}} \, d^{(1)}_{i_{1}} + 
\frac{1}{2!} \, 
i\partiall_{\alpha_{1}}\zeta^{\alpha_{1}}_{i_{1}} \, 
i\partiall_{\alpha_{2}}\zeta^{\alpha_{2}}_{i_{2}} \,
d^{(2)}_{i_{1}i_{2}}
+ \cdots + \frac{1}{n!}\,
\left( i\partiall_{\alpha_{1}}\zeta^{\alpha_{1}}_{i_{1}} \right)^{n_{1}} \, 
\cdots 
\left( i\partiall_{\alpha_{4}}\zeta^{\alpha_{4}}_{i_{4}} \right)^{n_{4}} \, 
d^{(n)}_{i_{1}\cdots i_{4}}
\label{expl}
\,, 
\end{align}
with the condition $n_{1}+\cdots +n_{4}=n$ and 
\begin{align}
d^{(0)} := & \nld\vert_{\{\xi_{i_{1}},\xi_{i_{2}},\cdots, \xi_{i_{4}}\}\to 0}
\,,\\
d^{(1)}_{i_{1}} := &  \frac{\partial}{\partial\xi_{i_{1}}}
\nld\vert_{\{\xi_{i_{1}},\xi_{i_{2}},\cdots, \xi_{i_{4}}\}\to 0}
\,,\\
\cdots 
\,, \nonumber \\
d^{(n)}_{i_{1}i_{2} i_{3} i_{4}} := & 
\frac{\partial^{n}}{\partial\xi_{i_{1}}^{n_{1}}\partial\xi_{i_{2}}^{n_{2}}
\cdots \partial\xi_{i_{4}}^{n_{4}}}
\nld\vert_{\{\xi_{i_{1}},\xi_{i_{2}},\cdots, \xi_{i_{4}}\}\to 0}
\,.
\end{align}
The pairs between Latin and between Greek indices in above equations denote 
the summation over the 
multiple variables $\xi_{i}~(i=1,2,3,4)$ and over the $4$-coordinates, 
respectively. In order to simplify the derivations in the following appendices, 
we skip the summation over the multiple variables. 

For the  partial derivative of the Lagrangian density with 
respect to the spinor field $\overline{\Psi}$ only the zero-order terms in 
Eqs.~(\ref{expr},\ref{expl}) contribute. Therefore we obtain in detail 
\begin{align}
\frac{\partial {\cal L}}{\partial\overline{\Psi}} = 
\frac{1}{2}\gamma_{\mu}\,i\partial^{\mu}\psi - m\Psi 
+ & \frac{1}{2}g_{\sigma} \sigma
\left[
	d^{(0)}\Psi+\nldr\Psi
\right]
 - \frac{1}{2}g_{\omega} \omega^{\mu}
\left[
	\gamma_{\mu}d^{(0)}\Psi+\gamma_{\mu}\nldr\Psi
\right]
- \frac{1}{2}g_{\rho} \rhovec^{\mu}
\left[
  \gamma_{\mu}\vec{\tau}\,d^{(0)}\Psi+
  \gamma_{\mu}\vec{\tau}\,\nldr\Psi
\right]
\nonumber\\
+ & \frac{1}{2}g_{\delta} \deltavec
\left[
	\tauvec\, d^{(0)} \Psi+\nldr\tauvec\Psi
\right]
\label{deriv0bar}
\,,
\end{align} 
and similar for the first-order derivative with respect to the 
spinor field $\Psi$
\begin{align}
\frac{\partial {\cal L}}{\partial\Psi} = 
-\frac{1}{2}\gamma_{\mu}\,i\partial^{\mu}\overline{\psi} - m\overline{\Psi} 
+ & \frac{1}{2}g_{\sigma}
\left[
	\overline{\Psi}\nldl\sigma+\overline{\Psi} d^{(0)} \sigma
\right]
- \frac{1}{2}g_{\omega} \omega^{\mu}
\left[
	\overline{\Psi}\nldl\gamma_{\mu}+
	\overline{\Psi}\gamma_{\mu} d^{(0)} 
\right]
- \frac{1}{2}g_{\rho} \rhovec^{\mu}
\left[
  \overline{\Psi}\nldl\gamma_{\mu}\vec{\tau}+
  \overline{\Psi}\gamma_{\mu}\vec{\tau}\,d^{(0)} 
\right]
\nonumber\\
+ & 
\frac{1}{2}g_{\delta} \deltavec
\left[
	\tauvec\overline{\Psi}\nldl+\tauvec\overline{\Psi} d^{(0)} 
\right]
\label{deriv0}
\,.
\end{align} 
Concerning the partial derivatives with respect to the first-order spinor fields 
$\partial_{\alpha}\overline{\Psi}$ and $\partial_{\alpha}\Psi$ only the 
first-order derivative terms in Eqs.~(\ref{expr},\ref{expl}) contribute 
and we get
\begin{align}
\frac{\partial {\cal L}}{\partial(\partial_{\alpha_{1}}\overline{\Psi})} = & 
-\frac{1}{2}\gamma^{\alpha_{1}}\,i\psi
+
\frac{1}{2}g_{\sigma}
d^{(1)} \, i\zeta^{\alpha_{1}}\, \Psi\sigma
 - \frac{1}{2}g_{\omega}
d^{(1)} \, i\zeta^{\alpha_{1}} \, \gamma^{\mu}\Psi\omega_{\mu}
 - \frac{1}{2}g_{\rho}
d^{(1)} \, i\zeta^{\alpha_{1}} \, \gamma^{\mu}\tauvec\Psi\rhovec_{\mu}
+
\frac{1}{2}g_{\delta}
d^{(1)} \, i\zeta^{\alpha_{1}}\, \tauvec\Psi\deltavec \,,
\label{deriv1bar}\\
\frac{\partial {\cal L}}{\partial(\partial_{\alpha_1}\Psi)} = & 
\frac{1}{2}i\overline{\psi}\gamma^{\alpha_{1}}
-
\frac{1}{2}g_{\sigma}
\sigma\overline{\Psi}\, i\zeta^{\alpha_{1}} \, d^{(1)}
 + \frac{1}{2}g_{\omega}
\omega_{\mu}\overline{\Psi}\gamma^{\mu} \, i\zeta^{\alpha_{1}} \, d^{(1)}
 + \frac{1}{2}g_{\rho}
\rhovec_{\mu}\overline{\Psi}\tauvec\gamma^{\mu} \, i\zeta^{\alpha_{1}} \, d^{(1)}
-
\frac{1}{2}g_{\delta}
\deltavec\tauvec\overline{\Psi} \, i\zeta^{\alpha_{1}} \, d^{(1)}
\label{deriv1}
\,.
\end{align} 
In a similar way as above one evaluates the derivatives of the Lagrangian density 
with respect to the second-order partial derivatives of the Dirac spinors. In 
this case obviously the second-order derivative terms in 
Eqs.~(\ref{expr},\ref{expl}) are of relevance, and the result reads
\begin{align}
\frac{\partial {\cal L}}{\partial(\partial_{\alpha_{1}\alpha_{2}}\overline{\Psi})} = &
\frac{1}{2}g_{\sigma} \, 
d^{(2)}
\frac{1}{2!} \,
i\zeta^{\alpha_{1}} \,i\zeta^{\alpha_{2}} \, \Psi\sigma
 - \frac{1}{2}g_{\omega} \, 
d^{(2)}
\frac{1}{2!} \,
i\zeta^{\alpha_{1}} \, i\zeta^{\alpha_{2}} \, 
\gamma^{\mu}\Psi\omega_{\mu}
 - \frac{1}{2}g_{\rho} \, 
d^{(2)}
\frac{1}{2!} \,
i\zeta^{\alpha_{1}} \, i\zeta^{\alpha_{2}} \, 
\gamma^{\mu}\tauvec\Psi\rhovec_{\mu} 
\nonumber\\
+ & \frac{1}{2}g_{\delta} \, 
d^{(2)}
\frac{1}{2!} \,
i\zeta^{\alpha_{1}} \,i\zeta^{\alpha_{2}} \, \tauvec\Psi\deltavec \,,
\label{deriv2bar}\\
\frac{\partial {\cal L}}{\partial(\partial_{\alpha_{1}\alpha_{2}}\Psi)} = &
\frac{1}{2}g_{\sigma} \, 
\sigma\overline{\Psi}
\frac{1}{2!} \, 
i\zeta^{\alpha_{1}} \, i\zeta^{\alpha_{2}} \, d^{(2)}
 - \frac{1}{2}g_{\omega} \, 
\omega_{\mu}\overline{\Psi}\gamma^{\mu}
\frac{1}{2!} \, 
i\zeta^{\alpha_{1}} \, i\zeta^{\alpha_{2}} \, d^{(2)}
 - \frac{1}{2}g_{\rho} \, 
\rhovec_{\mu}\overline{\Psi}\tauvec\gamma^{\mu}
\frac{1}{2!} \, 
i\zeta^{\alpha_{1}} \, i\zeta^{\alpha_{2}} \, d^{(2)}
\nonumber\\
+ & \frac{1}{2}g_{\delta} \, 
\deltavec\tauvec\overline{\Psi}
\frac{1}{2!} \, 
i\zeta^{\alpha_{1}} \, i\zeta^{\alpha_{2}} \, d^{(2)}
\label{deriv2}
\,.
\end{align} 

With the intermediate results of this appendix we can now derive the relevant 
equations of the NLD model, {\it i.e.}, the Dirac-equation for the spinor field 
$\Psi$ in Appendix~\ref{app4} as well as the conserved Noether current 
and the energy-momentum tensor in Appendix~\ref{app5}.
Furthermore, we will perform these derivations 
up to second order in the higher-order fields and for the isoscalar meson-nucleon
interaction Lagrangians only, since the evaluation of higher-order terms and 
of the other meson-nucleon vertices proceeds in a similar way. We will insert 
then the remaining terms, \textit{i.e.}, the higher-order derivatives as well as all 
meson-nucleon contributions in the final expressions. 

In the following the
terms containing the derivatives of the meson fields are not shown, since,
they do not contribute on the mean-field level.
 
\section{\label{app4}Derivation of the Dirac-equation in the NLD model}

For the derivation of the Dirac-equation we start with the Euler-Lagrange 
equations of motion, Eq.~(\ref{Euler0}), which read as
\begin{align}
\frac{\partial{\cal L}}{\partial\varphi_{r}}
+
\sum_{i=1}^{n} 
(-)^{i}
\partial_{\alpha_{1}\cdots\alpha_{i}}
\frac{\partial{\cal L}}
{\partial(\partial_{\alpha_{1}\cdots\alpha_{i}}\varphi_{r})}
= 0
\;.
\end{align}
Up to second order in the partial derivatives of the spinor field $\overline{\Psi}$ 
they reduce to
\begin{align}
\frac{\partial{\cal L}}{\partial\overline{\Psi}}
-
 \partial_{\alpha_{1}}\frac{\partial{\cal L}}{\partial(\partial_{\alpha_{1}}\overline{\Psi})}
+
 \partial_{\alpha_{1}}\partial_{\alpha_{2}}
\frac{\partial{\cal L}}{\partial(\partial_{\alpha_{1}}\partial_{\alpha_{2}}\overline{\Psi})} = 0
\label{Euler2}
\,.
\end{align}
We insert the various partial field derivatives, 
Eqs.~(\ref{deriv0bar}),~(\ref{deriv1bar}) and~(\ref{deriv2bar}), into the second order 
Euler-Lagrange equations, Eq.~(\ref{Euler2}), and obtain
\begin{align}
\gamma_{\mu}\,i\partialr^{\mu}\psi - m\Psi 
+ & \frac{1}{2}g_{\sigma} \sigma
\left[
	d^{(0)}\Psi+\nldr\Psi
\right]
 - \frac{1}{2}g_{\omega} \omega^{\mu}
\left[
	d^{(0)}\gamma_{\mu}\Psi+\gamma_{\mu}\nldr\Psi
\right]
\nonumber\\
- &
\frac{1}{2}g_{\sigma} \, 
\sigma \, d^{(1)} \, \zeta^{\alpha_{1}}i\partialr_{\alpha_{1}} \, \Psi 
+ \frac{1}{2}g_{\omega} \, 
\omega_{\mu} \, d^{(1)} \, \zeta^{\alpha} \, \gamma^{\mu}i\partialr_{\alpha_{1}}\Psi
\nonumber\\
+ &
\frac{1}{2}g_{\sigma} \, \sigma \,
d^{(2)}\, 
\frac{1}{2!}\, 
\zeta^{\alpha_{1}} \, \zeta^{\alpha_{2}} \, 
i\partialr_{\alpha_{1}}\, i\partialr_{\alpha_{2}}
\Psi 
- \frac{1}{2}g_{\omega} \, 
d^{(2)} \, \omega_{\mu} \, 
\frac{1}{2!} \, 
\zeta^{\alpha_{1}} \, \zeta^{\alpha_{2}} \, 
\gamma^{\mu}
i\partialr_{\alpha_{1}}\, i\partialr_{\alpha_{2}}
\Psi
= 0
\label{dirac1_app}
\,.
\end{align}
We rewrite Eq.~(\ref{dirac1_app}) such to separate the series contributions 
from the standard terms and obtain following expression
\begin{align}
\left\lbrace
\gamma_{\mu}\,i\partialr^{\mu} - m 
+ 
\right. & \left. 
\frac{1}{2}g_{\sigma}\, \sigma\, \nldr 
- \frac{1}{2}g_{\omega}\, \gamma^{\mu}\omega_{\mu }\nldr
\right.
\label{dirac2_app}\\
+ & \left.
\frac{1}{2}g_{\sigma} \, \sigma
\left[
d^{(0)} - d^{(1)}\, \zeta^{\alpha_{1}}i\partialr_{\alpha_{1}} + 
\frac{1}{2!}\, d^{(2)}\, 
\zeta^{\alpha_{1}}i\partialr_{\alpha_{1}} \, 
\zeta^{\alpha_{2}}i\partialr_{\alpha_{2}}
\right]
\right.
\nonumber\\
- & \left.
\frac{1}{2}g_{\omega} \, \omega_{\mu}
\left[
d^{(0)} - d^{(1)}\, \zeta^{\alpha_{1}}i\partialr_{\alpha_{1}} + 
\frac{1}{2!}\, d^{(2)}\, 
\zeta^{\alpha_{1}}i\partialr_{\alpha_{1}} \, 
\zeta^{\alpha_{2}}i\partialr_{\alpha_{2}} 
\right]\gamma^{\mu}
\right\rbrace \Psi = 0
\nonumber
\,.
\end{align}
In fact, if one would perform above procedure for all higher-order terms, 
one would obtain 
\begin{align}
& 
\left\lbrace
\gamma_{\mu}\,i\partialr^{\mu} - m 
+ 
\frac{1}{2}g_{\sigma}\, \sigma\, \nldr 
- \frac{1}{2}g_{\omega}\, \gamma^{\mu}\omega_{\mu }\nldr
\right.
\label{resu1}\\
+ & \left.
\frac{1}{2}g_{\sigma} \, \sigma
\left[
d^{(0)} - d^{(1)}\, \zeta^{\alpha_{1}}i\partialr_{\alpha_{1}} + 
\frac{1}{2!}\, d^{(2)}\, 
\zeta^{\alpha_{1}}i\partialr_{\alpha_{1}} \, 
\zeta^{\alpha_{2}}i\partialr_{\alpha_{2}}
+ \cdots + 
(-)^{n} \, \frac{1}{n!} \, d^{(n)}\, 
\zeta^{\alpha_{1}}i\partialr_{\alpha_{1}} \,
\cdots \,
\zeta^{\alpha_{n}}i\partialr_{\alpha_{n}}
\right]
\right.
\nonumber\\
- & \left.
\frac{1}{2}g_{\omega} \, \omega_{\mu}
\left[
d^{(0)} - d^{(1)}\, \zeta^{\alpha_{1}}i\partialr_{\alpha_{1}} + 
\frac{1}{2!}\, d^{(2)}\, 
\zeta^{\alpha_{1}}i\partialr_{\alpha_{1}} \, 
\zeta^{\alpha_{2}}i\partialr_{\alpha_{2}}
+ \cdots + 
(-)^{n} \, \frac{1}{n!} \, d^{(n)}\, 
\zeta^{\alpha_{1}}i\partialr_{\alpha_{1}} \,
\cdots \,
\zeta^{\alpha_{n}}i\partialr_{\alpha_{n}}
\right]\gamma^{\mu}
\right\rbrace \Psi = 0 \,.
\nonumber
\end{align}
The infinite series inside the brackets in Eq.~(\ref{resu1}) 
add together with the non-linear terms 
in the first line of Eq.~(\ref{resu1}) for each meson-nucleon vertex. One arrives 
to the following Dirac equation for the spinor field $\Psi$ in the NLD model 
\begin{align}
\left[
\gamma_{\mu}\,i\partial^{\mu} 
- g_{\omega}\, \gamma^{\mu}\omega_{\mu }\nldr
- g_{\rho}\, \gamma^{\mu}\tauvec\rhovec_{\mu }\nldr
- m + g_{\sigma}\, \sigma\, \nldr + g_{\delta}\, \tauvec\deltavec\, \nldr 
\right]\Psi = 0
\,.
\end{align}
This is the desired result, Eq.~(\ref{Dirac_nld}). Again the terms containing
the derivatives of meson fields are not show in the above equation, since,
they will not contribute in the final RMF expressions.

\section{\label{app5}Derivation of the Noether current in the NLD model}

As in the derivation of the Dirac equation, we consider in the following only 
terms up to second order in the field derivatives, and 
use Eqs.~(\ref{deriv1bar}),~(\ref{deriv1}),~(\ref{deriv2bar}) and~(\ref{deriv2}). 
We start from the general expression in Eq.~(\ref{current}) for the nucleonic 
degrees of freedom which up to second order takes the following form
\begin{align}
J^{\mu} = -i \left\lbrace
\left[ 
  \frac{\partial{\cal L}}{\partial(\partial_{\mu}\Psi)} 
  - 
  \partial_{\beta}
  \frac{\partial{\cal L}}{\partial(\partial_{\mu}\partial_{\beta}\Psi)}
  \right]\Psi   
- \overline{\Psi}\left[ 
  \frac{\partial{\cal L}}{\partial(\partial_{\mu}\overline{\Psi})} 
  - 
  \partial_{\beta}
  \frac{\partial{\cal L}}{\partial(\partial_{\mu}\partial_{\beta}\overline{\Psi})}
  \right]          
+ \left[ 
  \frac{\partial{\cal L}}{\partial(\partial_{\mu}\partial_{\beta}\Psi)}
  \right]\partial_{\beta}\Psi
- \partial_{\beta}\overline{\Psi}\left[ 
  \frac{\partial{\cal L}}{\partial(\partial_{\mu}\partial_{\beta}\overline{\Psi})}
  \right]  \right\rbrace
\,.
 \label{strom-1}
\end{align}
We rewrite now Eq. (\ref{strom-1}) by separating the terms between the different orders 
in the partial derivatives (the order is labeled by a subscript)
\begin{align}
J_{\mu} = {\cal O}^{(1)}_{\mu} + {\cal O}^{(2)}_{\mu}
\label{sepapp}
\,.
\end{align}
The first-order contribution to Eq.~(\ref{sepapp}) reads as
\begin{eqnarray}
{\cal O}^{(1)}_{\mu} = -i
\left( 
	\frac{\partial{\cal L}}{\partial(\partial^{\mu}\Psi)}\Psi
	-
	\bar{\Psi}\frac{\partial{\cal L}}{\partial(\partial^{\mu}\bar{\Psi})}
\right)
\;, \label{1-order}
\end{eqnarray}
while the second-order contribution to Eq.~(\ref{sepapp}) takes the 
following form
\begin{align}
{\cal O}^{(2)}_{\mu} = -i\left[ - 
\left( 
	\partial_{\beta}
	\frac{\partial{\cal L}}{\partial(\partial^{\mu}\partial_{\beta}\Psi)}
	\Psi
	-
	\bar{\Psi}
	\partial_{\beta}
	\frac{\partial{\cal L}}{\partial(\partial^{\mu}\partial_{\beta}\bar{\Psi})}
\right) 
+
\left( 
	\frac{\partial{\cal L}}{\partial(\partial^{\mu}\partial_{\beta}\Psi)}
	\partial_{\beta}\Psi
	-
	\partial_{\beta}\bar{\Psi}
	\frac{\partial{\cal L}}{\partial(\partial^{\mu}\partial_{\beta}\bar{\Psi})}
\right) \right]
\;. \label{2-order}
\end{align}
For the evaluation of the first-order contribution ${\cal O}^{(1)}_{\mu}$, 
Eq.~(\ref{1-order}),  we insert Eq.~(\ref{deriv1}) and its adjoint 
form, Eq.~(\ref{deriv1bar}), into Eq.~(\ref{1-order}), and obtain 
\begin{align}
{\cal O}^{(1)}_{\mu} = \overline{\Psi}\gamma_{\mu}\Psi 
- 
\frac{1}{2} \, g_{\sigma}
\left[
\sigma\overline{\Psi} \, \zeta_{\mu} \, d^{(1)} \, \Psi
+ 
\overline{\Psi} \, d^{(1)} \, \zeta_{\mu} \, \Psi \, \sigma
\right]
+ 
\frac{1}{2} g_{\omega}
\left[
\omega_{\alpha}\overline{\Psi}\gamma^{\alpha}\zeta_{\mu} \, d^{(1)} \, \Psi
+ 
\overline{\Psi} \, d^{(1)}\,\zeta_{\mu} \, \gamma^{\alpha}\Psi\omega_{\alpha}
\right]
\label{O-Eins}
\,.
\end{align}
The derivation of the second-order contribution ${\cal O}^{(2)}_{\mu}$, 
Eq.~(\ref{2-order}), proceeds in a similar way. We get
\begin{align}
{\cal O}^{(2)}_{\mu} = & 
- \frac{1}{2}\,g_{\sigma}\,
\sigma\,\overline{\Psi}\,
i\!\derivl_{\beta}\! \frac{1}{2!} \, \zeta_{\mu}\,\zeta^{\beta}\,
d^{(2)}\,
\Psi
+ \frac{1}{2}\,g_{\sigma}\,
\overline{\Psi}\,d^{(2)}
\! \frac{1}{2!} \, \zeta_{\mu}\,\zeta^{\beta}\,
i\!\derivr_{\beta}\,
\Psi\,\sigma
\nonumber\\
& +  
\frac{1}{2}\,g_{\sigma}\,
\sigma \,\overline{\Psi}
\,\frac{1}{2!} \, \zeta_{\mu}\,\zeta^{\beta}\,
d^{(2)}\, i\!\derivr_{\beta}\,
\Psi
- \frac{1}{2}\,g_{\sigma}\,
\overline{\Psi}\,d^{(2)}\,
i\!\derivl_{\beta}\! \frac{1}{2!} \, \zeta_{\mu}\,\zeta^{\beta}\,
\Psi\,\sigma
\nonumber\\
& + \frac{1}{2}\,g_{\omega}\,
\omega_{\delta}\,\overline{\Psi}\,
i\!\derivl_{\beta}\! \frac{1}{2!} \, \zeta_{\mu}\,\zeta^{\beta}\,
\gamma^{\delta} \, d^{(2)}\,
\Psi
- \frac{1}{2}\,g_{\omega}\,
\overline{\Psi}\, d^{(2)}
\! \frac{1}{2!} \, \zeta_{\mu} \,\zeta^{\beta} \,
\gamma^{\delta}\,
i\!\derivr_{\beta}\,
\Psi\,\omega_{\delta}
\nonumber\\
& -  
\frac{1}{2}\,g_{\omega}\,
\omega_{\delta} \,\overline{\Psi}
\, \gamma^{\delta} \,
\frac{1}{2!} \, \zeta_{\mu} \, \zeta^{\beta} \,
d^{(2)}\,i\!\derivr_{\beta}\,
\Psi
+ \frac{1}{2}\,g_{\omega}\,
\overline{\Psi}\, d^{(2)}\,
i\!\derivl_{\beta}\! \frac{1}{2!} \, \zeta_{\mu} \, \zeta^{\beta} \,
\gamma^{\delta}\,\Psi\,\omega_{\delta}
\,.
\label{O-Zwei-step1}
\end{align}
For each isoscalar meson-nucleon interaction we obtain now $4$ terms, which differ 
between each other in the direction where the partial derivative operators act.
We arrive to the following expression
\begin{align}
{\cal O}^{(2)}_{\mu} = & 
+ \frac{1}{2}\,g_{\sigma}\,
\sigma \,\overline{\Psi}
\,\frac{2}{2!} \, \zeta_{\mu} \, \zeta^{\beta} \,
d^{(2)}\, i\!\derivr_{\beta}\,
\Psi
- \frac{1}{2}\,g_{\sigma}\,
\overline{\Psi}\, d^{(2)}\,
i\!\derivl_{\beta}\! \frac{2}{2!} \, \zeta_{\mu} \, \zeta^{\beta} \,
\Psi\,\sigma
\nonumber\\
& -  \frac{1}{2}\,g_{\omega}\,
\omega_{\delta} \,\overline{\Psi}
\, \gamma^{\delta} \,
\frac{2}{2!} \, \zeta_{\mu} \, \zeta^{\beta} \,
d^{(2)}\,i\!\derivr_{\beta}\,
\Psi
+ \frac{1}{2}\,g_{\omega}\,
\overline{\Psi}\, d^{(2)}\,
i\!\derivl_{\beta}\! \frac{2}{2!} \, \zeta_{\mu} \, \zeta^{\beta} \,
\gamma^{\delta}\,\Psi\,\omega_{\delta} 
\label{O-Zwei-step2} \,.
\end{align}
The procedure is similar for the remaining higher-order derivative contributions. 
The evaluation procedure according 
Eqs.~(\ref{O-Zwei-step1}) and~(\ref{O-Zwei-step2}) for the third-order 
contribution, ${\cal O}^{(3)}_{\mu}$, would result in three terms for each vertex, 
for the fourth-order term, ${\cal O}^{(4)}_{\mu}$, in four terms for 
each vertex, and so forth. 

Therefore, the resummation of all higher-order terms according 
\begin{align}
J_{\mu} = {\cal O}^{(1)}_{\mu} + {\cal O}^{(2)}_{\mu} + {\cal O}^{(3)}_{\mu} 
+ \cdots + {\cal O}^{(n)}_{\mu}
\label{sepinf}
\end{align}
leads to infinite series for each meson-nucleon interaction. For instance, for 
the scalar-isoscalar meson-nucleon interaction we get
\begin{align}
-&  \frac{1}{2}\, g_{\sigma}\, 
\overline{\Psi}\left[
\frac{1}{1!}\, d^{(1)}\, \zeta^{\mu} + 
\frac{2}{2!}\, d^{(2)}\, 
i\partiall_{\alpha_{1}} \, \zeta^{\alpha_{1}}\, \zeta^{\mu}
+ \cdots + 
\frac{n}{n!}\, d^{(n)}\, 
i\partiall_{\alpha_{1}}\, \zeta^{\alpha_{1}}
i\partiall_{\alpha_{2}}\, \zeta^{\alpha_{2}} \cdots
i\partiall_{\alpha_{n-1}}\, \zeta^{\alpha_{n-1}}
\, \zeta^{\mu}
\right]\Psi
\nonumber\\
+& \frac{1}{2}\, g_{\sigma}\, 
\overline{\Psi}\left[
-\frac{1}{1!}\, \zeta^{\mu} \, d^{(1)} + 
\frac{2}{2!}\, \zeta^{\mu}
\, d^{(2)}\, \zeta^{\alpha_{1}} \, i\partialr_{\alpha_{1}} 
+ \cdots + 
(-)^{n}\, \frac{n}{n!}\, \zeta^{\mu} \, 
d^{(n)}\, 
\zeta^{\alpha_{1}}   \, i\partialr_{\alpha_{1}}\, 
\zeta^{\alpha_{2}}   \, i\partialr_{\alpha_{2}}\, \cdots
\zeta^{\alpha_{n-1}} \, i\partialr_{\alpha_{n-1}}
\right]\Psi
\label{exa1}
\,.
\end{align}
Note that the $n$-th term, ${\cal O}^{(n)}_{\mu}$, contains $(n-1)$ partial 
derivatives and that it appears $n$-times. These terms can be also 
resummed to infinite series. Indeed, by considering, for instance, 
the Taylor-expansion of the operator $\nldr$
\begin{align}
\nldr =  1 - d^{(1)} \, 
\zeta_{\alpha}\, i\partialr^{\alpha} + 
\frac{1}{2!} \, 
d^{(2)} 
\, \zeta_{\alpha}\, i\partialr^{\alpha} \,
   \zeta_{\beta}\,  i\partialr^{\beta} 
 + \cdots
\label{taylor}
\,, 
\end{align}
we obtain for the derivative of $\nldr$ with respect to the operator-like 
argument $i\partialr^{\mu}$
\begin{align}
\calor^{\mu} := 
\frac{\partial\nldr}{\partial(i\partialr_{\mu})} = 
-d^{(1)} \, \zeta^{\mu} + \frac{2}{2!}\, d^{(2)}\, \zeta^{\mu} \, 
\zeta_{\beta}\, i\partialr^{\beta} 
 + \cdots
\label{dtaylor}
\,. 
\end{align}
In a similar way we obtain $\calol^{\mu}=(\calor^{\mu})^{\dagger}$. As in the case 
of the scalar operators $\nldr$ and $\nldl$, the new vector operators 
$\calor^{\mu}$ and $\calol^{\mu}$ act from the right side to the spinor 
field $\Psi$ and from the left 
side to $\overline{\Psi}$, respectively. Below we will see that in the RMF 
approximation to nuclear matter both vector operators will give the derivative of 
the scalar operator $\nld$ with respect to the single-particle $4$-momentum $p^{\mu}$. 

Collecting now all the contributions from Eqs.~(\ref{O-Eins}) 
and~(\ref{O-Zwei-step2}) under consideration of Eq.~(\ref{dtaylor}) we arrive to 
compact forms, \textit{e.g.}, the scalar-isoscalar meson-nucleon vertex from 
Eq.~(\ref{exa1}) can be resummed as follows
\begin{align}
-&  \frac{1}{2}\, g_{\sigma}\, 
\overline{\Psi}\left[
\frac{1}{1!}\, d^{(1)}\, \zeta^{\mu} + 
\frac{2}{2!}\, d^{(2)}\, 
i\partiall_{\alpha_{1}} \, \zeta^{\alpha_{1}}\, \zeta^{\mu}
+ \cdots + 
\frac{n}{n!}\, d^{(n)}\, 
i\partiall_{\alpha_{1}}\, \zeta^{\alpha_{1}}
i\partiall_{\alpha_{2}}\, \zeta^{\alpha_{2}} \cdots
i\partiall_{\alpha_{n-1}}\, \zeta^{\alpha_{n-1}}
\, \zeta^{\mu}
\right]\Psi
\nonumber\\
+& \frac{1}{2}\, g_{\sigma}\, 
\overline{\Psi}\left[
-\frac{1}{1!}\, \zeta^{\mu} \, d^{(1)} + 
\frac{2}{2!}\, \zeta^{\mu}
\, d^{(2)}\, \zeta^{\alpha_{1}} \, i\partialr_{\alpha_{1}} 
+ \cdots + 
(-)^{n}\, \frac{n}{n!}\, \zeta^{\mu} \, 
d^{(n)}\, 
\zeta^{\alpha_{1}}   \, i\partialr_{\alpha_{1}}\, 
\zeta^{\alpha_{2}}   \, i\partialr_{\alpha_{2}}\, \cdots
\zeta^{\alpha_{n-1}} \, i\partialr_{\alpha_{n-1}}
\right]\Psi
\nonumber\\
= & 
\frac{1}{2}\, g_{\sigma}\, \sigma\,
\overline{\Psi}\, \calor^{\mu}\, \Psi
-\frac{1}{2}\, g_{\sigma}\, 
\overline{\Psi}\, \calol^{\mu} \, \Psi\sigma
\label{exa}
\,.
\end{align}
The evaluation method for the isovector channels of the NLD interaction Lagrangian proceeds 
in the same way. In total, including all degrees of freedom we obtain following compact 
expression for the conserved baryon current within the NLD formalism
\begin{align}
J^{\mu} = \overline{\Psi}\gamma^{\mu}\Psi 
- & \frac{1}{2}\, g_{\sigma}\, 
\left[
\overline{\Psi}\, \calol^{\mu} \Psi\sigma - \sigma\overline{\Psi}\, \calor^{\mu}\Psi
\right]
+ \frac{1}{2}\, g_{\omega}\, 
\left[
\overline{\Psi}\, \calol^{\mu} \gamma^{\alpha}\Psi\omega_{\alpha} - 
\omega_{\alpha}\overline{\Psi}\gamma^{\alpha}\, \calor^{\mu}\Psi
\right]
\nonumber\\
+ & \frac{1}{2}\, g_{\rho}\, 
\left[
\overline{\Psi}\, \calol^{\mu} \gamma^{\alpha}\tauvec\Psi\rhovec_{\alpha} - 
\rhovec_{\alpha}
\overline{\Psi}\tauvec\gamma^{\alpha}\, \calor^{\mu}\Psi
\right]
- \frac{1}{2}\, g_{\delta}\, 
\left[
\overline{\Psi}\, \calol^{\mu} \tauvec \Psi\deltavec 
- \deltavec\,\overline{\Psi}\, \tauvec\, \calor^{\mu}\Psi
\right]
\label{strom}
\,,
\end{align}
which obeys the continuity equation $\partial_{\mu}\, J^{\mu}=0$.

Now we apply the RMF approximation to the general expression 
for the Noether current, Eq.~(\ref{strom}). All bilinear operator-like terms 
are replaced by their expectation values relative to the nuclear matter ground state. 
Furthermore, we use for the spinor field the plane-wave \textit{ansatz}, 
Eq.~(\ref{plane_wave}), in order to evaluate the operators $\nldr,~\nldl$, 
$\calor^{\mu}$ and $\calol^{\mu}$. Taking also into account the equations  $i\partialr^{\mu}\Psi=p^{\mu}\Psi$ and 
$\overline{\Psi}i\partiall^{\mu}=-\overline{\Psi}p^{\mu}$ the 
current $J^{\mu}$, Eq.~(\ref{strom}), takes following form in the RMF approximation
\begin{align}
J^{\mu} = \langle \overline{\Psi}\gamma^{\mu}\Psi \rangle
+ g_{\sigma}\, 
\langle\overline{\Psi}\, \big( \partial_{p}^{\mu}\nld\big) \Psi\rangle\sigma 
- g_{\omega}\, 
\langle
\overline{\Psi}\, 
\big( \partial_{p}^{\mu}\nld\big) \gamma^{\alpha}\Psi\rangle\omega_{\alpha} 
- g_{\rho}\, 
\langle\overline{\Psi}\, 
\big( \partial_{p}^{\mu}\nld\big) \gamma^{\alpha}\tauvec\Psi\rangle
\rhovec_{\alpha} 
+ g_{\delta}\, 
\langle\overline{\Psi}\, \big( \partial_{p}^{\mu}\nld\big)\tauvec \Psi\rangle\deltavec
\label{strom_nm}
\,,
\end{align}
with $\partial_{p}^{\mu}=\frac{\partial}{\partial p_{\mu}}$. This is 
the desired result, see Eq.~(\ref{current_NLD}). 

The derivation of the energy-momentum tensor proceeds in the same way as for 
the current, therefore, we skip further derivations. 

\end{widetext}


\end{appendix}


\end{document}